\documentclass[reprint,onecolumn,notitlepage,preprintnumbers,nofootinbib,amsmath,amssymb,aps,prd,floatfix,superscriptaddress,longbibliography,svgnames]{revtex4-1}

\usepackage{amsmath}
\usepackage{amsfonts}
\usepackage{amssymb}
\usepackage{graphicx}
\usepackage{bm}
\usepackage{subfigure}
\usepackage{url}
\usepackage[hyperindex]{hyperref}
\usepackage{color}
\usepackage[ddmmyy,24hr]{datetime}
\usepackage{bigdelim}
\usepackage{booktabs}
\usepackage{dcolumn}
\usepackage{multirow}
\usepackage{subfigure}
\usepackage{cancel}
\usepackage{stackrel}
\usepackage{paralist}
\usepackage{xspace}
\usepackage{slashed}
\usepackage{cancel}
\usepackage{todonotes}
\usepackage{enumerate}
\usepackage{float}
\usepackage{fullpage}
\usepackage{ulem}
\newcommand{\nua}[1]{\ensuremath{\rlap{\kern-2.5pt\ensuremath{\overset{\scriptscriptstyle(-)}{\phantom{\nu}}}}{\ensuremath{{\nu}_{#1}}}}}

\usepackage{enumitem}

\newcommand{\cenns}{CE$\nu$NS\xspace}

\begin{document}
\title{Impact of the Dresden-II and COHERENT neutrino scattering data on neutrino electromagnetic properties and electroweak physics}
%\title{Neutrino Electromagnetic Properties from COHERENT Elastic Neutrino Nucleus Scattering}

\author{M. Atzori Corona}
\email{mattia.atzori.corona@ca.infn.it}
\affiliation{Dipartimento di Fisica, Universit\`{a} degli Studi di Cagliari,
	Complesso Universitario di Monserrato - S.P. per Sestu Km 0.700,
	09042 Monserrato (Cagliari), Italy}
\affiliation{Istituto Nazionale di Fisica Nucleare (INFN), Sezione di Cagliari,
	Complesso Universitario di Monserrato - S.P. per Sestu Km 0.700,
	09042 Monserrato (Cagliari), Italy}

\author{M. Cadeddu}
\email{matteo.cadeddu@ca.infn.it}
\affiliation{Istituto Nazionale di Fisica Nucleare (INFN), Sezione di Cagliari,
	Complesso Universitario di Monserrato - S.P. per Sestu Km 0.700,
	09042 Monserrato (Cagliari), Italy}

\author{N. Cargioli}
\email{nicola.cargioli@ca.infn.it}
\affiliation{Dipartimento di Fisica, Universit\`{a} degli Studi di Cagliari,
	Complesso Universitario di Monserrato - S.P. per Sestu Km 0.700,
	09042 Monserrato (Cagliari), Italy}
\affiliation{Istituto Nazionale di Fisica Nucleare (INFN), Sezione di Cagliari,
	Complesso Universitario di Monserrato - S.P. per Sestu Km 0.700,
	09042 Monserrato (Cagliari), Italy}

\author{F. Dordei}
\email{francesca.dordei@cern.ch}
\affiliation{Istituto Nazionale di Fisica Nucleare (INFN), Sezione di Cagliari,
	Complesso Universitario di Monserrato - S.P. per Sestu Km 0.700,
	09042 Monserrato (Cagliari), Italy}

\author{C. Giunti}
\email{carlo.giunti@to.infn.it}
\affiliation{Istituto Nazionale di Fisica Nucleare (INFN), Sezione di Torino, Via P. Giuria 1, I--10125 Torino, Italy}

\author{Y.F. Li}
\email{liyufeng@ihep.ac.cn}
\affiliation{Institute of High Energy Physics,
	Chinese Academy of Sciences, Beijing 100049, China}
\affiliation{School of Physical Sciences, University of Chinese Academy of Sciences, Beijing 100049, China}

\author{C. A. Ternes}
\email{ternes@to.infn.it}
\affiliation{Istituto Nazionale di Fisica Nucleare (INFN), Sezione di Torino, Via P. Giuria 1, I--10125 Torino, Italy}

\author{Y.Y. Zhang}
\email{zhangyiyu@ihep.ac.cn}
\affiliation{Institute of High Energy Physics,
	Chinese Academy of Sciences, Beijing 100049, China}
\affiliation{School of Physical Sciences, University of Chinese Academy of Sciences, Beijing 100049, China}

\date{\dayofweekname{\day}{\month}{\year} \ddmmyydate\today, \currenttime}

\begin{abstract}
Coherent elastic neutrino-nucleus scattering (\cenns) represents a powerful tool to investigate key electroweak physics parameters and neutrino properties since its first observation in 2017 by the COHERENT experiment exploiting the spallation neutron source at Oak Ridge National Laboratory. In light of the recent detection of such a process with antineutrinos produced by the Dresden-II reactor scattering off a germanium detector, we revisit the limits so far set on the neutrino magnetic moments, charge radii and millicharges as well as on the weak mixing angle. In order to do so, we also include the contribution of elastic neutrino-electron scattering, whose effect becomes non negligible in some beyond the Standard Model theories. By using different hypotheses for the germanium quenching factor and the reactor antineutrino flux, we provide a measurement of the weak mixing angle at the low-energy scale of the Dresden-II reactor experiment and, thanks to a combined analysis with the latest cesium iodide and argon data set released by the COHERENT Collaboration, we deliver updated limits for the neutrino electromagnetic properties. Interestingly, we are able to set a new best upper limit on the electron neutrino charge radius and significantly improve the other \cenns-related limits on the neutrino electric charge and magnetic moment.
\end{abstract}

\maketitle  
%\newpage

\section{Introduction}

Until recently, coherent elastic neutrino-nucleus scattering (\cenns) has been observed only exploiting neutrinos coming from the spallation neutron source (SNS) at the Oak Ridge Nation Laboratory by the COHERENT Collaboration~\cite{Akimov:2017ade}. Indeed, by making use of neutrinos produced by pion-decay-at-rest ($\pi$DAR) at the SNS, the \cenns process has been observed in 2017 using cesium-iodide (CsI)~\cite{Akimov:2017ade,Akimov:2018vzs} as well as in argon (Ar) in 2020~\cite{COHERENT:2020iec,COHERENT:2020ybo}. The CsI analysis has also been updated in 2021 with a refined quenching factor (QF) determination and more statistics~\cite{Akimov:2021dab}.
The \cenns process is a pure neutral current interaction which happens when low energy neutrinos elastically scatter off atomic nuclei with a small momentum transfer between the incoming neutrino and the target nucleus, such that the neutrino interacts coherently with the entire nucleus~\cite{Freedman:1973yd}. When this happens, the cross section becomes roughly proportional to the square of the number of neutrons participating in the interaction.
The \cenns process proved to be a powerful tool to test new physics interactions beyond the Standard Model (SM)~\cite{Coloma:2017ncl,Liao:2017uzy,Lindner:2016wff,Giunti:2019xpr,Denton:2018xmq,AristizabalSierra:2018eqm,CONUS:2022qbb} as well as to perform stringent tests of nuclear physics, astrophysics, neutrino properties and electroweak interactions~\cite{Cadeddu:2020lky,Miranda:2020tif,Cadeddu:2020nbr,Cadeddu:2021ijh,Banerjee:2021laz, Cadeddu:2017etk,Papoulias:2019lfi,Coloma:2017ncl,Lindner:2016wff,Giunti:2019xpr,Denton:2018xmq,AristizabalSierra:2018eqm,Cadeddu:2018dux,Papoulias:2017qdn,Cadeddu:2019eta,Papoulias:2019txv,Khan:2019cvi,Dutta:2019eml,AristizabalSierra:2018eqm,Cadeddu:2018izq,Dutta:2019nbn,Abdullah:2018ykz,Ge:2017mcq,Miranda:2021kre,Flores:2020lji}.

An alternative source of neutrinos to $\pi$DAR are antineutrinos produced at nuclear power reactors. As continuous and well-localized sources, they offer the advantage of very intense fluxes of low-energy antineutrinos ($E<10$~MeV), with the drawback of a larger background that cannot be removed exploiting the pulsed feature of sources like the SNS. Due to the increased experimental challenge, the CONNIE~\cite{CONNIE:2021ngo} and CONUS~\cite{CONUS:2020skt,CONUS:2021dwh} Collaborations have only managed to put stringent limits on \cenns observation with reactor antineutrinos. However, recently a tantalizing evidence of \cenns using reactor antineutrinos has been reported in Ref.~\cite{Colaresi:2022obx} using an ultra-low noise 2.924 kg p-type point-contact germanium detector, called NCC-1701, located 10.39~meters away from the Dresden-II boiling water reactor. The data released corresponds to 96.4 days of effective exposure. Thanks to the much lower energy of reactor antineutrinos and the low energy threshold of such a detector, namely 0.2 $\mathrm{keV_{ee}}$, these data provide complementary information with respect to $\pi$DAR sources, with negligible dependence on the neutron distribution inside the target nucleus. This feature makes the bounds extracted using reactor antineutrinos robust against possible variations of the neutron distribution root mean square (rms) radius, that is experimentally poorly known, with the drawback that no information on the latter can be extracted~\cite{Cadeddu:2021ijh}.

In this paper, we analyse the new Dresden-II data, revisiting the limits so far set using \cenns on the neutrino magnetic moments, charge radii and millicharges as well as on the weak mixing angle. In order to do so, we also perform a combined analysis with the latest CsI and Ar data set released by the COHERENT Collaboration, using different hypotheses for the germanium quenching factor and the reactor antineutrino flux. We will also introduce the contribution of the elastic neutrino-electron scattering, that is observed to be non negligible when some neutrino electromagnetic properties beyond the SM (BSM) are taken into account, namely for the electric charges and magnetic moments, whose contributions are significantly enhanced at low recoil energies.

The paper is organized as follows.
In Sec.~\ref{sec:cs}, we present the theoretical framework with particular emphasis on the influence of neutrino electromagnetic properties and the weak mixing angle on the \cenns cross section and we will discuss the effect of elastic neutrino-electron scattering on the constraints presented in this work. In Sec.~\ref{sec:method} the methods and inputs used for the data analysis are described. In Sec.~\ref{sec:result}, the combined constraints of the Dresden-II data with the COHERENT CsI and Ar data set are presented.
Finally, we draw our conclusions in Sec.~\ref{sec:conclusions}.

\section{Theoretical framework}
\label{sec:cs}

In this section, the \cenns differential cross section in the SM will be introduced, together with the modifications necessary to include the contribution of possible neutrino charge radii, electric charges and magnetic moments. Moreover, we will also briefly summarize the phenomenology behind the elastic neutrino-electron scattering. \\

The \cenns differential cross section as a function of the nuclear kinetic recoil energy $T_\mathrm{nr}$ for a neutrino $\nu_{\ell}$ ($\ell=e,\mu,\tau$) that scatters off a nucleus $\mathcal{N}$ is given by~\cite{Drukier:1984vhf,Barranco:2005yy,Patton:2012jr}
\begin{equation}
	\dfrac{d\sigma_{\nu_{\ell}\text{-}\mathcal{N}}}{d T_\mathrm{nr}}
	(E,T_\mathrm{nr})
	=
	\dfrac{G_{\text{F}}^2 M}{\pi}
	\left( 1 - \dfrac{M T_\mathrm{nr}}{2 E^2} \right)
	(Q^{V}_{\ell, \mathrm{SM}})^2,
	\label{cs-std}
\end{equation}
where $G_{\text{F}}$ is the Fermi constant, $E$ is the neutrino energy, $M$ the nuclear mass, and
\begin{equation}\label{Qsm}
	Q^{V}_{\ell, \mathrm{SM}}=\left[g_{V}^{p}\left(\nu_{\ell}\right) Z F_{Z}\left(|\vec{q}|^{2}\right)+g_{V}^{n} N F_{N}\left(|\vec{q}|^{2}\right)\right]^{}
\end{equation}
is the weak charge of the nucleus. Here, $Z$ and $N$ are the numbers of protons and neutrons in the nucleus, respectively.
In this analysis, we set $(Z,N)_{\mathrm{Cs}}=(55,78)$, $(Z,N)_{\mathrm{I}}=(53,74)$ and $(Z,N)_{\mathrm{Ar}}=(18,22)$. Actually, one should consider that atmospheric argon is contaminated by a small percentage of $^{36}\mathrm{Ar}$ and $^{38}\mathrm{Ar}$, namely $\textit{f}(^{36}\mathrm{Ar}) = 0.33\%$ and $\textit{f}(^{38}\mathrm{Ar}) = 0.06\%$~\cite{DarkSide-20k:2021nia}. However, since the amount of these contaminants is very small and the uncertainties are large, in practice one gets the same results considering $\textit{f}(^{40}\mathrm{Ar}) = 100\%$. For Ge we use $(Z,N)_{\mathrm{{}^{70, 72, 73, 74, 76}Ge}}=(32,(38,40,41,42,44))$ with the corresponding natural abundances of 0.2057, 0.2745, 0.0775, 0.3650, 0.0773~\cite{BerglundWieser+2011+397+410}.
 The neutrino-proton, $g_{V}^{p}$, and neutrino-neutron, $g_{V}^{n}$, vector couplings correspond to $g_{V}^{p}(\nu_{e}) = 0.0382$, $g_{V}^{p}(\nu_{\mu}) = 0.0300$ and $g_{V}^{n} = -0.5117$, when taking into account radiative corrections in the $\overline{\mathrm{MS}}$ scheme~\cite{Cadeddu:2020lky,Erler:2013xha,ParticleDataGroup:2020ssz}.
The proton, $F_{Z}\left(|\vec{q}|^{2}\right)$, and neutron, $F_{N}\left(|\vec{q}|^{2}\right)$, nuclear form factors represent the Fourier transforms of the corresponding nucleon distribution in the nucleus and describe the loss of coherence for large values of the momentum transfer $|\vec{q}|$.
We use an analytic expression, namely the Helm parameterization~\cite{Helm:1956zz}, for the form factors, that is practically equivalent to the other two well known parameterizations, i.e., the symmetrized Fermi~\cite{Piekarewicz:2016vbn} and Klein-Nystrand~\cite{Klein:1999qj} ones. However, it is important to note that while the form factors are key ingredients in the analysis of COHERENT data, in the energy window of the Dresden-II experiment the form factor of both protons and neutrons is practically equal to unity, making the particular choice of the parameterization completely insignificant.
The proton rms radii can be obtained from the muonic atom spectroscopy and electron scattering data~\cite{Fricke:1995zz,Angeli:2013epw,Fricke2004} as explained in Ref.~\cite{Cadeddu:2020lky}, and correspond to
\begin{align}\label{Rp}
	R_{p}(\mathrm{Cs})=4.821(5)~\mathrm{fm}, \quad R_{p}(\mathrm{I})=4.766(8) ~\mathrm{fm},
	\quad
	R_{p}(\mathrm{Ar})=3.448(2)~\mathrm{fm}, \quad R_{p}(\mathrm{Ge})=4.073(1) ~\mathrm{fm}.
\end{align}
On the other hand,
there is poor knowledge of the values of the ${}^{133}\text{Cs}$, ${}^{127}\text{I}$, ${}^{40}\text{Ar}$, and $\text{Ge}$
neutron rms radii using electroweak probes~\cite{Cadeddu:2017etk,Papoulias:2019lfi,Cadeddu:2018dux,Huang:2019ene,Papoulias:2019txv,Khan:2019cvi,Cadeddu:2020lky,Cadeddu:2019eta,Payne:2019wvy}.
The values of these neutron rms radii can, however, be estimated with theoretical calculations
based on different nuclear models~\cite{Hoferichter:2020osn,Cadeddu:2020lky,Cadeddu:2021ijh}.
Here, we consider the following values obtained from the recent nuclear shell model
estimate of the corresponding neutron skins
(i.e. the differences between the neutron and the proton rms radii)
in Ref.~\cite{Hoferichter:2020osn}
\begin{align}\label{Rn}
	R_{n}(^{133}\mathrm{Cs})\simeq5.09~\mathrm{fm}, \quad\quad R_{n}(^{127}\mathrm{I})\simeq5.03~\mathrm{fm},\quad\quad
	R_{n}(^{40}\mathrm{Ar})\simeq3.55~\mathrm{fm}, \quad\quad R_{n}(\mathrm{Ge})\simeq4.15\mathrm{-}4.28~\mathrm{fm},
\end{align}
where for Ge a neutron skin of 0.08-0.17 fm has been considered~\cite{Hoferichter:2020osn}. 
Concerning the COHERENT data~\cite{Akimov:2021dab,COHERENT:2020iec,COHERENT:2020ybo},
we take into account the effect of the uncertainty of the values of the neutron rms radii by considering
3.4\% and 2\% uncertainties for the CsI and Ar \cenns rates, respectively.

\subsection{Neutrino-electron elastic scattering}
\label{sec:essec}

Neutrino-electron elastic scattering (ES) is a concurrent process to \cenns. In the SM, its contribution to the total event rate at low recoil energies is very small and thus it is usually neglected in \cenns analyses. However, in certain BSM scenarios the ES contribution could increase significantly, making it important to include it since stronger constraints can be obtained~\cite{Coloma:2022avw}. For the Ar data set, the COHERENT Collaboration performed a selection exploiting the so-called $f_{90}$ parameter~\cite{COHERENT:2020iec,COHERENT:2020ybo}, namely the ratio between the integrated photomultiplier amplitude in the first 90 ns with respect to the total amplitude. This parameter permits to perform strong pulse shape discrimination between nuclear recoils due to \cenns and electron recoils due to ES, such that the latter contribution becomes completely negligible. However, there is no similar feature that can be exploited in the COHERENT CsI data set, nor in the Dresden-II one, making it important to fit also for the ES contribution.

The SM neutrino-electron elastic scattering cross section per atom $\mathcal{A}$ is obtained multiplying the ES cross section per electron with the effective electron charge of the target atom $Z_{\text{eff}}^{\mathcal{A}}(T_{e})$~\cite{Coloma:2022avw}, namely
\begin{equation}
\dfrac{d\sigma_{\nu_{\ell}-\mathcal{A}}^{\text{ES}}}{d T_{\text{e}}}
(E,T_{\text{e}})
=
%\sum_{i=1}^{Z}
Z_{\text{eff}}^{\mathcal{A}}(T_{e})
\,
\dfrac{G_{\text{F}}^2 m_{e}}{2\pi}
\left[
\left( g_{V}^{\nu_{\ell}} + g_{A}^{\nu_{\ell}} \right)^2
+
\left( g_{V}^{\nu_{\ell}} - g_{A}^{\nu_{\ell}} \right)^2
\left( 1 - \dfrac{T_{e}}{E} \right)^2
-
\left( (g_{V}^{\nu_{\ell}})^2 - (g_{A}^{\nu_{\ell}})^2 \right)
\dfrac{m_{e} T_{e}}{E^2}
\right]
,
\label{ES-SM}
\end{equation}
where $m_{e}$ is the electron mass, $T_e$ is the electron recoil energy, and the neutrino-flavour dependent electron couplings are
\begin{align}
\null & \null
g_{V}^{\nu_{e}}
=
2\sin^{2} \theta_{W} + 1/2
,
\quad
\null && \null
g_{A}^{\nu_{e}}
=
1/2
,
\label{gnue}
\\
\null & \null
g_{V}^{\nu_{\mu,\tau}}
=
2\sin^{2} \theta_{W} - 1/2
,
\quad
\null && \null
g_{A}^{\nu_{\mu,\tau}}
=
- 1/2
.
\label{gnumu}
\end{align}
For antineutrinos one must substitute
$g_{A} \to -g_{A}$. Here, $\theta_{W}$ is the weak mixing angle, also known as the Weinberg angle, whose value at zero momentum transfer is $\sin^{2} \theta_{W}=0.23857$~\cite{ParticleDataGroup:2020ssz} in the $\overline{\mathrm{MS}}$ scheme.
The $Z_{\text{eff}}^{\mathcal{A}}(T_{e})$ term~\cite{Mikaelyan:2002nv,Fayans:2000ns} quantifies the number of electrons that can be ionized by a certain energy deposit $T_e$.
It is needed to correct the cross section derived under the Free Electron Approximation (FEA) hypothesis, where electrons are considered to be free and at rest~\cite{Kouzakov:2014lka,Kouzakov:2017hbc,Chen:2014ypv,PhysRevD.100.073001}. It is given for Cs, I and Ge in Tabs.~\ref{tab:electroneffchargeCsI} and~\ref{tab:electroneffcharge}~\cite{Thompsonxray}, respectively. 
% Please add the following required packages to your document preamble:
% \usepackage{multirow}
% Please add the following required packages to your document preamble:
% \usepackage{multirow}
In the sub-keV regime, as in the case of Dresden-II, energies are comparable with those of atomic scales and a correction to the FEA analogous to the  $Z_{\text{eff}}^{\mathcal{A}}(T_{e})$ term is mandatory. An alternative approach, that takes into account the many-electron dynamics in atomic ionization is obtained by exploiting an ab-initio approach in the framework of the multi-configuration relativistic random phase approximation (MCRRPA)~\cite{PhysRevA.25.634,PhysRevA.26.734,Chen:2013lba}, which is able to give an improved description of the atomic many-body effects. On the other hand, FEA, in particular when corrected by the stepping function $Z_{\text{eff}}^{\mathcal{A}}(T_{e})$, is known to provide a very good approximation at higher energies, as in the case of COHERENT CsI.
Throughout this paper, we will discuss the validity of our results concerning this issue.

\begin{table}[ht]
\renewcommand{\arraystretch}{1.45}
\begin{tabular}{c|llllc|lll}
\multirow{13}{*}{\begin{tabular}[c]{@{}c@{}}$Z_{\rm eff}^{\rm Cs}$=\end{tabular}} & 55, & & $T_{e}>$ 35.99 keV          & \multirow{13}{*}{} & \multirow{13}{*}{\begin{tabular}[c]{@{}l@{}}$Z_{\rm eff}^{\rm I}$ =\end{tabular}} & 53, & & $T_{e}>$ 33.17 keV           \\
                                                                             & 53, & &35.99 keV\ $\geq\ T_{e} >$5.71 keV &                    &                                                                             & 51, & & 33.17 keV\ $\geq\ T_{e} >$5.19 keV  \\
                                                                             & 51, & &5.71 keV\ $\geq\ T_{e} >$5.36 keV  &                    &                                                                             & 49, & & 5.19 keV\ $\geq\ T_{e} >$4.86 keV   \\
                                                                             & 49, & &5.36 keV\ $\geq\ T_{e} >$5.01 keV  &                    &                                                                             & 47, & & 4.86 keV\ $\geq\ T_{e} >$4.56 keV   \\
                                                                             & 45, & &5.01 keV\ $\geq\ T_{e} >$1.21 keV  &                    &                                                                             & 43, & &4.56 keV\ $\geq\ T_{e} >$1.07 keV   \\
                                                                             & 43, & &1.21 keV\ $\geq\ T_{e} >$1.07 keV  &                    &                                                                             & 41, & &1.07 keV\ $\geq\ T_{e} >$0.93 keV   \\
                                                                             & 41, & &1.07 keV\ $\geq\ T_{e} >$1 keV     &                    &                                                                             & 39, & &0.93 keV\ $\geq\ T_{e} >$0.88 keV   \\
                                                                             & 37, & &1 keV\ $\geq\ T_{e} >$0.74 keV     &                    &                                                                             & 35, & &0.88 keV\ $\geq\ T_{e} >$0.63 keV   \\
                                                                             & 33, & &0.74 keV\ $\geq\ T_{e} >$0.73 keV  &                    &                                                                             & 31, & &0.63 keV\ $\geq\ T_{e} >$0.62 keV   \\
                                                                             & 27, & &0.73 keV\ $\geq\ T_{e} >$0.23 keV  &                    &                                                                             & 25, & &0.62 keV\ $\geq\ T_{e} >$0.19 keV   \\
                                                                             & 25, & &0.23 keV\ $\geq\ T_{e} >$0.17 keV  &                    &                                                                             & 23, & & 0.19 keV\ $\geq\ T_{e} >$0.124 keV  \\
                                                                             & 23, & & 0.17 keV\ $\geq\ T_{e} >$0.16 keV  &                    &                                                                             & 21, & &0.124 keV\ $\geq\ T_{e} >$0.123 keV \\
                                                                             & 19, & &$T_{e} <$ 0.16 keV        &                    &                                                                             & 17, & &$T_{e} <$ 0.123 keV       
\end{tabular}
\caption{The effective electron charge of the target atom, $Z_{\text{eff}}^{\mathcal{A}}(T_{e})$, for Cs and I.}
\label{tab:electroneffchargeCsI}
\end{table}

\begin{table}[ht]
\renewcommand{\arraystretch}{1.45}
    \centering
\begin{tabular}{c|lll}

\multirow{10}{*}{$Z_{\rm eff}^{\rm Ge}$=} & 32, & & $T_{e} >$ 11.103 keV          \\
                                 & 30, & & 11.103 keV\ $\geq T_{e} >$1.4146 keV \\
                                 & 28, & & 1.4146 keV\ $\geq T_{e} >$1.2481 keV \\
                                 & 26, & & 1.2481 keV\ $\geq T_{e} >$1.217 keV  \\
                                 & 22, & & 1.217 keV\ $\geq T_{e} >$0.1801 keV  \\
                                 & 20, & & 0.1801 keV\ $\geq T_{e} >$0.1249 keV \\
                                 & 18, & & 0.1249 keV\ $\geq T_{e} >$0.1208 keV \\
                                 & 14, & & 0.1208 keV\ $\geq T_{e} >$0.0298 keV \\
                                 & 10, & & 0.0298 keV\ $\geq T_{e} >$0.0292 keV \\
                                 & 4,  & & $T_{e} \leq$ 0.0292 keV      
\end{tabular}

\caption{The effective electron charge of the target atom, $Z_{\text{eff}}^{\mathcal{A}}(T_{e})$, for Ge.}
\label{tab:electroneffcharge}
\end{table}

%%%%%%%%%%%%%%%%%%%%%%%%%%%%%%%%%%%%%%%%%%%%%%%%%%%%%%%%%%%

\subsection{Neutrino charge radii}

In the SM, the neutrino charge radii (CR) are the only electromagnetic properties of neutrinos that are different from zero. The contribution of the SM neutrino CR
is taken into account as one of the radiative corrections to $g_{V}^{p}(\nu_{\ell})$ and corresponds to
\cite{Bernabeu:2000hf,Bernabeu:2002nw,Bernabeu:2002pd}
\begin{equation}
\langle{r}_{\nu_{\ell}}^2\rangle_{\text{SM}}
=
-
\frac{G_{\text{F}}}{2\sqrt{2}\pi^2}
\left[
3-2\ln\left(\frac{m_{\ell}^2}{m^2_{W}}\right)
\right]
,
\label{G050}
\end{equation}
where $m_{W}$ and $m_{\ell}$ are the $W$ boson and charged lepton masses
($\ell = e, \mu, \tau$) respectively,
and we use the conventions in Refs.~\cite{Giunti:2014ixa,Cadeddu:2018dux,Cadeddu:2019eta}.
The SM neutrino CR are diagonal in the flavor basis, due to the conservation of generation lepton numbers.
Numerically, the predicted values of
$\langle{r}_{\nu_{e}}^2\rangle_{\text{SM}}$
and
$\langle{r}_{\nu_{\mu}}^2\rangle_{\text{SM}}$,
that can be probed with \cenns data,
are
\begin{align}
\null & \null
\langle{r}_{\nu_{e}}^2\rangle_{\text{SM}}
=
- 0.83 \times 10^{-32} \, \text{cm}^2
,
\label{reSM}
\\
\null & \null
\langle{r}_{\nu_{\mu}}^2\rangle_{\text{SM}}
=
- 0.48 \times 10^{-32} \, \text{cm}^2
.
\label{rmSM}
\end{align}
%The current 90\% CL experimental bounds for $\langle{r}_{\nu_{e}}^2\rangle$ and $\langle{r}_{\nu_{\mu}}^2\rangle$ obtained in laboratory experiments that do not involve CE$\nu$NS are listed in Table~I of Ref.~\cite{Cadeddu:2018dux}. Since they are only about one order of magnitude larger than the Standard Model predictions, they may be the first neutrino electromagnetic properties measured by new experiments in a near future.
Here, we want to constrain possible BSM effects that could modify the SM value of the neutrino CR.
Thus, we consider the general case in which neutrinos can have both diagonal and
off-diagonal, also referred to as transition, CR in the flavor basis that can be generated by BSM physics.
The differential \cenns cross section that takes into account the contribution of the neutrino charge radii in addition to the SM neutral-current weak interaction is
\begin{equation}
\label{cs-chr}
\dfrac{d\sigma^{\mathrm{CR}}_{\nu_{\ell}\text{-}\mathcal{N}}}{d T_\mathrm{nr}}
(E,T_\mathrm{nr}) = \dfrac{G_{\text{F}}^2 M}{\pi}
\left( 1 - \dfrac{M T_\mathrm{nr}}{2 E^2} \right)
\left\{
\left[
\left( \tilde{g}_{V}^{p} - \tilde{Q}_{\ell\ell} \right)
Z F_{Z}(|\vec{q}|^2) + g_{V}^{n} N F_{N}(|\vec{q}|^2)
\right]^2
+ Z^2 F_{Z}^2(|\vec{q}|^2) \sum_{\ell'\neq\ell}|\tilde{Q}_{\ell\ell'}|^2
\right\},
\end{equation}
where $\tilde{g}_{V}^{p}=0.0186$
is the neutrino-proton coupling without the contribution of the SM neutrino CR.
The effects of the charge radii
$\langle{r}_{\nu_{\ell\ell'}}^2\rangle$
in the cross section are expressed as~\cite{Kouzakov:2017hbc}
\begin{equation}
\tilde{Q}_{\ell\ell'}
=
\dfrac{ \sqrt{2} \pi \alpha }{ 3 G_{\text{F}} }
\, \langle{r}_{\nu_{\ell\ell'}}^2\rangle
,
\label{Qchr}
\end{equation}
where $\alpha$ is the electromagnetic fine-structure constant. 
The diagonal CR of  flavor neutrinos
contribute to the cross section coherently with the
neutrino-proton neutral current interaction,
generating an effective shift of $\sin^2\!\vartheta_{W}$.
In the case of
$\bar\nu_{\ell}\text{-}\mathcal{N}$ scattering,
we have
$g_{V}^{p,n} \to - g_{V}^{p,n}$
and
$\langle{r}_{\nu_{\ell\ell'}}\rangle
\to
\langle{r}_{\bar\nu_{\ell\ell'}}\rangle = - \langle{r}_{\nu_{\ell\ell'}}\rangle$.
Therefore,
the CR of flavor neutrinos and antineutrinos
contribute with the same sign to the shift of
$\sin^2\!\vartheta_{W}$
in the CE$\nu$NS cross section.

There are five CR that can be determined with the 
CE$\nu$NS data:
the two diagonal charge radii
$\langle{r}_{\nu_{ee}}^2\rangle$
and
$\langle{r}_{\nu_{\mu\mu}}^2\rangle$,
that sometimes are denoted with the simpler notation
$\langle{r}_{\nu_{e}}^2\rangle$
and
$\langle{r}_{\nu_{\mu}}^2\rangle$
in connection to the SM CR in Eqs.~(\ref{G050})--(\ref{rmSM}),
and the absolute values of the three off-diagonal CR
$\langle{r}_{\nu_{e\mu}}^2\rangle=\langle{r}_{\nu_{\mu e}}^2\rangle^{*}$,
$\langle{r}_{\nu_{e\tau}}^2\rangle$, and
$\langle{r}_{\nu_{\mu\tau}}^2\rangle$. \\

In the presence of the neutrino charge radii, the neutrino-electron elastic scattering cross section in Eq.~\eqref{ES-SM},  is modified to~\cite{Kouzakov:2017hbc}
\begin{equation}
\left(
\dfrac{d\sigma_{\nu_{\ell}-\mathcal{A}}^{\text{ES,CR}}}{d T_{\text{e}}}
\right)_{\text{SM}+\tilde{Q}}
=
\left(
\dfrac{d\sigma_{\nu_{\ell}-\mathcal{A}}^{\text{ES,CR}}}{d T_{\text{e}}}
\right)_{\text{SM}+\tilde{Q}_{\ell\ell}}
+
\sum_{\ell'\neq\ell}
\left(
\dfrac{d\sigma_{\nu_{\ell}-\mathcal{A}}^{\text{ES,CR}}}{d T_{\text{e}}}
\right)_{\tilde{Q}_{\ell\ell'}}
,
\label{ES-Qt}
\end{equation}
where
$
(d\sigma_{\nu_{\ell}-\mathcal{A}}^{\text{ES,CR}} / d T_{\text{e}} )_{\text{SM}+\tilde{Q}_{\ell\ell}}
$
is given by Eq.~\eqref{ES-SM} with
\begin{equation}
g_{V}^{\nu_{\ell}}
\to
g_{V}^{\nu_{\ell}} + \tilde{Q}_{\ell\ell}
,
\label{chrgVshift}
\end{equation}
and
\begin{equation}
\left(
\dfrac{d\sigma_{\nu_{\ell}-\mathcal{A}}^{\text{ES,CR}}}{d T_{\text{e}}}
\right)_{\tilde{Q}_{\ell\ell'}}
=
Z_{\text{eff}}^{\mathcal{A}}(T_{e})
\,
\dfrac{\pi \alpha^2 m_{e}}{9}
\left[
1
+
\left( 1 - \dfrac{T_{e}}{E} \right)^2
-
\dfrac{m_{e} T_{e}}{E^2}
\right]
|\langle{r}_{\nu_{\ell\ell'}}^2\rangle|^2
,
\label{ES-Qtfc}
\end{equation}
for $\ell'\neq\ell$. In this scenario, the FEA approach corrected by the stepping function as used in this work slightly overestimates the cross section with respect to MCRRPA for $T_e\lesssim1$ keV, but they rapidly converge for $T_e>1$ keV~\cite{Chen:2014ypv}, causing a negligible difference.

%%%%%%%%%%%%%%%%%%%%%%%%%%%%%%%%%%%%%%%%%%%%%%%%%%%%%%%%%%%%%%%%%

\subsection{Neutrino magnetic moments}
\label{sec:magnetic}

The neutrino magnetic moment (MM) is the most investigated neutrino electromagnetic
property, both theoretically and experimentally.
Indeed, its existence is predicted by many BSM theories,
especially those that include right-handed neutrinos, see the reviews in Refs.~\cite{Giunti:2014ixa,Giunti:2015gga}.
The differential \cenns cross section that takes into account the contribution of the neutrino magnetic moment is given by adding to the SM cross section in Eq.~(\ref{cs-std}) the MM contribution, namely
\begin{equation}
\dfrac{d\sigma_{\nu_{\ell}\text{-}\mathcal{N}}^{\text{MM}}}{d T_\mathrm{nr}}
(E,T_\mathrm{nr})
=
\dfrac{ \pi \alpha^2 }{ m_{e}^2 }
\left( \dfrac{1}{T_\mathrm{nr}} - \dfrac{1}{E} \right)
Z^2 F_{Z}^2(|\vec{q}|^2)
\left| \dfrac{\mu_{\nu_{\ell}}}{\mu_{\text{B}}} \right|^2
,
\label{cs-mag}
\end{equation}
where $\mu_{\nu_{\ell}}$ is the effective MM of the flavor neutrino $\nu_{\ell}$
in elastic scattering (see Ref.~\cite{Giunti:2014ixa}),
and $\mu_{\text{B}}$ is the Bohr magneton.

\noindent In the case of neutrino-electron scattering, the cross section in presence of neutrino magnetic moments receives an additional contribution equal to
\begin{equation}
\dfrac{d\sigma_{\nu_{\ell}\text{-}\mathcal{A}}^{\text{ES, MM}}}{d T_\mathrm{e}}
(E,T_\mathrm{e})
=
Z_{\text{eff}}^{\mathcal{A}}(T_{\text{e}}) \dfrac{ \pi \alpha^2 }{ m_{e}^2 }
\left( \dfrac{1}{T_\mathrm{e}} - \dfrac{1}{E} \right)
\left| \dfrac{\mu_{\nu_{\ell}}}{\mu_{\text{B}}} \right|^2,
\label{es-mag}
\end{equation}
with $Z_{\text{eff}}^{\mathcal{A}}(T_{e})$ detailed in Tabs.~\ref{tab:electroneffchargeCsI} and~\ref{tab:electroneffcharge}. As in the case of neutrino charge radii, the cross section obtained with the corrected FEA is slightly larger than the MCRRPA one only for $T_e\lesssim$ 1 keV~\cite{Chen:2014ypv}.

%%%%%%%%%%%%%%%%%%%%%%%%%%%%%%%%%%%%%%%%%%%%%%%%%%%%%%%%%%%%%%%%%%%%%%%%%%%%%%%%%%%%%

\subsection{Neutrino electric charges}
\label{sec:nuec}

As already shown in many experimental and theoretical studies (for a review see Ref.~\cite{Giunti:2014ixa}), \cenns process is sensitive not only to the neutrino CR, but also to the existence of neutrino electric charges (EC). Indeed, even if neutrinos are considered as neutral particles, in some BSM theories they can acquire small electric charges, usually referred to as millicharges.
The differential \cenns cross section taking into account the contribution of the neutrino electric charges in addition to SM neutral-current weak interactions is similar to that derived for the neutrino charge radii,
with $g_{V}^{p}$ and $g_{V}^{n}$ given in Sec.~\ref{sec:cs} and $\tilde{Q}_{\ell\ell'}$ replaced by $Q_{\ell\ell'}$~\cite{Kouzakov:2017hbc,Giunti:2014ixa}
\begin{equation}
Q_{\ell\ell'}
=
\dfrac{ 2 \sqrt{2} \pi \alpha }{ G_{\text{F}} q^2 }
\, q_{\nu_{\ell\ell'}}
,
\label{Qech}
\end{equation}
where $q_{\nu_{\ell\ell'}}$ is the neutrino EC and $ q^2 = - 2 M T_{\mathrm{nr}} $ is the squared four-momentum transfer. Given the extremely low momentum transfer and low-energy thresholds of reactor experiments, the $q^2$ dependence in the denominator of Eq.~(\ref{Qech}) helps to set more stringent constraints using the data of Dresden-II with respect to COHERENT, as we will show in Sec.~\ref{sec:result}.
As in the case of neutrino CR, the contribution of
neutrinos and antineutrinos to the neutrino EC will also shift $\sin^2\!\vartheta_{W}$ with the same sign, since the electric charges of neutrino and antineutrino are opposite as well as the weak neutral current couplings.

If neutrinos have electric charges,
the neutrino-electron elastic scattering cross section in Eq.~\eqref{ES-SM} becomes~\cite{Kouzakov:2017hbc}
\begin{equation}
\left(
\dfrac{d\sigma_{\nu_{\ell}-\mathcal{A}}^{\text{ES,EC}}}{d T_{\text{e}}}
\right)_{\text{SM}+Q}
=
\left(
\dfrac{d\sigma_{\nu_{\ell}-\mathcal{A}}^{\text{ES,EC}}}{d T_{\text{e}}}
\right)_{\text{SM}+Q_{\ell\ell}}
+
\sum_{\ell'\neq\ell}
\left(
\dfrac{d\sigma_{\nu_{\ell}-\mathcal{A}}^{\text{ES,EC}}}{d T_{\text{e}}}
\right)_{Q_{\ell\ell'}}
,
\label{ES-EC}
\end{equation}
where
$
( d\sigma_{\nu_{\ell}-\mathcal{A}}^{\text{ES,EC}} / d T_{\text{e}} )_{\text{SM}+Q_{\ell\ell}}
$
is given by Eq.~\eqref{ES-SM} with
\begin{equation}
g_{V}^{\nu_{\ell}}
\to
g_{V}^{\nu_{\ell}} + Q_{\ell\ell}
,
\label{chrgVshift2}
\end{equation}
and
\begin{equation}
\left(
\dfrac{d\sigma_{\nu_{\ell}-\mathcal{A}}^{\text{ES,EC}}}{d T_{\text{e}}}
\right)_{Q_{\ell\ell'}}
=
Z_{\text{eff}}^{\mathcal{A}}(T_{e})
\,
\dfrac{\pi \alpha^2}{m_{e} T_{\text{e}}^2}
\left[
1
+
\left( 1 - \dfrac{T_{\text{e}}}{E} \right)^2
-
\dfrac{m_{e} T_{\text{e}}}{E^2}
\right]
|q_{\nu_{\ell\ell'}}|^2
,
\label{ES-ECfc}
\end{equation}
for $\ell'\neq\ell$.
In neutrino-electron elastic scattering $ |q^2| = 2 m_e T_{e} $,
which is much smaller than the CE$\nu$NS $|q^2|$.
Therefore, the analysis of the COHERENT CsI and Dresden-II data taking into account
ES scattering allows us to enhance substantially
the sensitivity to neutrino millicharges. Let us note that, for neutrino millicharges, the MCRRPA cross section for $T_e\lesssim$ 1 keV is more than one order of magnitude bigger than that obtained with the corrected FEA~\cite{Chen:2014ypv}. In this respect, we can consider our Dresden-II ES limits as conservative and tighter limits are expected if the MCRRPA approach is used.

%%%%%%%%%%%%%%%%%%%%%%%%%%%%%%%%%%%%%%%%%%%%%%%%%%%%%%%%%%%%%%%%%%
%%%%%%%%%%%%%%%%%%%%%%%%%%%%%%%%%%%%%%%%%%%%%%%%%%%%%%%%%%%%%%%%%%

\section{Data analysis strategy}\label{sec:method}

In this section we will summarize the prescriptions followed for the analysis of the COHERENT and Dresden-II data set.

%%%%%%%%%%%%%%%%%%%%%%%%%%%%%%%%%%%%%%%%%%%%%%%%%%%%%%%%%%%%%%%%%%%

\subsection{COHERENT}

For the analysis of the COHERENT CsI and Ar data we follow closely the strategy explained in detail in Ref.~\cite{Corona:2022wlb}. We obtained information on all the quantities used from Refs.~\cite{COHERENT:2020iec,COHERENT:2020ybo} for the Ar data
and from Ref.~\cite{Akimov:2021dab} for the CsI data.
The total differential neutrino flux, $d N_{\nu}/d E$, is given by the sum of the three neutrino components produced by the pion decay at rest. Namely, the first prompt component is coming directly from the pion decay
($\pi^{+} \rightarrow \mu^{+}+\nu_{\mu}$), while the second two delayed components are coming from the subsequent muon decay ($\mu^{+} \rightarrow e^{+}+\nu_{e}+\bar{\nu}_{\mu}$). The neutrino flux depends on the number $r$ of neutrinos produced for each proton-on-target (POT), the number of protons-on-target $N_{\mathrm{POT}}$ and the baseline $L$ between the source and the detector.
For the COHERENT Ar detector, called CENNS-10, we use $r=0.09$, $N_{\mathrm{POT}}=13.8 \times 10^{22}$ and $L=27.5~\mathrm{m}$~\cite{COHERENT:2020ybo}.
For the COHERENT CsI detector, we use $r=0.0848$, $N_{\mathrm{POT}}=3.198 \times 10^{23}$ and $L=19.3~\mathrm{m}$~\cite{Akimov:2018vzs}. The prompt $\nu_{\mu}$'s component arrives within about $1~\mu\text{s}$ from the on-beam trigger,
whereas the delayed $\nu_{e}$'s and $\bar\nu_{\mu}$'s arrive in a time interval which can extend up to about $10~\mu\text{s}$.
The inclusion of the time evolution of the COHERENT data is thus important to distinguish the two neutrino components.

In each nuclear-recoil energy-bin $i$, the theoretical \cenns event number $N^\mathrm{CE \nu NS}_{i}$  is given by
\begin{equation}\label{N_cevns}
N_{i}^{\mathrm{CE}\nu\mathrm{NS}}(\mathcal{N})
=
N({\rm{tg}})
\int_{T_{\mathrm{nr}}^{i}}^{T_{\mathrm{nr}}^{i+1}}
\hspace{-0.3cm}
d T_{\mathrm{nr}}\,
A(T_{\mathrm{nr}})
\int_{0}^{T^{\prime\text{max}}_{\text{nr}}}
\hspace{-0.3cm}
dT'_{\text{nr}}
\,
R(T_{\text{nr}},T'_{\text{nr}})
\int_{E_{\text{min}}(T'_{\text{nr}})}^{E_{\text{max}}}
\hspace{-0.3cm}
d E
\sum_{\nu=\nu_{e}, \nu_{\mu}, \bar{\nu}_{\mu}}
\frac{d N_{\nu}}{d E}(E)
\frac{d \sigma_{\nu-\mathcal{N}}}{d T'_{\mathrm{nr}}}(E, T'_{\mathrm{nr}})
,
\end{equation}
where
$\mathcal{N}$= Cs, I or Ar, and $N_{i}^{\mathrm{CE}\nu\mathrm{NS}}(\mathrm{CsI})=N_{i}^{\mathrm{CE}\nu\mathrm{NS}}(\mathrm{Cs})+N_{i}^{\mathrm{CE}\nu\mathrm{NS}}(\mathrm{I})$. Moreover,
$T_{\text{nr}}$ is the reconstructed nuclear recoil kinetic energy,
$T'_{\text{nr}}$ is the true nuclear recoil kinetic energy,
$A(T_{\text{nr}})$ is the energy-dependent detector efficiency,
$R(T_{\text{nr}},T'_{\text{nr}})$ is the energy resolution function,
$T^{\prime\text{max}}_{\text{nr}} \simeq 2 E_{\text{max}}^2 / M$,
$E_{\text{max}} = m_\mu/2 \sim 52.8$~MeV,
$E_{\text{min}}(T'_{\text{nr}}) \simeq \sqrt{MT'_\text{nr}/2}$,
$m_\mu$ being the muon mass,
and
$N(\mathrm{tg})$ the number of target atoms in the detector, where the targets are tg = CsI or Ar.
The number of target atoms in each detector is given by
$N(\mathrm{tg}) = N_{\mathrm{A}} M_{\mathrm{det}} / M_{\mathrm{\mathrm{tg}}}$, 
where $N_{\mathrm{A}}$ is the Avogadro number, 
$M_{\mathrm{det}}$ is the detector active mass
($M_{\mathrm{det}}=24.4~\mathrm{kg}$ for Ar
and
$M_{\mathrm{det}}=14.6~\mathrm{kg}$ for CsI),
and
$M_{\mathrm{\mathrm{tg}}}$ is the molar mass
($M_{\mathrm{Ar}} = 39.96 ~\mathrm{g/mol}$
and
$M_{\mathrm{CsI}} = 259.8 ~\mathrm{g/mol}$).
Finally, the differential \cenns cross section
$d \sigma_{\nu-\mathcal{N}} / d T_{\mathrm{nr}}$
has been discussed in Section~\ref{sec:cs}.

Differently from Ref.~\cite{Corona:2022wlb}, in the CsI analysis we also include the contribution of the electron-neutrino scattering, as stated in Sec.~\ref{sec:cs}.
In each electron-recoil energy-bin $i$, the theoretical ES event number $N^\mathrm{ES}_{i}$  is given by
\begin{equation}\label{N_es}
N_{i}^{\mathrm{ES}}(\mathcal{A})
=
N(\mathrm{tg})
\int_{T_{\mathrm{e}}^{i}}^{T_{\mathrm{e}}^{i+1}}
\hspace{-0.3cm}
d T_{\mathrm{e}}\,
A(T_{\mathrm{e}})
\int_{0}^{T^{\prime\text{max}}_{\text{e}}}
\hspace{-0.3cm}
dT'_{\text{e}}
\,
R(T_{\text{e}},T'_{\text{e}})
\int_{E_{\text{min}}(T'_{\text{e}})}^{E_{\text{max}}}
\hspace{-0.3cm}
d E
\sum_{\nu=\nu_{e}, \nu_{\mu}, \bar{\nu}_{\mu}}
\frac{d N_{\nu}}{d E}(E)
\frac{d \sigma^{\rm{ES}}_{\nu-\mathcal{A}}}{d T'_{\mathrm{e}}}(E, T'_{\mathrm{e}})
,
\end{equation}
where
$\mathcal{A}$ = Cs or I, and $N_{i}^{\mathrm{ES}}(\mathrm{CsI})=N_{i}^{\mathrm{ES}}(\mathrm{Cs})+N_{i}^{\mathrm{ES}}(\mathrm{I})$, $E_{\text{min}}(T'_{\text{e}}) = (T'_\text{e}+\sqrt{T_\text{e}^{'2} + 2m_e T'_\text{e}})/2$, and $T^{\prime\text{max}}_{\text{e}} = 2 E_{\text{max}}^2 / (2E_{\text{max}}+m_e)$.

It is important to consider that the energy actually observed in the detector is the electron-equivalent recoil energy $T_{e}$, which is transformed into the nuclear recoil energy $T_{\mathrm{nr}}$ in the \cenns rate by inverting the relation
\begin{equation}\label{Qf}
	T_{\mathrm{e}}=f_{Q}\left(T_{\mathrm{nr}}\right) T_{\mathrm{nr}},
\end{equation}
where $f_{Q}$ is the quenching factor~\cite{COHERENT:2021pcd}. 

In order to include also the timing information we separated the theoretical \cenns event numbers
$N^\text{\cenns}_{i}$ in Eq.~\eqref{N_cevns} in time bins that are calculated from the exponential decay laws of
the generating pions and muons. With this procedure we obtained the theoretical \cenns event numbers
$N^\text{\cenns}_{ij}$, where $i$ is the index of the energy bins and $j$ is the index of the time bins.

We performed the analysis of the COHERENT CsI data using the Poissonian least-squares function~\cite{Baker:1983tu,ParticleDataGroup:2020ssz}, given that in some energy-time bins the number of events is very small, namely
\begin{align}\nonumber
	\chi^2_{\mathrm{CsI}, \mathrm{CE}\nu\mathrm{NS}+\rm{ES}}
	=
	\null & \null
	2
	\sum_{i=1}^{9}
	\sum_{j=1}^{11}
	\left[
	    \sum_{z=1}^{4}\left[( 1 + \eta_{z} ) N_{ij}^{z}\right] + \eta_{5} N_{ij}^{5} -
		N_{ij}^{\text{exp}}
		+ N_{ij}^{\text{exp}} \ln\left(\frac{N_{ij}^{\text{exp}}}{\sum_{z=1}^{4}\left[( 1 + \eta_{z} ) N_{ij}^{z}\right] + \eta_{5} N_{ij}^{5}}\right)
	\right]\\
	&+ \sum_{z=1}^{5}
	\left(
	\dfrac{ \eta_{z} }{ \sigma_{z} }
	\right)^2
	,
	\label{chi2coherentCsI}
\end{align}
where
the indices
$z=1,2,3,4,5$ for $N_{ij}^{z}$ stand, respectively, for CE$\nu$NS+ES, namely $N_{ij}^{1}=N_{ij}^{\mathrm{CE}\nu\mathrm{NS}}+N_{ij}^{\mathrm{ES}}$,
beam-related neutron ($N_{ij}^{2}=N_{ij}^{\text{BRN}}$),
neutrino-induced neutron ($N_{ij}^{3}=N_{ij}^{\text{NIN}}$),
steady-state ($N_{ij}^{4}=N_{ij}^{\text{SS}}$) backgrounds, and \cenns only
($N_{ij}^{5}=N_{ij}^{\mathrm{CE}\nu\mathrm{NS}}$).
In our notation,
$N_{ij}^{\text{exp}}$ is the experimental event number obtained from coincidence (C) data,
$N_{ij}^{\text{\cenns}}$ is the predicted number of \cenns events
that depends on the physics model under consideration,
$N_{ij}^{\text{BRN}}$ is the estimated BRN background,
$N_{ij}^{\text{NIN}}$ is the estimated NIN background,
$N_{ij}^{\text{SS}}$ is the SS background obtained from the anti-coincidence (AC) data and, $N_{ij}^{\text{ES}}$ is the contribution of the electron scattering that also depends on the physics model under consideration. Clearly, when summing the \cenns and ES contributions, both event numbers as well as the background contributions must be determined either in nuclear-recoil or in electron-recoil energy bins.
We took into account the systematic uncertainties with the nuisance parameters $\eta_{z}$
and
the corresponding uncertainties
$\sigma_{\text{CE}\nu\text{NS+ES}}=0.11$, $\sigma_{\text{\cenns}}=0.05$,
$\sigma_{\text{BRN}}=0.25$,
$\sigma_{\text{NIN}}=0.35$,
$\sigma_{\text{SS}}=0.021$. The uncertainty $\sigma_{\text{CE}\nu\text{NS+ES}}$ does not include the form factor and quenching factor related uncertainties that are affecting only \cenns and are implemented thanks to an additional contribution $\sigma_{\text{\cenns}}$. 

In this work, to appreciate the impact of the ES contribution, we will sometimes fit the CsI data set for \cenns only. In this case, the least-squares function $\chi^2_{\mathrm{CsI}, \mathrm{CE}\nu\mathrm{NS}}$ is obtained removing the ES contribution for $z=1$ in Eq.~(\ref{chi2coherentCsI}). \\

We performed the analysis of the COHERENT Ar data using the least-squares function
\begin{align}
	\chi^2_{\mathrm{Ar}, \mathrm{CE}\nu\mathrm{NS}}
	=
	\null & \null
	\sum_{i=1}^{12}
	\sum_{j=1}^{10}
	\left(
	\dfrac
	{
		N_{ij}^{\text{exp}}
		- \sum_{z=1}^{4}( 1 + \eta_{z} 
		+ \sum_{l}\eta^{\mathrm{sys}}_{zl,ij} ) N_{ij}^{z}
	}
	{ \sigma_{ij} }
	\right)^2
	+ \sum_{z=1}^{4}
	\left(
	\dfrac{ \eta_{z} }{ \sigma_{z} }
	\right)^2
	+ \sum_{z,l}\left(
	\epsilon_{zl}
	\right)^2
	,
	\label{chi2coherentAr}
\end{align}
where $z=1,2,3,4$ stands for the theoretical prediction of \cenns, SS, Prompt Beam-Related Neutron (PBRN) and Delayed Beam-Related Neutron (DBRN) backgrounds, and $N_{ij}^{\text{exp}}$ is the number of observed events in each energy and time bin.
The statistical uncertainty ${\sigma_{ij}}$ is given by
\begin{equation}\label{sigma_stat}
	(\sigma_{ij}^\mathrm{} )^2 =
	( \sigma_{ij}^\mathrm{exp} )^2
	+ ( \sigma_{ij}^\text{SS} )^2,
\end{equation}
where 
$\sigma_{ij}^\mathrm{exp} = \sqrt{N_{ij}^{\text{exp}}}$ and 
$\sigma_{ij}^\mathrm{SS} = \sqrt{ {N_{ij}^{\text{SS}}}/{5} }$.
The factor 1/5 is due to the 5 times longer sampling time of the SS background with respect to the signal time window.   
The nuisance parameters $\eta_z$ quantify the systematic uncertainties of the event rate for the theoretical prediction of \cenns, SS, PBRN, and DBRN backgrounds,
with the corresponding uncertainties
$\sigma_\mathrm{CE\nu NS}=0.13$, 
$\sigma_\mathrm{PBRN}=0.32$, 
$\sigma_\mathrm{DBRN}=1$, and
$\sigma_\mathrm{SS}=0.0079$.
We considered also the systematic uncertainties of the shapes of \cenns and PBRN spectra
using the information in the COHERENT data release~\cite{COHERENT:2020ybo}.
This is done in Eq.~\eqref{chi2coherentAr} through
the nuisance parameters $\epsilon_{zl}$ and the terms $\eta^{\mathrm{sys}}_{zl,ij}$ given by
\begin{equation}\label{simga_sys}
	\eta^{\mathrm{sys}}_{zl,ij} = \epsilon_{zl} \, \frac{N_{zl,ij}^\mathrm{sys}-N_{zl,ij}^\mathrm{CV}}
	{N_{zl,ij}^\mathrm{CV}},
\end{equation}
where $l$ is the index of the source of the systematic uncertainty.
Here, $N_{zl,ij}^\mathrm{sys}$ and $N_{zl,ij}^\mathrm{CV}$ are, respectively,
$1\sigma$ probability distribution functions (PDFs) described in Tab.~3 of Ref.~\cite{COHERENT:2020ybo}
and the central-value (CV) SM predictions described in Tab.~2 of Ref.~\cite{COHERENT:2020ybo}.

%%%%%%%%%%%%%%%%%%%%%%%%%%%%%%%%%%%%%%%%%%%%%%%%%%%%%%%%%%%%%%%%%%%

\subsection{Dresden-II}
\label{sec:dresden}

For the analysis of the NCC-1701 data obtained using antineutrinos produced by the Dresden-II reactor, we use the data release and related information in Ref.~\cite{Colaresi:2022obx}.

In order to derive the antineutrino spectra $d N_{\overline{\nu}}/d E$ from the Dresden-II reactor we have considered three different parametrizations, obtained by combining four different predictions for specific energy ranges.
In particular, the neutrino spectra are built by combining the expected spectra for antineutrino energies above 2 MeV from either Ref.~\cite{Mueller:2011nm} or Ref.~\cite{Estienne:2019ujo}, that we indicate as HM and EF, respectively, with the low energy part determined by Ref.~\cite{Vogel:1989iv} and Refs.~\cite{Kopeikin:1999tc,Kopeikin:2012zz}, that we indicate as VE and K, respectively. In this way, three different combinations are obtained, to which we will refer to as HMVE, EFK, and HMK. These spectra are obtained from the weighted average of the antineutrino fluxes from four main fission isotopes, namely $^{235}\mathrm{U}$, $^{239}\mathrm{Pu}$, $^{238}\mathrm{U}$ and $^{241}\mathrm{Pu}$. In the K prediction~\cite{Kopeikin:1999tc,Kopeikin:2012zz}, the contribution at low energies from radiative neutron capture on $^{238}\mathrm{U}$ is also taken into account. The latter has the effect to enhance the spectrum for neutrino energies below $\sim1$ MeV. In all cases, we set the spectra to zero above 10 MeV.
The neutrino spectra for reactor antineutrinos have been normalized to the antineutrino flux estimate reported in Ref.~\cite{Colaresi:2022obx} and corresponding to $\Phi_{\rm{est}}=4.8\times 10^{13}\ \mathrm{cm^{-2} s^{-1}}$, that has been determined considering a reactor power $P=2.96\ \mathrm{GW}_{\rm{th}}$ and a reactor-detector distance of $L=10.39\ \mathrm{m}$~\cite{Colaresi:2022obx}.

In the energy region of interest of Dresden-II, $0.2 \, \mathrm{keV}_{\rm{ee}} < T_{\rm{e}} < 1.5 \, \mathrm{keV}_{\rm{ee}}$,
the background comes from the elastic scattering of epithermal neutrons and the electron capture in ${}^{71}$Ge. The epithermal neutron contribution, which is the dominant one in the \cenns recoil-energy region, $T_{\rm{e}} \lesssim 0.5 \, \mathrm{keV}_{\rm{ee}}$, is described by an exponential function with decay constant $T_{\mathrm{epith}}$ plus a constant term $N_{\mathrm{epith}}$, while the electron capture peaks from ${}^{71}$Ge, namely the L1-, L2- and M-shell peaks, are described each by a Gaussian function. The latter is parametrized by an amplitude $A_i$, the centroid $T_i$ and the standard deviation $\sigma_i$, where $i=$ L1, L2 and M.
Thus, the expected event rate of background is given by
\begin{equation}
    \frac{d N^\mathrm{bkg}}{ d T_{\rm{e}} }=N_{\mathrm{epith}}+A_{\mathrm{epith}} e^{-T_{\mathrm{e}} / T_{\mathrm{epith}}}+\sum_{i=\mathrm{L1, L2, M}} \frac{A_{i}}{\sqrt{2 \pi} \sigma_{i}} e^{-\frac{\left(T_{\mathrm{e}}-T_{i}\right)^{2}}{2 \sigma_{i}^{2}}}.
\end{equation}
Following Ref.~\cite{Colaresi:2022obx}, the total amount of free parameters for the background prediction reduces to: $N_{\rm epith}$, $A_{\rm epith}$, $T_{\rm epith}$, $A_{\rm L1}$, $E_{\rm L1}$, $\sigma_{\rm L1}$ and $\beta_{\rm M/L1}$. 
In fact, the amplitude of the L2 shell contribution can be expressed in terms of the amplitude of the L1 shell ($A_{\rm L1}$), in particular $A_{\rm L2}/A_{\rm L1}=0.008$, and $\sigma_{\rm L2}=\sigma_{\rm L1}$. The centroid of the L2 Gaussian can be safely set to the nominal value $T_{\mathrm{L2}}=1.142\ \mathrm{keV}$. The standard deviation of the M-shell contribution can be fixed to the electronic noise uncertainty, which is $\sigma_n=68.5\ \mathrm{eV}$ for the Rx-ON (reactor operation period) data. The centroid of the M-shell Gaussian is fixed to its nominal value $T_{\rm M}=0.158\ \mathrm{keV}$, being smaller than the experimental threshold whereas its amplitude is left free to vary in the fit with a constraint corresponding to the experimentally determined ratio
$\beta_{\rm M/L1} = A_\mathrm{M}/A_{\mathrm{L_1}}=0.16\pm0.03$.

\begin{figure}
    \centering
        \subfigure[]{\label{fig:qf_Dresden}
    	\includegraphics[height=0.48\textwidth]{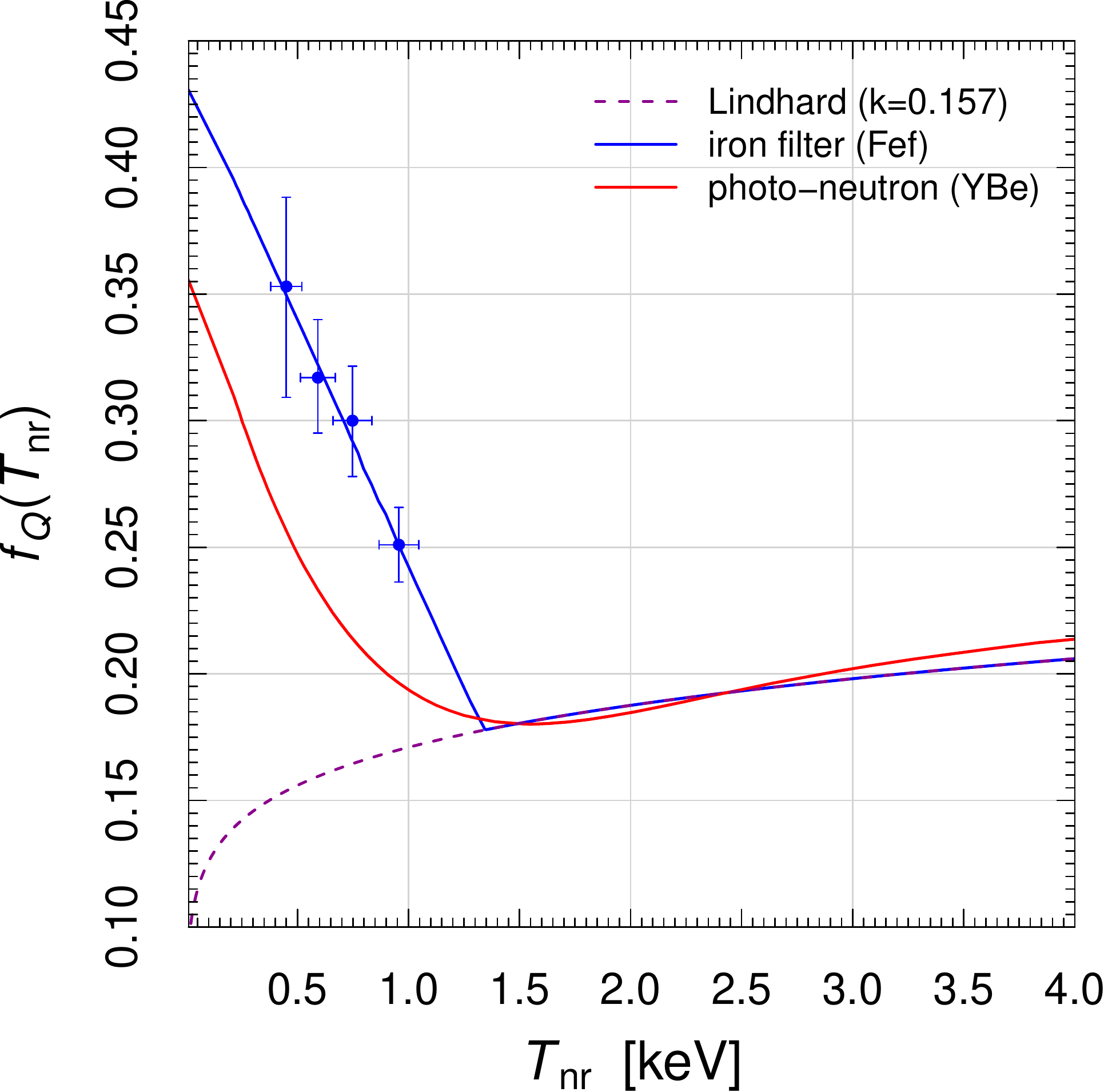}}
    	%\subfigure[]{\label{fig:spe_Dresden}
    	%\includegraphics[height=0.48\textwidth]{figure/spectrum_Dresden.pdf}}
    	\subfigure[]{\label{fig:spe_Dresden}
    	\includegraphics[height=0.48\textwidth]{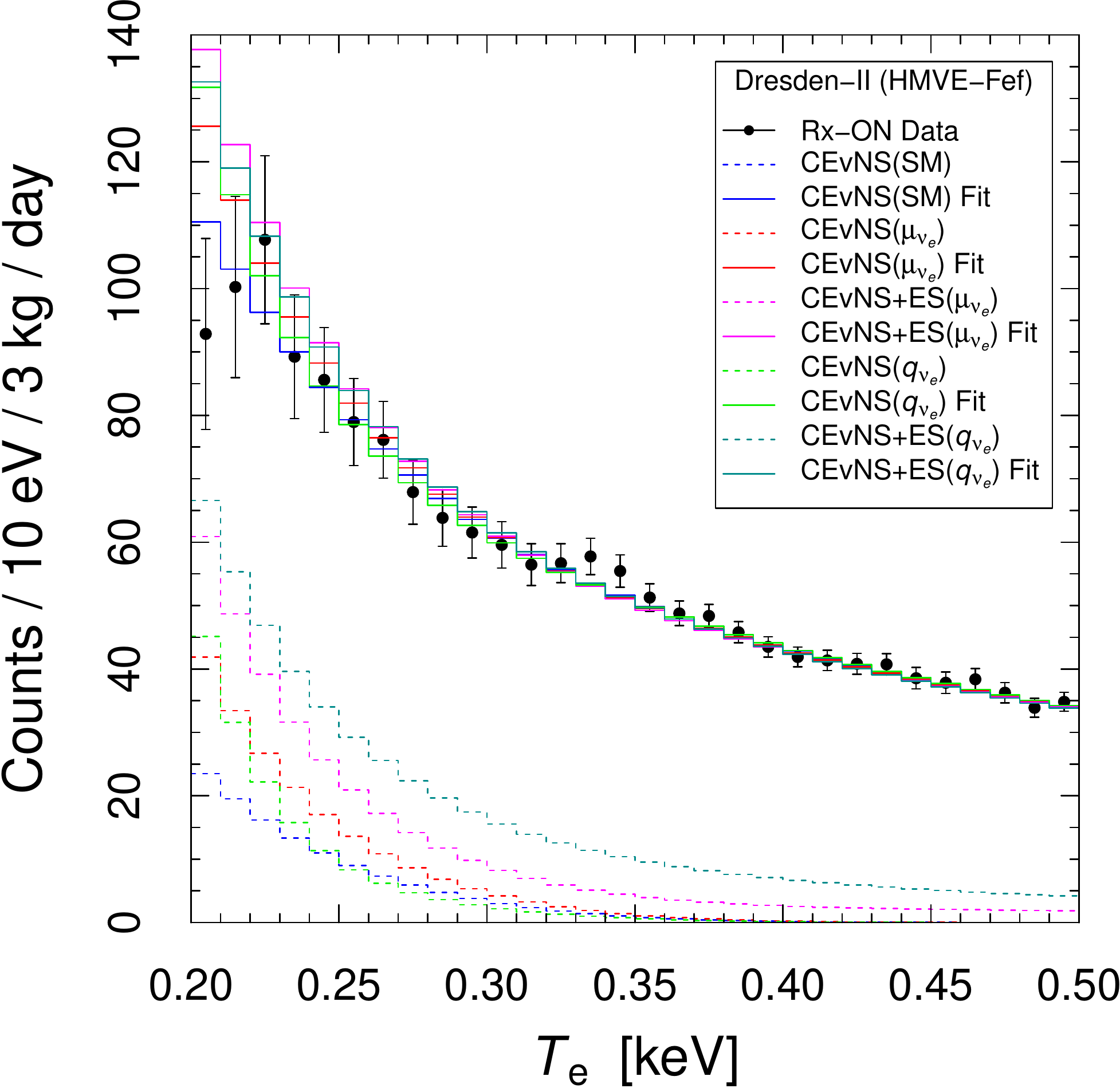}}
    \caption{
    \subref{fig:qf_Dresden}
    The germanium quenching factor models provided in Ref.~\cite{Colaresi:2022obx} for the analysis of the Dresden-II data: iron filter (Fef) given by the solid blue line
    that fits the blue data points~\cite{Collar:2021fcl}
    and is extended with the Lindhard model with $k=0.157$~\cite{osti_4701226} (dashed purple line)
    for $T_{\text{nr}} \gtrsim 1.35~\text{keV}$;
    photo-neutron (YBe) given by the solid red line.
    \subref{fig:spe_Dresden}
Illustration of different \cenns and ES predictions for the Dresden-II spectrum
compared with the Rx-ON data.
\cenns(SM) is the Standard Model \cenns prediction.
\cenns($\mu_{\nu_{e}}$) and \cenns+ ES($\mu_{\nu_{e}}$)
are, respectively, the \cenns(SM) plus \cenns and \cenns+ ES
predictions induced by $\mu_{\nu_{e}}=3\times10^{-10}~\mu_{\text{B}}$.
\cenns($q_{\nu_{e}}$)
is the \cenns(SM) plus \cenns
prediction induced by $q_{\nu_{e}}=1.5\times10^{-9}~e$.
\cenns+ ES($q_{\nu_{e}}$)
is the \cenns(SM) plus \cenns+ ES
prediction induced by $q_{\nu_{e}}=1.3\times10^{-11}~e$.
The dashed histograms represent the predictions
obtained considering the HMVE reactor antineutrino flux
and Fef quenching.
The solid histograms show the corresponding predictions plus fitted background.
    }
    \label{}
\end{figure}

The theoretical \cenns event-number $N^\mathrm{CE \nu NS}_{i}$ in each electron-recoil energy-bin $i$ is given by
\begin{equation}\label{N_cevns_dresden}
N^\mathrm{\mathrm{CE}\nu\mathrm{NS}}_i(\mathcal{N})
=
N({\rm{Ge}})
\int_{T_{\mathrm{e}}^{i}}^{T_{\mathrm{e}}^{i+1}}
\hspace{-0.3cm}
d T_{\mathrm{e}}\,
\int_{T^{\prime\text{min}}_{\text{nr}}}^{T^{\prime\text{max}}_{\text{nr}}}
\hspace{-0.3cm}
dT'_{\text{nr}}
\,
R(T_{\text{e}},T'_{\text{e}}(T'_{\text{nr}}))
\int_{E_{\text{min}}(T'_{\text{nr}})}^{E_{\text{max}}}
\hspace{-0.3cm}
d E
\frac{d N_{\overline{\nu}}}{d E}(E)
\frac{d \sigma_{\overline{\nu}-\mathcal{N}}}{d T'_{\mathrm{nr}}}(E, T'_{\mathrm{nr}})
,
\end{equation}
where
$\mathcal{N}$= $^{A}_{Z}\rm{Ge}$ with $A=70,72,73,74,76$, and $N_{i}^{\mathrm{CE}\nu\mathrm{NS}}(\mathrm{Ge})=\sum_A \textit{f}(^{A}_{Z}\mathrm{Ge}) N_{i}^{\mathrm{CE}\nu\mathrm{NS}}(^{A}_{Z}\mathrm{Ge})$, where $\textit{f}(^{A}_{Z}\rm{Ge})$ are the isotopic abundances introduced in Sec.~\ref{sec:cs}. Moreover, $N(\rm{Ge})=2.43\times10^{25}$ is the number of germanium atoms, $T^{\prime\text{min}}_{\text{nr}} \simeq 2.96$~eV is the minimum average ionization energy in Ge,  $R(T_{\text{e}},T'_{\text{e}}(T'_{\text{nr}}))$ is the detector energy resolution function, $T'_{\text{e}}(T'_{\text{nr}})=f_Q(T'_\mathrm{nr})T'_\mathrm{nr}$ is the ionization energy where $f_Q$ is the germanium quenching factor.
For the latter, following the data release in Ref.~\cite{Colaresi:2022obx} we consider two models based on experimental measurements. The first one determined from photo-neutron source measurements, so-called YBe~\cite{Collar:2021fcl}, and the second one derived from iron-filtered monochromatic neutrons, so-called Fef, that consists in a simple linear fit of the four data points for  $T_{\text{nr}} \lesssim 1.35~\text{keV}$ and is extended above this range with the standard Lindhard model with $k=0.157$~\cite{osti_4701226}.
The different quenching factors are shown in Fig.~\ref{fig:qf_Dresden} together with the experimental points used for the determination of the Fef model.
The differences in the constraints derived using the two quenching factors are used as an estimate of the related uncertainty.
The detector energy-resolution function is described as a truncated Gaussian
\begin{equation}
    R\left(T_{\mathrm{e}}, T_{\mathrm{e}}^\prime(T'_{\text{nr}})  \right)=\left(\frac{2}{1+\operatorname{Erf}\left(\frac{T_{\mathrm{e}}^\prime(T'_{\text{nr}})}{\sqrt{2} \sigma_{e}^\prime}\right)}\right) \frac{1}{\sqrt{2 \pi} \sigma_{e}^\prime} e^{-\frac{\left(T_{\mathrm{e}}-T_{\mathrm{e}}^\prime(T'_{\text{nr}})\right)^{2}}{2 {\sigma_{e}^\prime}^{2}}},
\end{equation}
with a standard deviation equal to $\sigma_{e}^\prime=\sqrt{\sigma_{n}^2+\eta F_f T_{\mathrm{e}} }$, where the average energy of electron-hole formation is $\eta=2.96~\mathrm{eV}$ and the Fano factor is $F_f=0.11$ for Ge~\cite{Colaresi:2022obx}.
Finally, in Eq.~(\ref{N_cevns_dresden}) the experimental acceptance does not appear since the data points provided in the data release are already corrected for it. \\

Similarly to the CsI analysis, we also include the contribution of the electron-antineutrino scattering.
In each electron-recoil energy-bin $i$, the theoretical ES event number $N^\mathrm{ES}_{i}$  is given by
\begin{equation}\label{N_ES_Ge}
N_{i}^{\mathrm{ES}}
=
N(\mathrm{Ge})
\int_{T_{\mathrm{e}}^{i}}^{T_{\mathrm{e}}^{i+1}}
\hspace{-0.3cm}
d T_{\mathrm{e}}\,
\int_{T^{\prime\text{min}}_{\text{e}}}^{T^{\prime\text{max}}_{\text{e}}}
\hspace{-0.3cm}
dT'_{\text{e}}
\,
R(T_{\text{e}},T'_{\text{e}})
\int_{E_{\text{min}}(T'_{\text{e}})}^{E_{\text{max}}}
\hspace{-0.3cm}
d E
\frac{d N_{\overline{\nu}}}{d E}(E)
\frac{d \sigma^{\rm{ES}}_{\overline{\nu}-\mathrm{Ge}}}{d T'_{\mathrm{e}}}(E, T'_{\mathrm{e}})
,
\end{equation}
with the difference that in the energy resolution the quenching factor must be set to unity.

We performed the analysis of the Dresden-II Ge data using the least-squares function
%\begin{equation}
 %   \chi^2_{\rm CE\nu NS}=\sum_{i=1}^{130}\Big(\dfrac{N^i_{\rm bkg}+\alpha N^i_{\rm CE\nu NS}-N^i_{\rm exp}}{\sigma_{\rm exp}}\Big)^2+\Big(\dfrac{\beta-\beta^{th}}{\sigma_{\beta^{th}}}\Big)^2+\Big(\dfrac{\eta-1}{\sigma_\eta}\Big)^2,
%\end{equation}
\begin{equation}
\label{eq:chi2ge}
    \chi^2_{\rm Ge, CE\nu NS+ES}=\sum_{i=1}^{130}\Big(\dfrac{N_i^{\rm bkg}+\alpha (N_i^{\rm CE\nu NS}+N_i^{\rm ES})-N_i^{\rm exp}}{\sigma_{\rm exp}}\Big)^2+\Big(\dfrac{\beta-\beta_{\rm M/L1}}{\sigma_{\beta_{\rm M/L1}}}\Big)^2+\Big(\dfrac{\alpha-1}{\sigma_\alpha}\Big)^2,
\end{equation}
where $N_i^{\rm bkg}$, $N_i^{\rm CE\nu NS}$ and $N_i^{\rm ES}$ are the predictions in the $i$-th electron recoil energy bin for the background, the CE$\nu$NS signal and the ES signal, respectively, and $N_i^{\rm exp}$ is the experimental number of events in the $i$-th bin. The nuisance parameter $\alpha$ takes into account the uncertainty on the neutrino flux (with $\sigma_\alpha=2\%$), while $\beta_{\rm M/L1}$ is a prior for the M- to $\mathrm{L}1$-shells ratio, with $\beta_{\rm M/L1}=0.16$ and $\sigma_{\beta_{\rm M/L1}}=0.03$. In this work, to appreciate the impact of the ES contribution, we will sometimes fit the Ge data set for \cenns only. In this case, the least-squares function $\chi^2_{\rm Ge, CE\nu NS}$ is obtained by removing the ES contribution from Eq.~(\ref{eq:chi2ge}).

In Fig.~\ref{fig:spe_Dresden} we show the \cenns and ES predictions for the Dresden-II spectrum
compared with the Rx-ON data under different hypotheses and with or without the inclusion of the background.
In this way, one can compare the SM \cenns prediction, CE$\nu$NS(SM), with the predictions obtained in presence of a possible neutrino MM, considering $\mu_{\nu_{e}}=3\times10^{-10}~\mu_{\text{B}}$, and a possible neutrino EC, considering $q_{\nu_{e}}=1.5\times10^{-9}~e$.
Moreover, in Fig.~\ref{fig:spe_Dresden} we also illustrate the impact of including the neutrino-electron elastic scattering, for the same neutrino MM value as before and for a much smaller neutrino EC, namely
$q_{\nu_{e}}=1.3\times10^{-11}~e$, given that, as we already pointed out in Sec.~\ref{sec:nuec}, the ES process is very sensitive to a possible neutrino millicharge.

%%%%%%%%%%%%%%%%%%%%%%%%%%%%%%%%%%%%%%%%%%%%%%%%%%%%%%%%%%%%%%%%%%%
%%%%%%%%%%%%%%%%%%%%%%%%%%%%%%%%%%%%%%%%%%%%%%%%%%%%%%%%%%%%%%%%%%%
%%%%%%%%%%%%%%%%%%%%%%%%%%%%%%%%%%%%%%%%%%%%%%%%%%%%%%%%%%%%%%%%%%%

\section{Results}
\label{sec:result}

In this section, we present the results of the fit using the COHERENT CsI and Ar data set and their combination, as well as the analysis of the Dresden-II data and its combination with COHERENT for the neutrino charge radii, electric charge and magnetic moment. We also present the DRESDEN-II results on the weak mixing angle.

%%%%%%%%%%%%%%%%%%%%%%%%%%%%%%%%%%%%%%%%%%%%%%%%%%%%%%%%%%%%%%%%%%%

\subsection{Weak mixing angle}
The weak mixing angle, $\vartheta_{\text{W}}$, is a fundamental parameter in the theory of EW interactions. So far, many experiments measured it at different energies~\cite{ParticleDataGroup:2020ssz},  since its value can be significantly modified in some BSM theories~\cite{Cadeddu:2021dqx}. In particular, low-energy determinations of $\vartheta_{\text{W}}$ offer a unique role, complementary to those at high-energy, being highly sensitive to extra $Z$ ($Z'$) bosons predicted  in grand unified theories,  technicolor models, supersymmetry and string theories~\cite{Safronova_2018}. This underscores  the need  for improved  experimental determinations of $\vartheta_{\text{W}}$ in the low-energy regime, where most of the measurements still suffer from large uncertainties. \\

As shown in Ref.~\cite{Cadeddu:2020lky}, the uncertainty obtained for the weak mixing angle from the old  CsI 2017 COHERENT data set combined with the Ar one is still very large when compared to the other determinations at low-momentum transfer. Moreover, as shown in Ref.~\cite{Cadeddu:2021ijh}, the COHERENT weak mixing angle determination is strongly correlated with the value chosen for the poorly known $R_n(\mathrm{Cs})$ and $R_n(\mathrm{I})$, making it necessary to fit for these parameters simultaneously in order to obtain a model-independent measurement of $\sin^2{\vartheta_{\text{W}}}$.  By performing a combined analysis with the so-called atomic parity violation (APV) experimental result using Cs atoms, as demonstrated in Ref.~\cite{Cadeddu:2018izq}, it has been possible to put rather stringent constraints on the weak mixing angle while keeping into account the correlation with $R_n(\mathrm{Cs})$.

\begin{figure}
\centering
\subfigure[]{\label{fig:s2tw-dresden}
\includegraphics[width=0.48\textwidth]{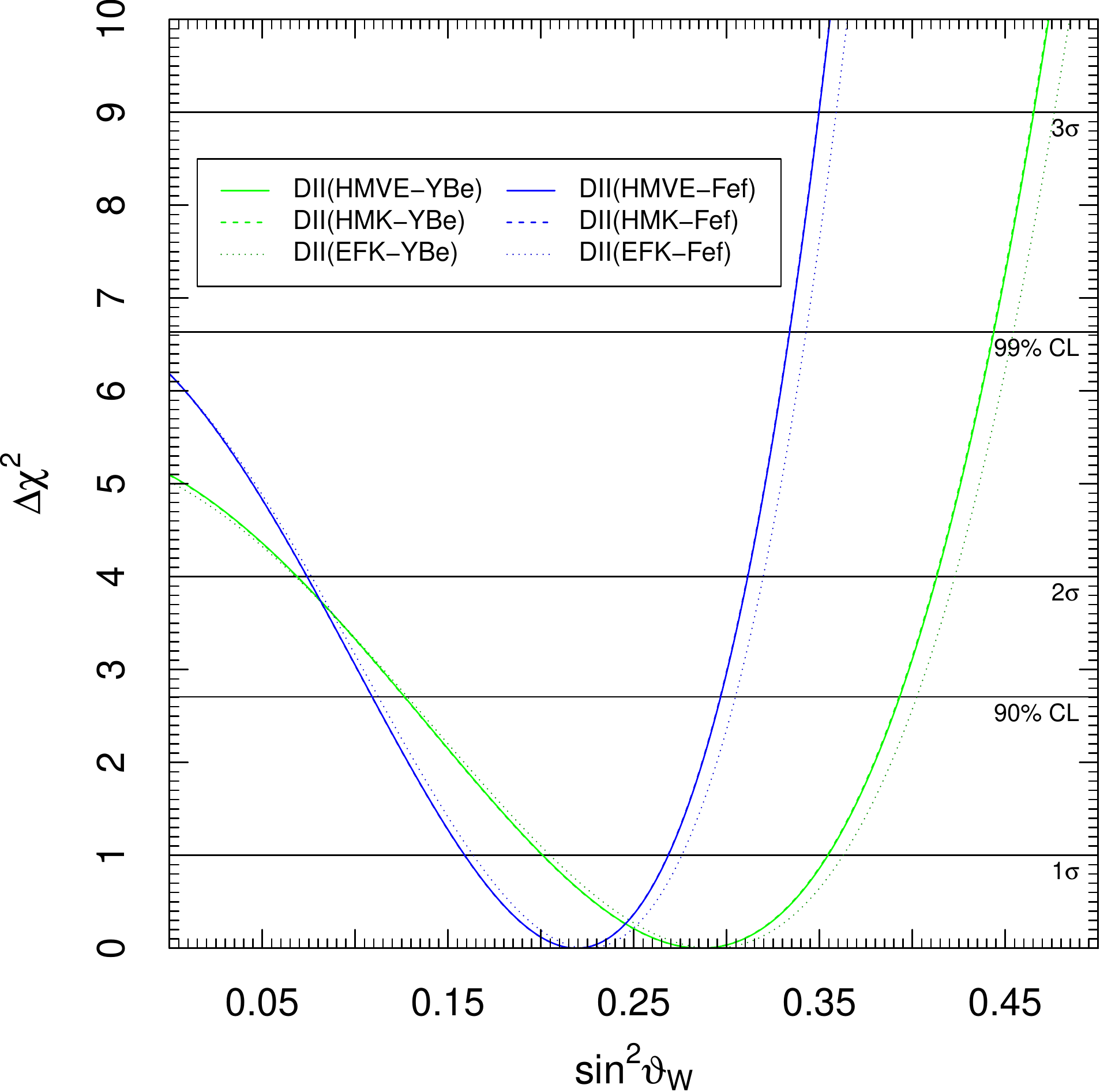}}
\subfigure[]{\label{fig:running}
\includegraphics*[width=0.48\textwidth]{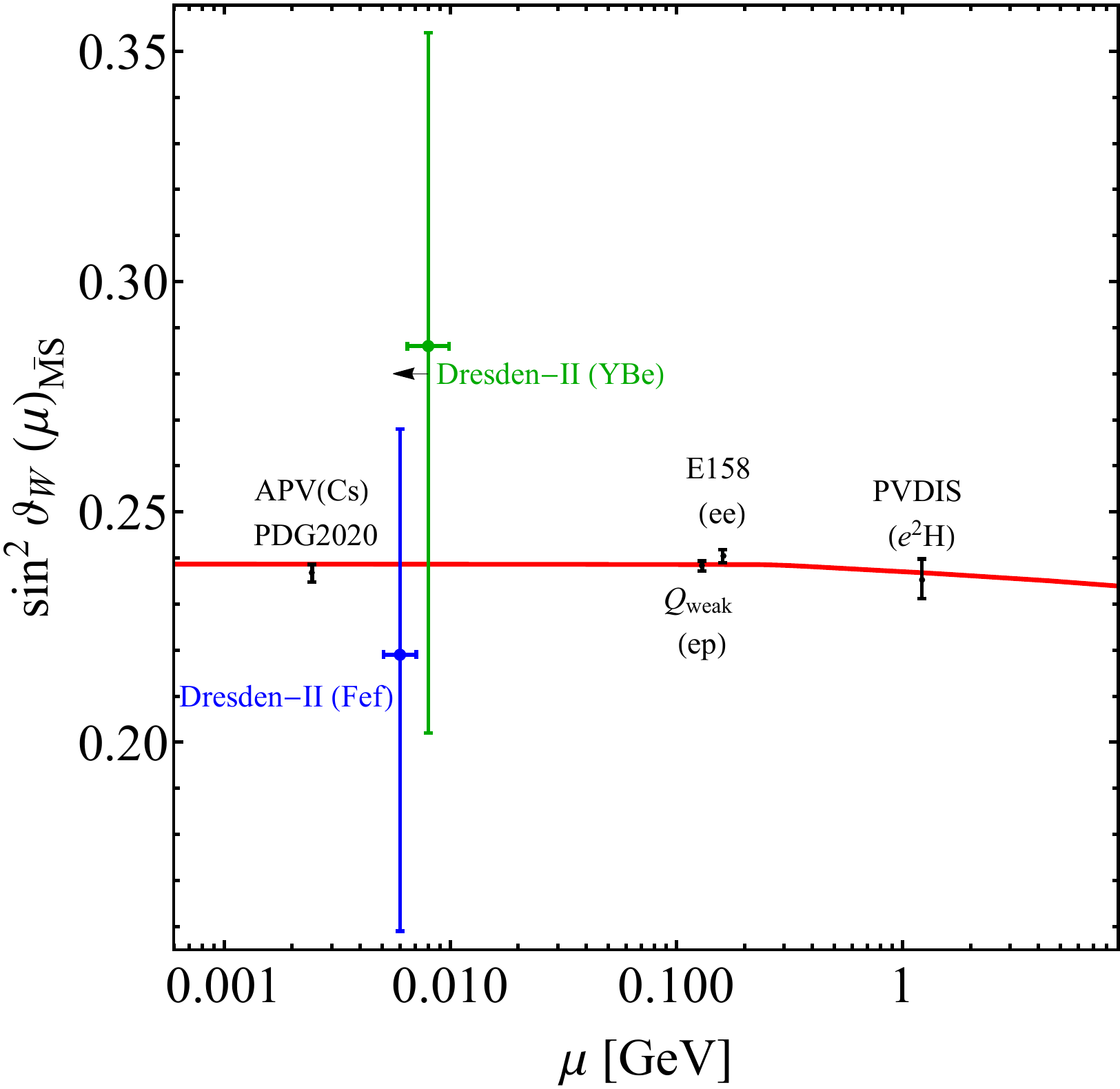}
}
\caption{ (a) Marginal $\Delta\chi^2$'s
for
$\sin^2 \vartheta_{\text{W}}$
obtained from the analysis of the 
Dresden-II data
assuming the HMVE, HMK, or EFK reactor antineutrino flux and
the Fef (blue) or YBe (green) quenching.
(b)
Variation of $\sin^2 \vartheta_{\text{W}}$ with the energy scale $\mu$. The SM prediction is shown as the red solid curve, together with
experimental determinations in black 
from APV on cesium~\cite{Wood:1997zq,Dzuba:2012kx}, which has a typical momentum transfer given by $\langle Q\rangle\simeq$~2.4 MeV, M{\o}ller scattering~\cite{Anthony:2005pm} (E158), deep inelastic scattering of polarized electrons on deuterons~\cite{Wang:2014bba} ($ e^2H $ PVDIS), and the result from the proton's weak charge
at $\langle Q^2 \rangle= 0.0248$ GeV$^2$~\cite{Androic:2018kni} ($ \text{Q}_{\text{weak}} $). The results derived in this paper using the Dresden-II data are shown in blue and green for the Fef and YBe quenching factor, respectively. For clarity we displayed the YBe point horizontally to the right, as indicated by the arrow.}
\end{figure}

This strong correlation between $\sin^2{\vartheta_{\text{W}}}$ and the neutron distribution rms radius applies to all EW determinations of the weak mixing angle exploiting nuclei
that have been done so far, see e.g. Ref.~\cite{Corona:2021yfd}.
On the contrary, as pointed out in Sec.~\ref{sec:cs}, in the analysis of the Dresden-II data the form factor of both protons and neutrons is practically equal to unity, making the particular choice of the value of $R_n(\mathrm{Ge})$ completely irrelevant. 
Here, we show the result of a fit of the Dresden-II data aimed at the determination of the value of the weak mixing angle using three different antineutrino flux parameterizations, indicated as HMVE, HMK and EFK, and two different germanium QF functional forms, indicated as Fef and YBe. 
The results of these fits are summarized in Tab.~\ref{tab:tab22-s2tw-DII} for all the six combinations of neutrino fluxes and QFs, and are depicted in Fig.~\ref{fig:s2tw-dresden}. The impact of the different antineutrino fluxes is minimal. On the contrary, the impact of the different QFs is non-negligible, being the YBe results shifted to larger values of the weak mixing angle and also less precise. Focusing thus only on the HMVE flux, our results are
\begin{equation}
  \sin^2{\vartheta_{\text{W}}}(\mathrm{Dresden-II\,Fef}) = 0.219^{+0.06}_{-0.05}\,(1\sigma),
  ^{+0.11}_{-0.08}\,(90\%),
  ^{+0.14}_{-0.09}\,(2\sigma),
\end{equation}
\begin{equation}
  \sin^2{\vartheta_{\text{W}}}(\mathrm{Dresden-II\,YBe}) = 0.286^{+0.08}_{-0.07}\,(1\sigma),
  ^{+0.16}_{-0.11}\,(90\%),
  ^{+0.22}_{-0.13}\,(2\sigma),
\end{equation}
for the Fef and YBe quenching factors, respectively.
These results are also depicted in Fig.~\ref{fig:running}, where a summary of the weak mixing angle measurements as a function of the energy scale $\mu$ is shown along with the SM predicted running of $\sin^2 \vartheta_{\text{W}}$, calculated in the $\overline{\text{MS}}$ scheme~\cite{Tanabashi:2018oca, Erler:2004in,Erler:2017knj}.

\begin{table}
\begin{center}
%\resizebox{\textwidth}{!}
{
\begin{tabular}{cccccc}
Dresden-II
&
$\sin^{2}\theta_{W}^{\text{b.f.}}$
&
$1\sigma$
&
$90\%$
&
$2\sigma$
%&
%$99\%$
&
$3\sigma$
\\
\hline
HMVE-Fef
&
$0.219$
&
$( 0.159 , 0.268 )$
&
$( 0.110 , 0.296 )$
&
$( 0.0742 , 0.311 )$
%&
%$< 0.334 $
&
$< 0.349 $
\\
%\hline
HMK-Fef
&
$0.219$
&
$( 0.159 , 0.268 )$
&
$( 0.110 , 0.296 )$
&
$( 0.0742 , 0.311 )$
%&
%$< 0.334 $
&
$< 0.349 $
\\
%\hline
EFK-Fef
&
$0.226$
&
$( 0.164 , 0.275 )$
&
$( 0.113 , 0.304 )$
&
$( 0.0772 , 0.319 )$
%&
%$< 0.343 $
&
$< 0.358 $
\\
%\hline
HMVE-YBe
&
$0.286$
&
$( 0.202 , 0.354 )$
&
$( 0.127 , 0.393 )$
&
$( 0.0693 , 0.413 )$
%&
%$< 0.444 $
&
$< 0.465 $
\\
%\hline
HMK-YBe
&
$0.286$
&
$( 0.201 , 0.353 )$
&
$( 0.127 , 0.392 )$
&
$( 0.0693 , 0.412 )$
%&
%$< 0.443 $
&
$< 0.464 $
\\
%\hline
EFK-YBe
&
$0.293$
&
$( 0.206 , 0.362 )$
&
$( 0.129 , 0.402 )$
&
$( 0.0683 , 0.423 )$
%&
%$< 0.453 $
&
$< 0.476 $
\\
\hline
\end{tabular}

}
\end{center}
\caption{ \label{tab:tab22-s2tw-DII}
Best-fit value and bounds on $\sin^{2}\theta_{W}$
obtained
from the analysis of the Dresden-II data
assuming the HMVE, HMK, or EFK reactor antineutrino flux and
the Fef or YBe quenching.
}
\end{table}

We repeated all of the above measurements including also the ES contribution in the Dresden-II data set. However, no effect is found due to ES on the weak mixing angle, thus the results are independent of its inclusion.

Other bounds on $\sin^2{\vartheta_{\text{W}}}$ have also been obtained exploiting the \cenns data from COHERENT and Dresden-II in Ref.~\cite{AristizabalSierra:2022axl}. Although the results with the Fef QF appear to be more stringent than those presented in this work, the results are not comparable with ours because we fit the complete Dresden-II data set, whereas the analysis of Ref.~\cite{AristizabalSierra:2022axl} is a fit of the \cenns residual events obtained by fixing the background contribution as determined by the Dresden-II Collaboration. Such an analysis neglects the systematic uncertainties related to the background leading thus to more stringent constraints.
%%%%%%%%%%%%%%%%%%%%%%%%%%%%%%%%%%%%%%%%%%%%%%%%%%%%%%%%%%%%%%%%%%%

\begin{table}
\begin{center}
%\resizebox{\textwidth}{!}
{
\begin{tabular}{ccccc}
&
$1\sigma$
&
$90\%$
&
$2\sigma$
%&
%$99\%$
&
$3\sigma$
\\
\hline
\multicolumn{5}{c}{\bf CsI}
\\
%\hline
$\langle{r}_{\nu_{ee}}^2\rangle$
&
$( -62 , 10 )$
&
$( -68 , 14 )$
&
$( -70 , 16 )$
%&
%$( -74 , 19 )$
&
$( -77 , 22 )$
\\
%\hline
$\langle{r}_{\nu_{\mu\mu}}^2\rangle$
&
$( -37.9 , 0.5 )$
&
$( -57.4 , 2.9 )$
&
$( -59.2 , 4.4 )$
%&
%$( -61.9 , 6.8 )$
&
$( -64.0 , 8.6 )$
\\
%\hline
$|\langle{r}_{\nu_{e\mu}}^2\rangle|$
&
$< 26 $
&
$< 30 $
&
$< 31 $
%&
%$< 33 $
&
$< 34 $
\\
%\hline
$|\langle{r}_{\nu_{e\tau}}^2\rangle|$
&
$< 36 $
&
$< 41 $
&
$< 43 $
%&
%$< 46 $
&
$< 49 $
\\
%\hline
$|\langle{r}_{\nu_{\mu\tau}}^2\rangle|$
&
$< 27 $
&
$< 30 $
&
$< 32 $
%&
%$< 34 $
&
$< 36 $
\\
\hline
\multicolumn{5}{c}{\bf Ar}
\\
%\hline
$\langle{r}_{\nu_{ee}}^2\rangle$
&
$( -79 , 29 )$
&
$( -88 , 38 )$
&
$( -93 , 43 )$
%&
%$( -102 , 52 )$
&
$( -110 , 59 )$
\\
%\hline
$\langle{r}_{\nu_{\mu\mu}}^2\rangle$
&
$( -59.2 , 8.6 )$
&
$( -64.9 , 14.6 )$
&
$( -67.6 , 17.3 )$
%&
%$( -71.8 , 21.5 )$
&
$( -74.8 , 24.5 )$
\\
%\hline
$|\langle{r}_{\nu_{e\mu}}^2\rangle|$
&
$< 33 $
&
$< 36 $
&
$< 38 $
%&
%$< 41 $
&
$< 44 $
\\
%\hline
$|\langle{r}_{\nu_{e\tau}}^2\rangle|$
&
$< 54 $
&
$< 63 $
&
$< 68 $
%&
%$< 77 $
&
$< 84 $
\\
%\hline
$|\langle{r}_{\nu_{\mu\tau}}^2\rangle|$
&
$< 34 $
&
$< 40 $
&
$< 42 $
%&
%$< 47 $
&
$< 50 $
\\
\hline
\multicolumn{5}{c}{\bf CsI + Ar}
\\
%\hline
$\langle{r}_{\nu_{ee}}^2\rangle$
&
$( -66 , 11 )$
&
$( -69 , 14 )$
&
$( -71 , 16 )$
%&
%$( -74 , 20 )$
&
$( -77 , 22 )$
\\
%\hline
$\langle{r}_{\nu_{\mu\mu}}^2\rangle$
&
$( -54.7 , 0.8 )$
&
$( -57.7 , 3.2 )$
&
$( -59.2 , 4.7 )$
%&
%$( -61.6 , 6.8 )$
&
$( -63.1 , 8.3 )$
\\
%\hline
$|\langle{r}_{\nu_{e\mu}}^2\rangle|$
&
$< 28 $
&
$< 30 $
&
$< 31 $
%&
%$< 33 $
&
$< 34 $
\\
%\hline
$|\langle{r}_{\nu_{e\tau}}^2\rangle|$
&
$< 38 $
&
$< 42 $
&
$< 44 $
%&
%$< 47 $
&
$< 50 $
\\
%\hline
$|\langle{r}_{\nu_{\mu\tau}}^2\rangle|$
&
$< 28 $
&
$< 31 $
&
$< 32 $
%&
%$< 34 $
&
$< 36 $
\\
\hline
\end{tabular}

}
\end{center}
\caption{ \label{tab:tab22-chr5-CsI}
Bounds on the neutrino charge radii in units of $10^{-32}~\text{cm}^2$
obtained from the analysis of the COHERENT CsI and Ar data.
}
\end{table}

\begin{table}
\begin{center}
%\resizebox{\textwidth}{!}
{
\begin{tabular}{ccccc}
&
$1\sigma$
&
$90\%$
&
$2\sigma$
%&
%$99\%$
&
$3\sigma$
\\
\hline
\multicolumn{5}{c}{\bf Dresden-II (HMVE-Fef)}
\\
%\hline
$\langle{r}_{\nu_{ee}}^2\rangle$
&
$( -54 , 2 )$
&
$( -56 , 4 )$
&
$( -58 , 5 )$
%&
%$( -60 , 7 )$
&
$( -61 , 8 )$
\\
%\hline
$|\langle{r}_{\nu_{e\mu}}^2\rangle|$, $|\langle{r}_{\nu_{e\tau}}^2\rangle|$
&
$< 28 $
&
$< 30 $
&
$< 32 $
%&
%$< 33 $
&
$< 35 $
\\
\hline
\multicolumn{5}{c}{\bf Dresden-II (HMK-Fef)}
\\
%\hline
$\langle{r}_{\nu_{ee}}^2\rangle$
&
$( -54 , 2 )$
&
$( -57 , 4 )$
&
$( -58 , 5 )$
%&
%$( -60 , 7 )$
&
$( -61 , 8 )$
\\
%\hline
$|\langle{r}_{\nu_{e\mu}}^2\rangle|$, $|\langle{r}_{\nu_{e\tau}}^2\rangle|$
&
$< 28 $
&
$< 30 $
&
$< 31 $
%&
%$< 34 $
&
$< 35 $
\\
\hline
\multicolumn{5}{c}{\bf Dresden-II (EFK-Fef)}
\\
%\hline
$\langle{r}_{\nu_{ee}}^2\rangle$
&
$( -55 , 2 )$
&
$( -57 , 5 )$
&
$( -58 , 6 )$
%&
%$( -60 , 8 )$
&
$( -62 , 9 )$
\\
%\hline
$|\langle{r}_{\nu_{e\mu}}^2\rangle|$, $|\langle{r}_{\nu_{e\tau}}^2\rangle|$
&
$< 28 $
&
$< 31 $
&
$< 32 $
%&
%$< 34 $
&
$< 36 $
\\
\hline
\multicolumn{5}{c}{\bf Dresden-II (HMVE-YBe)}
\\
%\hline
$\langle{r}_{\nu_{ee}}^2\rangle$
&
$( -61 , 9 )$
&
$( -65 , 12 )$
&
$( -66 , 14 )$
%&
%$( -69 , 16 )$
&
$( -71 , 18 )$
\\
%\hline
$|\langle{r}_{\nu_{e\mu}}^2\rangle|$, $|\langle{r}_{\nu_{e\tau}}^2\rangle|$
&
$< 35 $
&
$< 38 $
&
$< 40 $
%&
%$< 43 $
&
$< 44 $
\\
\hline
\multicolumn{5}{c}{\bf Dresden-II (HMK-YBe)}
\\
%\hline
$\langle{r}_{\nu_{ee}}^2\rangle$
&
$( -61 , 9 )$
&
$( -65 , 12 )$
&
$( -66 , 14 )$
%&
%$( -69 , 16 )$
&
$( -71 , 18 )$
\\
%\hline
$|\langle{r}_{\nu_{e\mu}}^2\rangle|$, $|\langle{r}_{\nu_{e\tau}}^2\rangle|$
&
$< 35 $
&
$< 38 $
&
$< 40 $
%&
%$< 43 $
&
$< 44 $
\\
\hline
\multicolumn{5}{c}{\bf Dresden-II (EFK-YBe)}
\\
%\hline
$\langle{r}_{\nu_{ee}}^2\rangle$
&
$( -62 , 10 )$
&
$( -65 , 13 )$
&
$( -67 , 15 )$
%&
%$( -70 , 17 )$
&
$( -72 , 19 )$
\\
%\hline
$|\langle{r}_{\nu_{e\mu}}^2\rangle|$, $|\langle{r}_{\nu_{e\tau}}^2\rangle|$
&
$< 36 $
&
$< 39 $
&
$< 41 $
%&
%$< 44 $
&
$< 45 $
\\
\hline
\end{tabular}

}
\end{center}
\caption{ \label{tab:tab22-chr5-DII}
Bounds on the neutrino charge radii in units of $10^{-32}~\text{cm}^2$
obtained from the analysis of the Dresden-II data
assuming the HMVE, HMK, or EFK reactor antineutrino flux and
the Fef or YBe quenching.
}
\end{table}

\begin{table}
\begin{center}
%\resizebox{\textwidth}{!}
{
\begin{tabular}{ccccc}
&
$1\sigma$
&
$90\%$
&
$2\sigma$
%&
%$99\%$
&
$3\sigma$
\\
\hline
\multicolumn{5}{c}{\bf CsI + Ar + Dresden-II (HMVE-Fef)}
\\
%\hline
$\langle{r}_{\nu_{ee}}^2\rangle$
&
$( -52 , 3 )$
&
$( -56 , 5 )$
&
$( -58 , 6 )$
%&
%$( -60 , 7 )$
&
$( -61 , 9 )$
\\
%\hline
$\langle{r}_{\nu_{\mu\mu}}^2\rangle$
&
$( -55.6 , 1.8 )$
&
$( -58.2 , 4.0 )$
&
$( -59.8 , 5.1 )$
%&
%$( -61.7 , 7.1 )$
&
$( -63.1 , 8.7 )$
\\
%\hline
$|\langle{r}_{\nu_{e\mu}}^2\rangle|$
&
$< 28 $
&
$< 29 $
&
$< 30 $
%&
%$< 31 $
&
$< 32 $
\\
%\hline
$|\langle{r}_{\nu_{e\tau}}^2\rangle|$
&
$< 28 $
&
$< 31 $
&
$< 32 $
%&
%$< 34 $
&
$< 35 $
\\
%\hline
$|\langle{r}_{\nu_{\mu\tau}}^2\rangle|$
&
$< 29 $
&
$< 32 $
&
$< 33 $
%&
%$< 35 $
&
$< 36 $
\\
\hline
\multicolumn{5}{c}{\bf CsI + Ar + Dresden-II (HMK-Fef)}
\\
%\hline
$\langle{r}_{\nu_{ee}}^2\rangle$
&
$( -52 , 3 )$
&
$( -56 , 5 )$
&
$( -58 , 6 )$
%&
%$( -60 , 7 )$
&
$( -61 , 9 )$
\\
%\hline
$\langle{r}_{\nu_{\mu\mu}}^2\rangle$
&
$( -55.8 , 1.8 )$
&
$( -58.4 , 3.8 )$
&
$( -59.8 , 5.4 )$
%&
%$( -61.7 , 7.1 )$
&
$( -63.3 , 8.7 )$
\\
%\hline
$|\langle{r}_{\nu_{e\mu}}^2\rangle|$
&
$< 28 $
&
$< 29 $
&
$< 30 $
%&
%$< 31 $
&
$< 32 $
\\
%\hline
$|\langle{r}_{\nu_{e\tau}}^2\rangle|$
&
$< 28 $
&
$< 31 $
&
$< 32 $
%&
%$< 34 $
&
$< 35 $
\\
%\hline
$|\langle{r}_{\nu_{\mu\tau}}^2\rangle|$
&
$< 29 $
&
$< 31 $
&
$< 33 $
%&
%$< 35 $
&
$< 36 $
\\
\hline
\multicolumn{5}{c}{\bf CsI + Ar + Dresden-II (EFK-Fef)}
\\
%\hline
$\langle{r}_{\nu_{ee}}^2\rangle$
&
$( -53 , 3 )$
&
$( -58 , 5 )$
&
$( -58 , 6 )$
%&
%$( -60 , 8 )$
&
$( -62 , 9 )$
\\
%\hline
$\langle{r}_{\nu_{\mu\mu}}^2\rangle$
&
$( -55.8 , 1.8 )$
&
$( -58.4 , 4.0 )$
&
$( -59.3 , 4.9 )$
%&
%$( -61.5 , 6.9 )$
&
$( -62.8 , 8.4 )$
\\
%\hline
$|\langle{r}_{\nu_{e\mu}}^2\rangle|$
&
$< 28 $
&
$< 29 $
&
$< 30 $
%&
%$< 31 $
&
$< 32 $
\\
%\hline
$|\langle{r}_{\nu_{e\tau}}^2\rangle|$
&
$< 29 $
&
$< 32 $
&
$< 33 $
%&
%$< 35 $
&
$< 36 $
\\
%\hline
$|\langle{r}_{\nu_{\mu\tau}}^2\rangle|$
&
$< 29 $
&
$< 31 $
&
$< 33 $
%&
%$< 35 $
&
$< 36 $
\\
\hline
\multicolumn{5}{c}{\bf CsI + Ar + Dresden-II (HMVE-YBe)}
\\
%\hline
$\langle{r}_{\nu_{ee}}^2\rangle$
&
$( -60 , 7 )$
&
$( -63 , 10 )$
&
$( -65 , 12 )$
%&
%$( -67 , 14 )$
&
$( -69 , 15 )$
\\
%\hline
$\langle{r}_{\nu_{\mu\mu}}^2\rangle$
&
$( -54.3 , 0.74 )$
&
$( -57.3 , 3.2 )$
&
$( -58.9 , 4.3 )$
%&
%$( -60.6 , 6.0 )$
&
$( -62.2 , 7.8 )$
\\
%\hline
$|\langle{r}_{\nu_{e\mu}}^2\rangle|$
&
$< 28 $
&
$< 30 $
&
$< 31 $
%&
%$< 32 $
&
$< 33 $
\\
%\hline
$|\langle{r}_{\nu_{e\tau}}^2\rangle|$
&
$< 35 $
&
$< 37 $
&
$< 38 $
%&
%$< 41 $
&
$< 42 $
\\
%\hline
$|\langle{r}_{\nu_{\mu\tau}}^2\rangle|$
&
$< 28 $
&
$< 30 $
&
$< 32 $
%&
%$< 34 $
&
$< 35 $
\\
\hline
\multicolumn{5}{c}{\bf CsI + Ar + Dresden-II (HMK-YBe)}
\\
%\hline
$\langle{r}_{\nu_{ee}}^2\rangle$
&
$( -60 , 8 )$
&
$( -63 , 10 )$
&
$( -65 , 12 )$
%&
%$( -67 , 14 )$
&
$( -69 , 15 )$
\\
%\hline
$\langle{r}_{\nu_{\mu\mu}}^2\rangle$
&
$( -53.8 , 0.74 )$
&
$( -57.1 , 2.9 )$
&
$( -58.4 , 4.0 )$
%&
%$( -60.6 , 6.0 )$
&
$( -62.2 , 7.8 )$
\\
%\hline
$|\langle{r}_{\nu_{e\mu}}^2\rangle|$
&
$< 28 $
&
$< 30 $
&
$< 31 $
%&
%$< 32 $
&
$< 33 $
\\
%\hline
$|\langle{r}_{\nu_{e\tau}}^2\rangle|$
&
$< 35 $
&
$< 37 $
&
$< 38 $
%&
%$< 40 $
&
$< 42 $
\\
%\hline
$|\langle{r}_{\nu_{\mu\tau}}^2\rangle|$
&
$< 28 $
&
$< 30 $
&
$< 32 $
%&
%$< 34 $
&
$< 35 $
\\
\hline
\multicolumn{5}{c}{\bf CsI + Ar + Dresden-II (EFK-YBe)}
\\
%\hline
$\langle{r}_{\nu_{ee}}^2\rangle$
&
$( -61 , 8 )$
&
$( -64 , 11 )$
&
$( -65 , 12 )$
%&
%$( -67 , 14 )$
&
$( -69 , 16 )$
\\
%\hline
$\langle{r}_{\nu_{\mu\mu}}^2\rangle$
&
$( -54.0 , 0.74 )$
&
$( -57.3 , 2.9 )$
&
$( -58.4 , 4.0 )$
%&
%$( -60.6 , 6.0 )$
&
$( -62.2 , 7.6 )$
\\
%\hline
$|\langle{r}_{\nu_{e\mu}}^2\rangle|$
&
$< 28 $
&
$< 30 $
&
$< 31 $
%&
%$< 32 $
&
$< 33 $
\\
%\hline
$|\langle{r}_{\nu_{e\tau}}^2\rangle|$
&
$< 35 $
&
$< 37 $
&
$< 39 $
%&
%$< 41 $
&
$< 43 $
\\
%\hline
$|\langle{r}_{\nu_{\mu\tau}}^2\rangle|$
&
$< 28 $
&
$< 30 $
&
$< 31 $
%&
%$< 34 $
&
$< 35 $
\\
\hline
\end{tabular}

}
\end{center}
\caption{ \label{tab:tab22-chr5-tot}
Bounds on the neutrino charge radii in units of $10^{-32}~\text{cm}^2$
obtained from the combined analysis of the COHERENT CsI and Ar data
and the Dresden-II data
assuming the HMVE, HMK, or EFK reactor antineutrino flux and
the Fef or YBe quenching.
}
\end{table}

\begin{figure}
\begin{center}
\includegraphics*[height=0.48\textwidth]{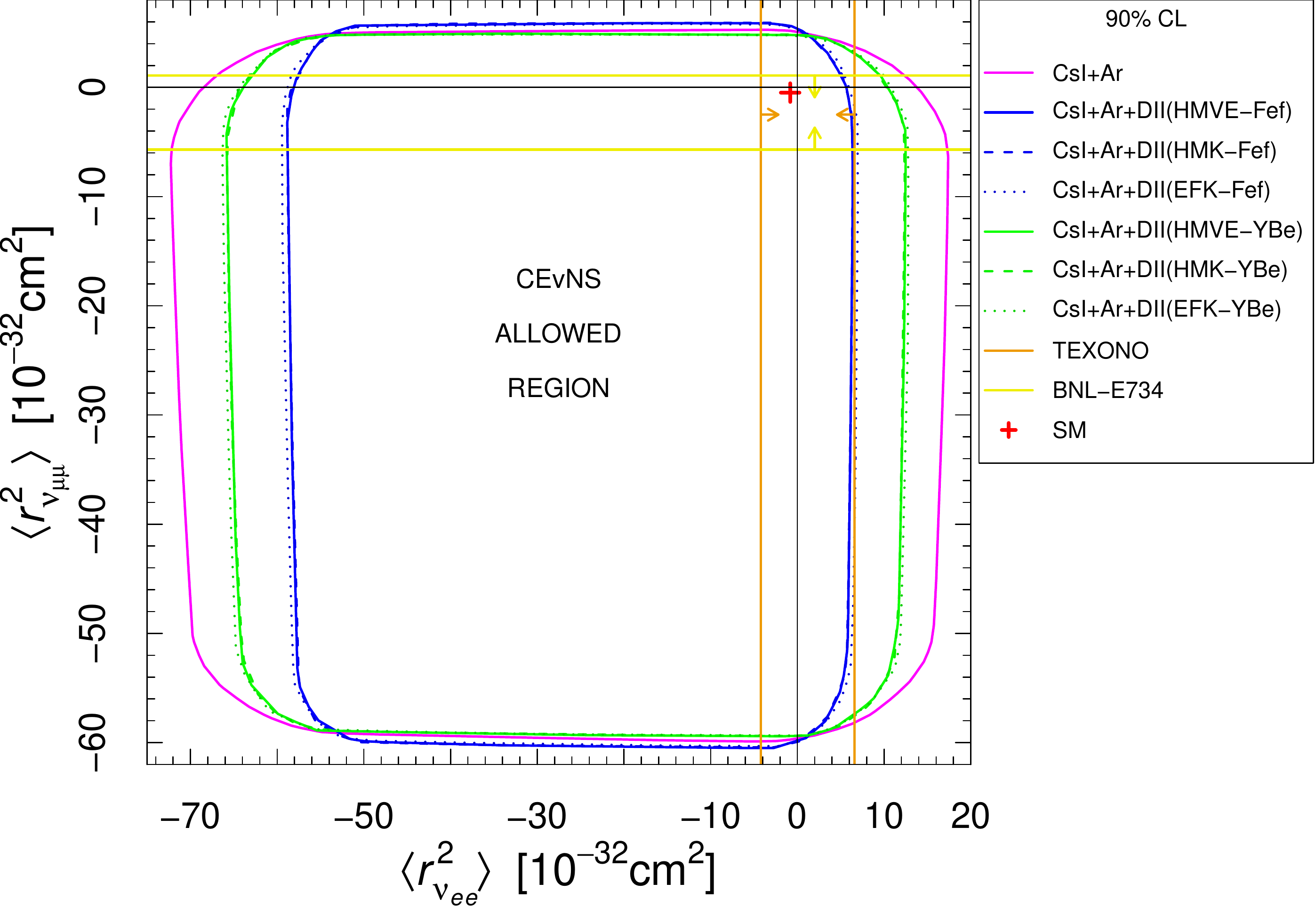}
\caption{ \label{fig:chr5-fig22-contour-ee-mm}
Contours of the 90\% C.L. allowed regions in the
$(\langle{r}_{\nu_{ee}}^2\rangle,\langle{r}_{\nu_{\mu\mu}}^2\rangle)$
plane
obtained from the analysis of the COHERENT
CsI and Ar data (magenta),
and from the combined analysis of the COHERENT data and
Dresden-II data
assuming the HMVE, HMK, or EFK reactor antineutrino flux and
the Fef (blue) or YBe (green) quenching.
The red cross near the origin indicates the Standard Model values
in Eqs.~(\ref{reSM}) and (\ref{rmSM}).
The orange and yellow lines delimit, respectively, the 90\% bounds on
$\langle{r}_{\nu_{ee}}^2\rangle$
and
$\langle{r}_{\nu_{\mu\mu}}^2\rangle$
obtained in the
TEXONO~\cite{Deniz:2009mu}
and
BNL-E734~\cite{Ahrens:1990fp}
experiments.
}
\end{center}
\end{figure}

\begin{figure}
\begin{center}
\includegraphics*[height=0.48\textwidth]{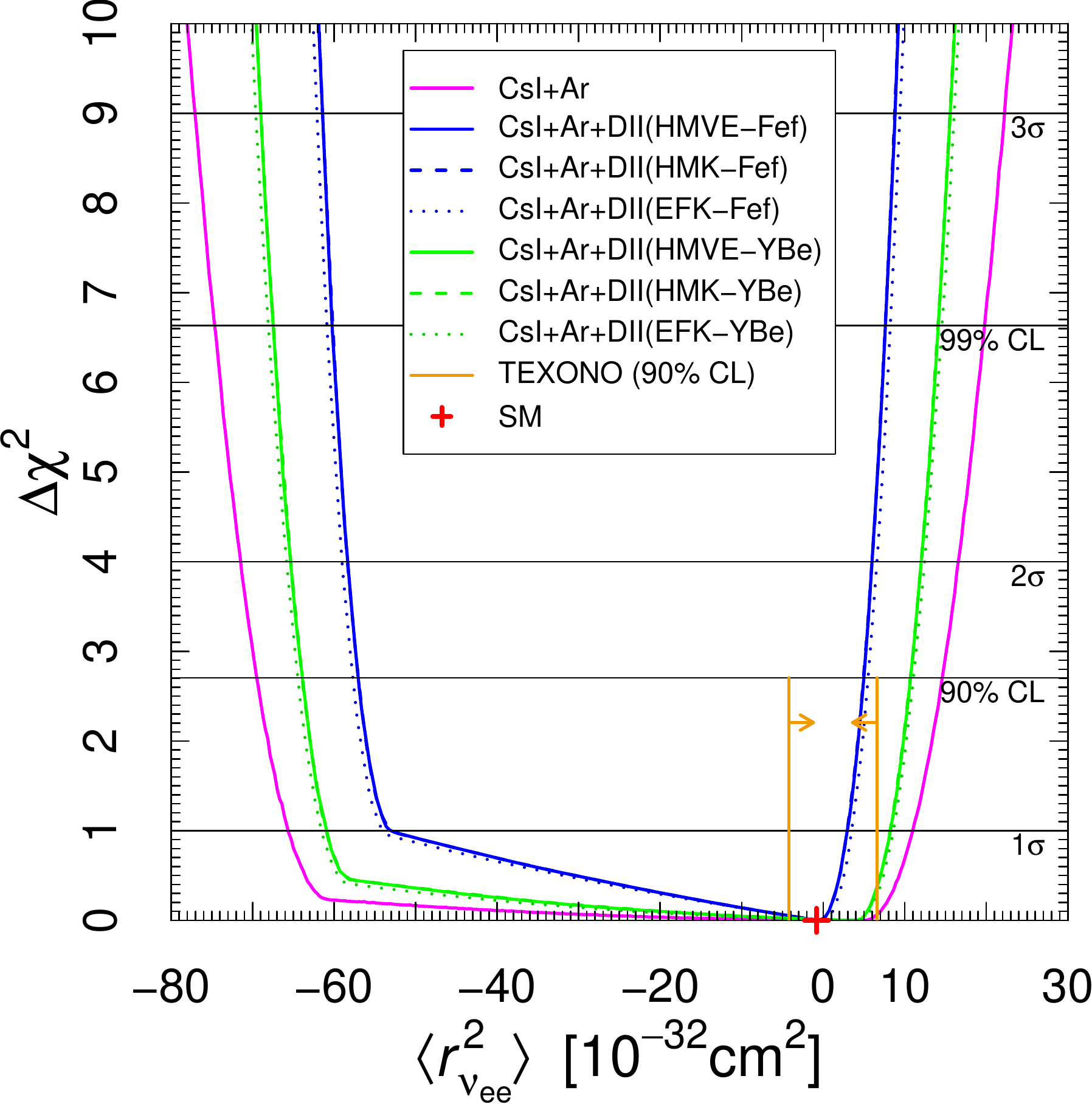}
\caption{ \label{fig:chr5-fig22-chi2-ee}
Marginal $\Delta\chi^2$'s
for
$\langle{r}_{\nu_{ee}}^2\rangle$
obtained from the analysis of the COHERENT
CsI and Ar data (magenta),
and from the combined analysis of the COHERENT data and
Dresden-II data
assuming the HMVE, HMK, or EFK reactor antineutrino flux and
the Fef (blue) or YBe (green) quenching.
The red cross near the origin indicates the Standard Model value
in Eq.~(\ref{reSM}).
The short vertical orange lines show the lower and upper 90\% bounds on
$\langle{r}_{\nu_{ee}}^2\rangle$
obtained in the
TEXONO~\cite{Deniz:2009mu}
experiment.
}
\end{center}
\end{figure}

\begin{figure}
\begin{center}
\includegraphics*[height=0.48\textwidth]{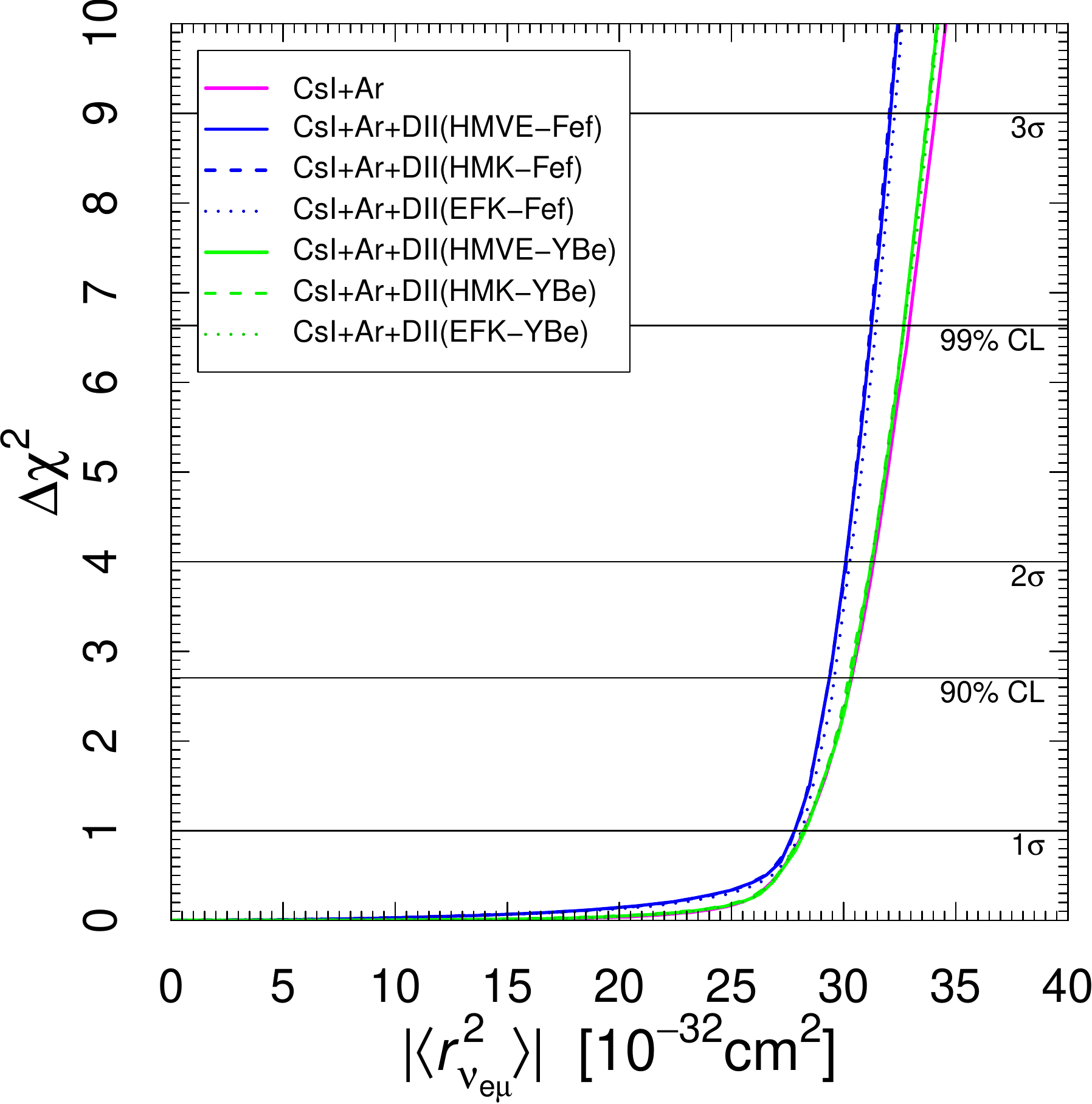}
\hfill
\includegraphics*[height=0.48\textwidth]{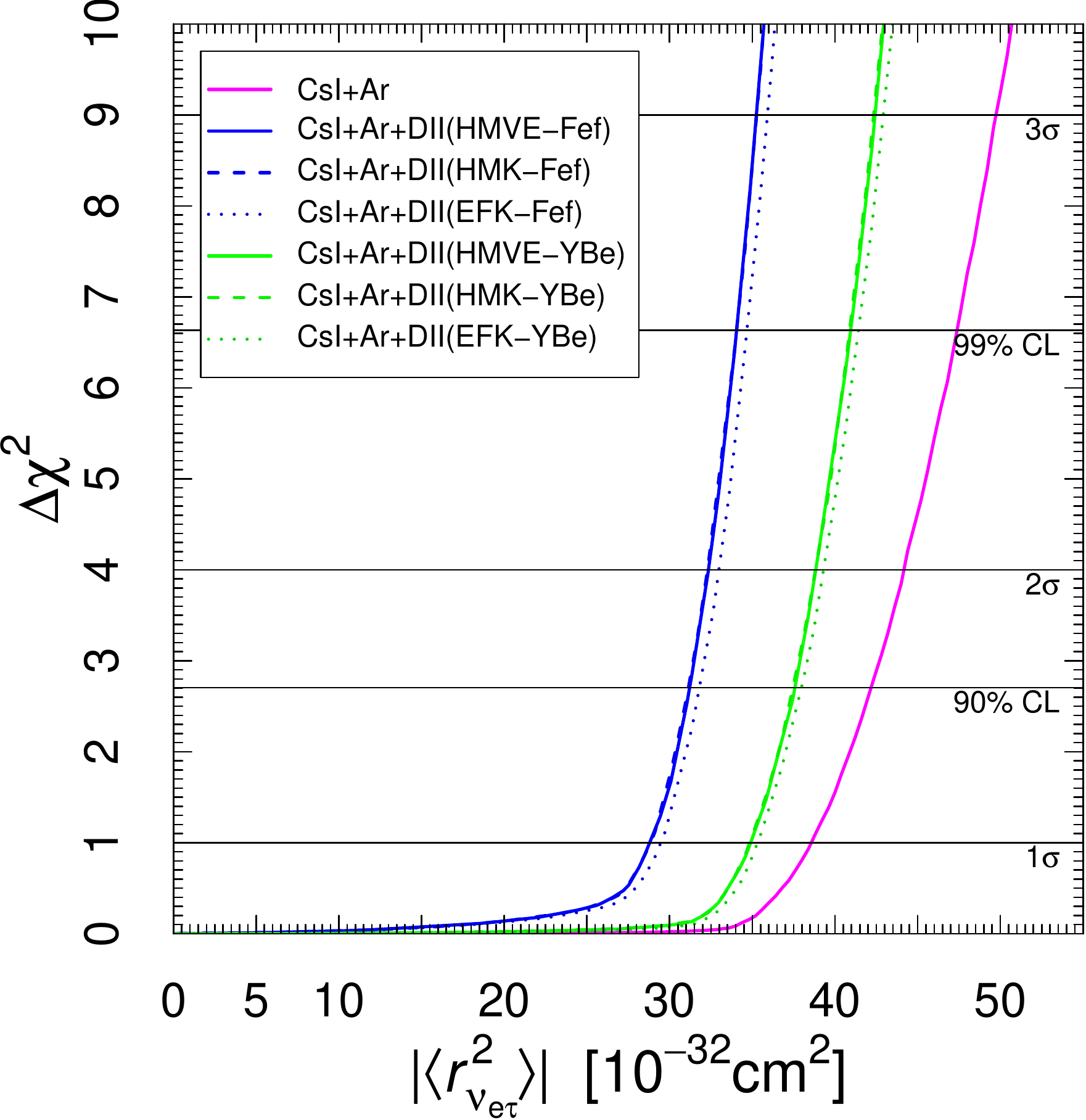}
\caption{ \label{fig:chr5-fig22-chi2-em-et}
Marginal $\Delta\chi^2$'s
for
$|\langle{r}_{\nu_{e\mu}}^2\rangle|$
and
$|\langle{r}_{\nu_{e\tau}}^2\rangle|$
obtained from the analysis of the COHERENT
CsI and Ar data (magenta),
and from the combined analysis of the COHERENT data and
Dresden-II data
assuming the HMVE, HMK, or EFK reactor antineutrino flux and
the Fef (blue) or YBe (green) quenching.
}
\end{center}
\end{figure}

\begin{table*}
\footnotetext{\label{f2}Corrected by a factor of two due to a different convention, see Ref.~\cite{Cadeddu:2018dux}.}
\footnotetext{\label{E734}Corrected in Ref.~\cite{Hirsch:2002uv}.}
\footnotetext{\label{QFspec}Using the Fef quenching factor.}
\renewcommand{\arraystretch}{1.2}
\begin{tabular}{llcll}
Process & Collaboration & Limit [$10^{-32} \, \text{cm}^2$] & C.L. & Ref.\\
\hline
\multirow{2}{*}{Reactor $\bar\nu_{e}$-$e$}
&Krasnoyarsk	&$|\langle{r_{\nu_{e}}^{2}}\rangle|<7.3$	&90\%	&\cite{Vidyakin:1992nf}\\
&TEXONO		&$-4.2<\langle{r_{\nu_{e}}^{2}}\rangle<6.6$	&90\%	&\cite{Deniz:2009mu}\textsuperscript{\ref{f2}}\\
\hline
\multirow{2}{*}{Accelerator $\nu_{e}$-$e$}
&LAMPF		&$-7.12<\langle{r_{\nu_{e}}^{2}}\rangle<10.88$	&90\%	&\cite{Allen:1992qe}\textsuperscript{\ref{f2}}\\
&LSND		&$-5.94<\langle{r_{\nu_{e}}^{2}}\rangle<8.28$	&90\%	&\cite{Auerbach:2001wg}\textsuperscript{\ref{f2}}\\
\hline
\multirow{2}{*}{Accelerator $\nu_{\mu}$-$e$ and $\bar\nu_{\mu}$-$e$}
&BNL-E734	&$-5.7<\langle{r_{\nu_{\mu}}^{2}}\rangle<1.1$	&90\%	&\cite{Ahrens:1990fp}\textsuperscript{\ref{f2},\ref{E734}}\\
&CHARM-II	&$|\langle{r_{\nu_{\mu}}^{2}}\rangle|<1.2$	&90\%	&\cite{Vilain:1994hm}\textsuperscript{\ref{f2}}\\
\hline
\multirow{2}{*}{COHERENT + Dresden-II }
& w/o transition CR	&$-7.1<\langle{r_{\nu_{e}}^{2}}\rangle<5$	&90\%	& This work\textsuperscript{\ref{QFspec}}\\
& w transition CR	&$-56<\langle{r_{\nu_{e}}^{2}}\rangle<5$	&90\%	&This work\textsuperscript{\ref{QFspec}}\\
\hline
\multirow{2}{*}{COHERENT + Dresden-II }
& w/o transition CR	&$-5.9<\langle{r_{\nu_{\mu}}^{2}}\rangle|<4.3$	&90\%	& This work\textsuperscript{\ref{QFspec}}\\
& w transition CR	&$-58.2<\langle{r_{\nu_{\mu}}^{2}}\rangle<4.0$	&90\%	&This work\textsuperscript{\ref{QFspec}}\\
\hline
\end{tabular}
\caption{\label{tab:limits}
Experimental limits for the neutrino charge radii.
}
\end{table*}

\subsection{Neutrino charge radii}

Bounds on the neutrino CR determined by combining the first COHERENT CsI~\cite{Akimov:2017ade} data set and the Ar~\cite{COHERENT:2020iec} data set have been discussed in Ref.~\cite{Cadeddu:2018dux,Cadeddu:2020lky}. In particular, in Ref.~\cite{Cadeddu:2018dux} we derived bounds on the neutrino CR using the 2017 CsI COHERENT data set and their determination of the QF, while in Ref.~\cite{Cadeddu:2020lky} we used the same data set but in combination with the more precise determination of the QF in Ref.~\cite{Collar:2019ihs}. In Ref.~\cite{Cadeddu:2020lky} we also derived constraints for the neutrino CR using the Ar data set exploiting only the \cenns nuclear recoil energy spectrum. \\

Here, we first update these limits by considering the latest CsI data release from COHERENT~\cite{Akimov:2021dab}, which presented more than doubled statistics and the refined QF determination, and their combination with the Ar data set, for which we follow the data release in Ref.~\cite{COHERENT:2020ybo}, that allowed us to also include the arrival time information. We start with the general case in which neutrinos are allowed to have both diagonal and off-diagonal CR. 
The results of these fits are summarized in Tab.~\ref{tab:tab22-chr5-CsI}.
The bounds obtained for the Ar data set are of the same order of magnitude, but as expected due to statistics, less stringent than those obtained from the COHERENT CsI data. Indeed, the latter clearly dominates the combined fit, where the addition of the Ar data only makes a little improvement.
%Thanks to the improvement of the CsI detection in the COHERENT experiment in 2021, the constraints on the neutrino charge radius are more stringent than the previous analysis in Ref.\cite{Cadeddu:2020lky}, especially for the constraints on the diagonal neutrino charge radius.

Similarly, we also fit the Dresden-II data set for the neutrino CR. In this case only $\langle{r}_{\nu_{ee}}^2\rangle$, $|\langle{r}_{\nu_{e\mu}}^2\rangle|$, and $|\langle{r}_{\nu_{e\tau}}^2\rangle|$ could be measured by the data. As explained in Sec.~\ref{sec:dresden}, we use three different antineutrino flux parameterizations, HMVE, HMK and EFK, and two different germanium QF functional forms, Fef and YBe. 
The results of these fits are summarized in Tab.~\ref{tab:tab22-chr5-DII} for all the six combinations of neutrino fluxes and QFs. %and the contours of the allowed regions of the neutrino charge radius are also shown on Fig.~\ref{fig:limit_charge_radius_dresden}. 
As it is possible to see, the three fluxes induce very small differences in the final bounds, while the QF plays a more important role. All in all, the bounds obtained from the Dresden-II data set are comparable with those obtained from the CsI and Ar data set, with a precision similar to the CsI data set.

Finally, in Tab.~\ref{tab:tab22-chr5-tot} we show the bounds on the neutrino CR obtained from the combined analysis of the COHERENT CsI and Ar data
and the Dresden-II data
assuming all the six combinations of neutrino fluxes and QFs. An improvement with respect to the results obtained fitting the COHERENT data set alone is visible.

\begin{figure}
\begin{center}
\includegraphics*[height=0.48\textwidth]{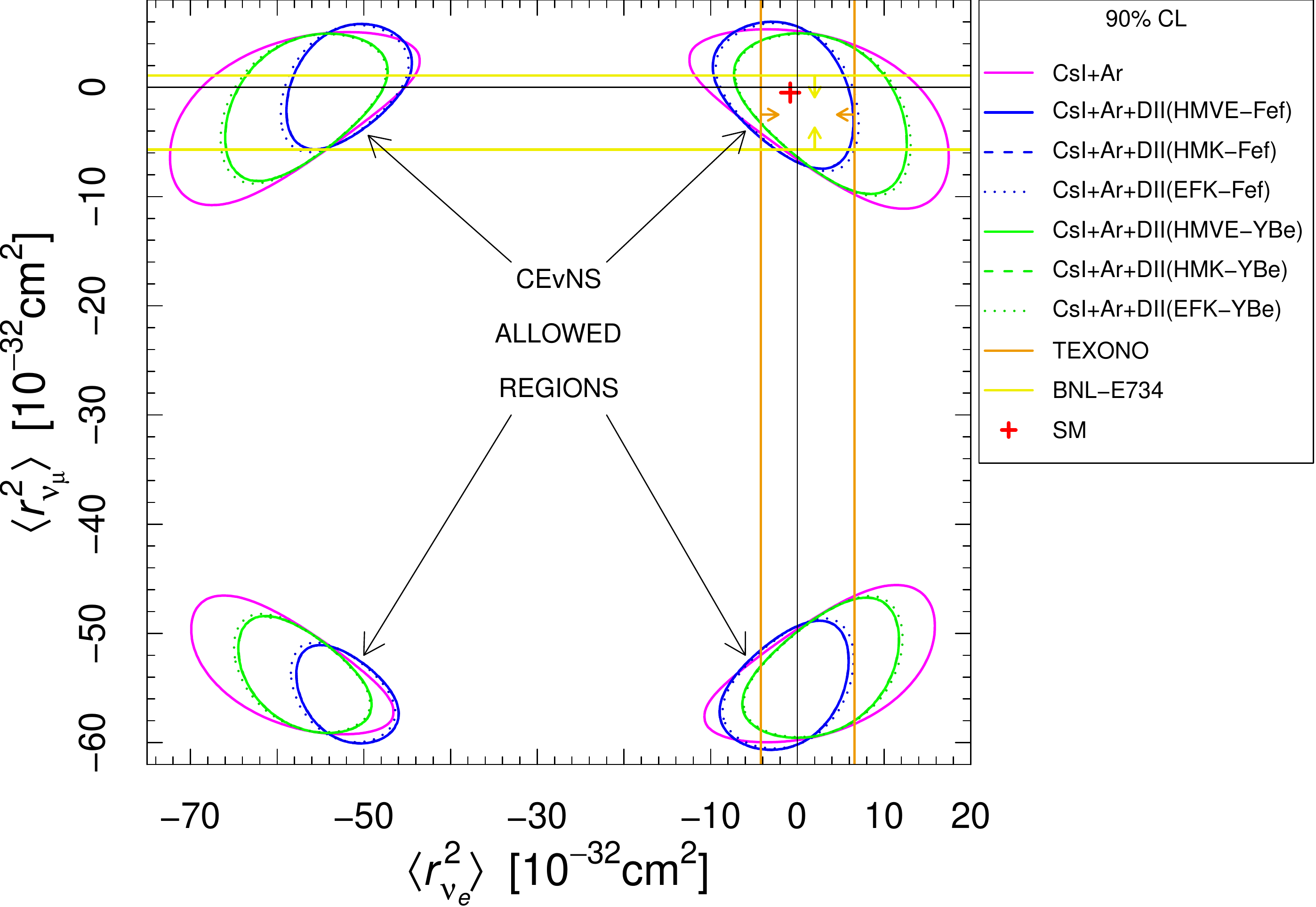}
\caption{ \label{fig:chr2-fig22-contour-ee-mm}
Contours of the 90\% C.L. allowed regions in the
$(\langle{r}_{\nu_{e}}^2\rangle,\langle{r}_{\nu_{\mu}}^2\rangle)$
plane
obtained from the analysis of the COHERENT
CsI and Ar data (magenta),
and from the combined analysis of the COHERENT data and
Dresden-II data
assuming the HMVE, HMK, or EFK reactor antineutrino flux and
the Fef (blue) or YBe (green) quenching,
in the absence of transition charge radii.
The red cross near the origin indicates the Standard Model values
in Eqs.~(\ref{reSM}) and (\ref{rmSM}).
The orange and yellow lines delimit, respectively, the 90\% bounds on
$\langle{r}_{\nu_{e}}^2\rangle$
and
$\langle{r}_{\nu_{\mu}}^2\rangle$
obtained in the
TEXONO~\cite{Deniz:2009mu}
and
BNL-E734~\cite{Ahrens:1990fp}
experiments.
}
\end{center}
\end{figure}

\begin{figure}
\begin{center}
\includegraphics*[height=0.48\textwidth]{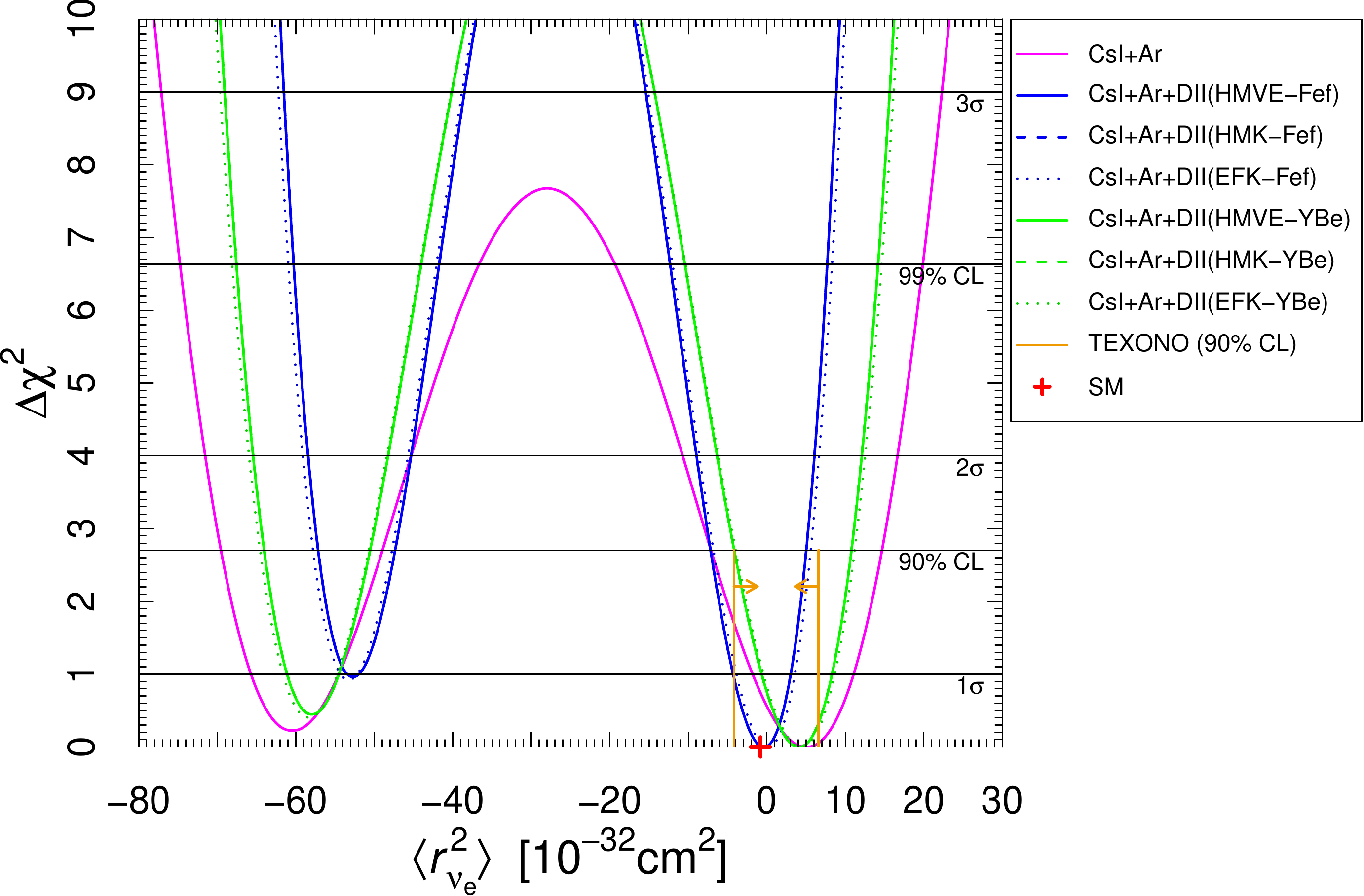}
\caption{ \label{fig:chr2-fig22-chi2-ee}
Marginal $\Delta\chi^2$'s
for
$\langle{r}_{\nu_{e}}^2\rangle$
obtained from the analysis of the COHERENT
CsI and Ar data (magenta),
and from the combined analysis of the COHERENT data and
Dresden-II data
assuming the HMVE, HMK, or EFK reactor antineutrino flux and
the Fef (blue) or YBe (green) quenching,
in the absence of transition charge radii.
The red cross near the origin indicates the Standard Model value
in Eq.~(\ref{reSM}).
The short vertical orange lines show the lower and upper 90\% bounds on
$\langle{r}_{\nu_{e}}^2\rangle$
obtained in the
TEXONO~\cite{Deniz:2009mu}
experiment.
}
\end{center}
\end{figure}

\begin{table}
\begin{center}
%\resizebox{\textwidth}{!}
{
\begin{tabular}{ccccc}
&
$1\sigma$
&
$90\%$
&
$2\sigma$
%&
%$99\%$
&
$3\sigma$
\\
\hline
\multicolumn{5}{c}{\bf CsI}
\\
%\hline
\multirow{2}{*}{$\langle{r}_{\nu_{e}}^2\rangle$}
&
$( -62.4 , -57.2 )$
&
$( -68.1 , -49.4 )$
&
$( -70.4 , -45.1 )$
%&
%$( -74.2 , -33.9 )$
&
\multirow{2}{*}{$( -76.8 , 21.6 )$}
\\
%$\langle{r}_{\nu_{e}}^2\rangle$
&
$( -2.9 , 10.1 )$
&
$( -8.6 , 13.8 )$
&
$( -12.4 , 15.8 )$
%&
%$( -23.6 , 19.0 )$
&

\\
\cline{2-5}
\multirow{2}{*}{$\langle{r}_{\nu_{\mu}}^2\rangle$}
&
\multirow{2}{*}{$( -7.0 , 0.5 )$}
&
$( -57.4 , -49.0 )$
&
$( -59.2 , -46.9 )$
%&
%$( -62.2 , -44.2 )$
&
$( -64.0 , -41.8 )$
\\
%$\langle{r}_{\nu_{\mu}}^2\rangle$
&

&
$( -9.7 , 2.9 )$
&
$( -11.2 , 4.4 )$
%&
%$( -13.9 , 6.8 )$
&
$( -16.0 , 8.6 )$
\\
\hline
\multicolumn{5}{c}{\bf Ar}
\\
\multirow{2}{*}{$\langle{r}_{\nu_{e}}^2\rangle$}
&
$( -79.3 , -37.7 )$
&
\multirow{2}{*}{$( -88.5 , 38.0 )$}
&
\multirow{2}{*}{$( -93.4 , 43.1 )$}
%&
%\multirow{2}{*}{$( -102.3 , 52.0 )$}
&
\multirow{2}{*}{$( -109.8 , 59.2 )$}
\\
%$\langle{r}_{\nu_{e}}^2\rangle$
&
$( -12.4 , 28.8 )$
&

&

%&
%
&

\\
\cline{2-5}
\multirow{2}{*}{$\langle{r}_{\nu_{\mu}}^2\rangle$}
&
$( -59.2 , -36.4 )$
&
\multirow{2}{*}{$( -64.9 , 14.6 )$}
&
\multirow{2}{*}{$( -67.6 , 17.3 )$}
%&
%\multirow{2}{*}{$( -71.8 , 21.5 )$}
&
\multirow{2}{*}{$( -75.1 , 24.5 )$}
\\
%$\langle{r}_{\nu_{\mu}}^2\rangle$
&
$( -13.9 , 8.6 )$
&

&

%&
%
&

\\
\hline
\multicolumn{5}{c}{\bf CsI + Ar}
\\
\multirow{2}{*}{$\langle{r}_{\nu_{e}}^2\rangle$}
&
$( -65.5 , -54.6 )$
&
$( -69.3 , -49.2 )$
&
$( -71.3 , -45.4 )$
%&
%$( -74.7 , -36.8 )$
&
\multirow{2}{*}{$( -77.0 , 22.1 )$}
\\
%$\langle{r}_{\nu_{e}}^2\rangle$
&
$( -1.7 , 10.9 )$
&
$( -6.9 , 14.4 )$
&
$( -10.6 , 16.4 )$
%&
%$( -19.3 , 19.8 )$
&

\\
\cline{2-5}
\multirow{2}{*}{$\langle{r}_{\nu_{\mu}}^2\rangle$}
&
$( -54.7 , -51.4 )$
&
$( -57.7 , -47.8 )$
&
$( -59.2 , -46.3 )$
%&
%$( -61.6 , -43.9 )$
&
$( -63.1 , -41.8 )$
\\
%$\langle{r}_{\nu_{\mu}}^2\rangle$
&
$( -6.4 , 0.8 )$
&
$( -8.8 , 3.2 )$
&
$( -10.3 , 4.7 )$
%&
%$( -13.0 , 6.8 )$
&
$( -14.8 , 8.6 )$
\\
\hline
\end{tabular}

}
\end{center}
\caption{ \label{tab:tab22-chr2-CsI}
Bounds on the diagonal neutrino charge radii in units of $10^{-32}~\text{cm}^2$
obtained from the analysis of the COHERENT CsI and Ar data
in the absence of transition charge radii.
}
\end{table}

\begin{table}
\begin{center}
%\resizebox{\textwidth}{!}
{
\begin{tabular}{ccccc}
&
$1\sigma$
&
$90\%$
&
$2\sigma$
%&
%$99\%$
&
$3\sigma$
\\
\hline
\multicolumn{5}{c}{\bf Dresden-II (HMVE-Fef)}
\\
%\hline
\multirow{2}{*}{$\langle{r}_{\nu_{e}}^2\rangle$}
&
$( -54.3 , -45.0 )$
&
$( -56.7 , -40.8 )$
&
$( -58.0 , -38.0 )$
%&
%$( -59.8 , -29.6 )$
&
\multirow{2}{*}{$( -61.1 , 8.4 )$}
\\
%$\langle{r}_{\nu_{e}}^2\rangle$
&
$( -7.4 , 1.6 )$
&
$( -11.6 , 4.0 )$
&
$( -14.7 , 5.4 )$
%&
%$( -23.0 , 7.1 )$
&

\\
\hline
\multicolumn{5}{c}{\bf Dresden-II (HMK-Fef)}
\\
%\hline
\multirow{2}{*}{$\langle{r}_{\nu_{e}}^2\rangle$}
&
$( -54.3 , -45.0 )$
&
$( -56.7 , -40.8 )$
&
$( -57.8 , -38.0 )$
%&
%$( -59.8 , -29.6 )$
&
\multirow{2}{*}{$( -61.1 , 8.4 )$}
\\
%$\langle{r}_{\nu_{e}}^2\rangle$
&
$( -7.4 , 1.6 )$
&
$( -11.6 , 4.0 )$
&
$( -14.7 , 5.4 )$
%&
%$( -23.0 , 7.1 )$
&

\\
\hline
\multicolumn{5}{c}{\bf Dresden-II (EFK-Fef)}
\\
%\hline
\multirow{2}{*}{$\langle{r}_{\nu_{e}}^2\rangle$}
&
$( -54.9 , -45.5 )$
&
$( -57.3 , -41.3 )$
&
$( -58.7 , -38.2 )$
%&
%$( -60.6 , -29.4 )$
&
\multirow{2}{*}{$( -62.0 , 9.3 )$}
\\
%$\langle{r}_{\nu_{e}}^2\rangle$
&
$( -7.2 , 2.3 )$
&
$( -11.4 , 4.7 )$
&
$( -14.4 , 6.0 )$
%&
%$( -23.2 , 8.0 )$
&

\\
\hline
\multicolumn{5}{c}{\bf Dresden-II (HMVE-YBe)}
\\
%\hline
\multirow{2}{*}{$\langle{r}_{\nu_{e}}^2\rangle$}
&
$( -61.5 , -48.5 )$
&
$( -64.8 , -42.4 )$
&
$( -66.4 , -37.5 )$
%&
%\multirow{2}{*}{$( -69.0 , 16.4 )$}
&
\multirow{2}{*}{$( -70.8 , 18.3 )$}
\\
%$\langle{r}_{\nu_{e}}^2\rangle$
&
$( -3.9 , 8.9 )$
&
$( -10.3 , 12.2 )$
&
$( -15.1 , 13.9 )$
%&
%
&

\\
\hline
\multicolumn{5}{c}{\bf Dresden-II (HMK-YBe)}
\\
%\hline
\multirow{2}{*}{$\langle{r}_{\nu_{e}}^2\rangle$}
&
$( -61.5 , -48.5 )$
&
$( -64.8 , -42.4 )$
&
$( -66.4 , -37.5 )$
%&
%\multirow{2}{*}{$( -69.0 , 16.4 )$}
&
\multirow{2}{*}{$( -70.8 , 18.1 )$}
\\
%$\langle{r}_{\nu_{e}}^2\rangle$
&
$( -3.9 , 8.9 )$
&
$( -10.3 , 12.2 )$
&
$( -15.1 , 13.7 )$
%&
%
&

\\
\hline
\multicolumn{5}{c}{\bf Dresden-II (EFK-YBe)}
\\
%\hline
\multirow{2}{*}{$\langle{r}_{\nu_{e}}^2\rangle$}
&
$( -62.2 , -49.0 )$
&
$( -65.5 , -42.6 )$
&
$( -67.2 , -37.5 )$
%&
%\multirow{2}{*}{$( -69.9 , 17.2 )$}
&
\multirow{2}{*}{$( -71.9 , 19.2 )$}
\\
%$\langle{r}_{\nu_{e}}^2\rangle$
&
$( -3.7 , 9.5 )$
&
$( -10.0 , 12.8 )$
&
$( -15.1 , 14.6 )$
%&
%
&

\\
\hline
\end{tabular}

}
\end{center}
\caption{ \label{tab:tab22-chr2-DII}
Bounds on the electron neutrino charge radius $\langle{r}_{\nu_{e}}^2\rangle$
in units of $10^{-32}~\text{cm}^2$
obtained
from the analysis of the Dresden-II data
assuming the HMVE, HMK, or EFK reactor antineutrino flux and
the Fef or YBe quenching
in the absence of transition charge radii.
}
\end{table}

\begin{table}
\begin{center}
%\resizebox{\textwidth}{!}
{
\begin{tabular}{ccccc}
&
$1\sigma$
&
$90\%$
&
$2\sigma$
%&
%$99\%$
&
$3\sigma$
\\
\hline
\multicolumn{5}{c}{\bf CsI + Ar + Dresden-II (HMVE-Fef)}
\\
%\hline
\multirow{2}{*}{$\langle{r}_{\nu_{e}}^2\rangle$}
&
$( -53.5 , -52.1 )$
&
$( -57.0 , -47.4 )$
&
$( -58.4 , -45.3 )$
%&
%$( -60.2 , -41.8 )$
&
$( -61.4 , -38.6 )$
\\
%$\langle{r}_{\nu_{e}}^2\rangle$
&
$( -4.2 , 2.9 )$
&
$( -7.1 , 5.0 )$
&
$( -8.9 , 5.9 )$
%&
%$( -12.2 , 7.6 )$
&
$( -15.4 , 8.8 )$
\\
\cline{2-5}
\multirow{2}{*}{$\langle{r}_{\nu_{\mu}}^2\rangle$}
&
$( -56.2 , -52.9 )$
&
$( -58.9 , -50.5 )$
&
$( -60.0 , -49.4 )$
%&
%$( -62.0 , -47.7 )$
&
$( -63.5 , -46.3 )$
\\
%$\langle{r}_{\nu_{\mu}}^2\rangle$
&
$( -3.9 , 2.3 )$
&
$( -5.9 , 4.3 )$
&
$( -7.0 , 5.4 )$
%&
%$( -8.7 , 7.3 )$
&
$( -10.0 , 8.9 )$
\\
\hline
\multicolumn{5}{c}{\bf CsI + Ar + Dresden-II (HMK-Fef)}
\\
%\hline
\multirow{2}{*}{$\langle{r}_{\nu_{e}}^2\rangle$}
&
$( -53.3 , -52.1 )$
&
$( -57.0 , -47.4 )$
&
$( -58.3 , -45.3 )$
%&
%$( -60.2 , -41.8 )$
&
$( -61.4 , -38.6 )$
\\
%$\langle{r}_{\nu_{e}}^2\rangle$
&
$( -4.2 , 2.9 )$
&
$( -7.1 , 5.0 )$
&
$( -8.9 , 5.9 )$
%&
%$( -12.2 , 7.6 )$
&
$( -15.4 , 8.7 )$
\\
\cline{2-5}
\multirow{2}{*}{$\langle{r}_{\nu_{\mu}}^2\rangle$}
&
$( -56.2 , -52.9 )$
&
$( -58.9 , -50.5 )$
&
$( -60.0 , -49.4 )$
%&
%$( -62.0 , -47.7 )$
&
$( -63.5 , -46.3 )$
\\
%$\langle{r}_{\nu_{\mu}}^2\rangle$
&
$( -3.9 , 2.3 )$
&
$( -5.9 , 4.3 )$
&
$( -7.0 , 5.4 )$
%&
%$( -8.7 , 7.3 )$
&
$( -10.0 , 8.9 )$
\\
\hline
\multicolumn{5}{c}{\bf CsI + Ar + Dresden-II (EFK-Fef)}
\\
%\hline
\multirow{2}{*}{$\langle{r}_{\nu_{e}}^2\rangle$}
&
$( -54.3 , -52.2 )$
&
$( -57.7 , -47.8 )$
&
$( -59.0 , -45.8 )$
%&
%$( -60.9 , -42.2 )$
&
$( -62.1 , -38.9 )$
\\
%$\langle{r}_{\nu_{e}}^2\rangle$
&
$( -3.8 , 3.5 )$
&
$( -6.7 , 5.5 )$
&
$( -8.5 , 6.6 )$
%&
%$( -11.9 , 8.3 )$
&
$( -15.1 , 9.4 )$
\\
\cline{2-5}
\multirow{2}{*}{$\langle{r}_{\nu_{\mu}}^2\rangle$}
&
$( -56.2 , -52.9 )$
&
$( -58.7 , -50.3 )$
&
$( -60.0 , -49.2 )$
%&
%$( -62.0 , -47.4 )$
&
$( -63.3 , -45.9 )$
\\
%$\langle{r}_{\nu_{\mu}}^2\rangle$
&
$( -4.1 , 2.1 )$
&
$( -6.1 , 4.0 )$
&
$( -7.2 , 5.4 )$
%&
%$( -8.9 , 7.3 )$
&
$( -10.3 , 8.7 )$
\\
\hline
\multicolumn{5}{c}{\bf CsI + Ar + Dresden-II (HMVE-YBe)}
\\
%\hline
\multirow{2}{*}{$\langle{r}_{\nu_{e}}^2\rangle$}
&
$( -61.0 , -54.6 )$
&
$( -63.9 , -50.6 )$
&
$( -65.4 , -48.4 )$
%&
%$( -67.5 , -44.1 )$
&
$( -69.0 , -40.3 )$
\\
%$\langle{r}_{\nu_{e}}^2\rangle$
&
$( -0.52 , 8.3 )$
&
$( -4.1 , 10.8 )$
&
$( -6.3 , 12.0 )$
%&
%$( -10.4 , 14.1 )$
&
$( -14.3 , 15.6 )$
\\
\cline{2-5}
\multirow{2}{*}{$\langle{r}_{\nu_{\mu}}^2\rangle$}
&
$( -54.7 , -51.6 )$
&
$( -57.6 , -48.8 )$
&
$( -58.9 , -47.4 )$
%&
%$( -61.1 , -45.2 )$
&
$( -62.4 , -43.5 )$
\\
%$\langle{r}_{\nu_{\mu}}^2\rangle$
&
$( -5.6 , 0.96 )$
&
$( -7.8 , 3.2 )$
&
$( -9.2 , 4.3 )$
%&
%$( -11.1 , 6.5 )$
&
$( -12.9 , 8.0 )$
\\
\hline
\multicolumn{5}{c}{\bf CsI + Ar + Dresden-II (HMK-YBe)}
\\
%\hline
\multirow{2}{*}{$\langle{r}_{\nu_{e}}^2\rangle$}
&
$( -61.0 , -54.6 )$
&
$( -63.9 , -50.6 )$
&
$( -65.3 , -48.4 )$
%&
%$( -67.5 , -44.1 )$
&
$( -69.0 , -40.1 )$
\\
%$\langle{r}_{\nu_{e}}^2\rangle$
&
$( -0.52 , 8.3 )$
&
$( -4.1 , 10.8 )$
&
$( -6.3 , 12.0 )$
%&
%$( -10.4 , 14.1 )$
&
$( -14.3 , 15.6 )$
\\
\cline{2-5}
\multirow{2}{*}{$\langle{r}_{\nu_{\mu}}^2\rangle$}
&
$( -54.7 , -51.6 )$
&
$( -57.6 , -48.8 )$
&
$( -58.9 , -47.4 )$
%&
%$( -61.1 , -45.2 )$
&
$( -62.6 , -43.5 )$
\\
%$\langle{r}_{\nu_{\mu}}^2\rangle$
&
$( -5.6 , 0.96 )$
&
$( -7.8 , 3.2 )$
&
$( -9.2 , 4.3 )$
%&
%$( -11.1 , 6.5 )$
&
$( -12.9 , 8.0 )$
\\
\hline
\multicolumn{5}{c}{\bf CsI + Ar + Dresden-II (EFK-YBe)}
\\
%\hline
\multirow{2}{*}{$\langle{r}_{\nu_{e}}^2\rangle$}
&
$( -61.6 , -54.7 )$
&
$( -64.5 , -50.8 )$
&
$( -65.8 , -48.5 )$
%&
%$( -68.0 , -44.2 )$
&
$( -69.5 , -40.3 )$
\\
%$\langle{r}_{\nu_{e}}^2\rangle$
&
$( -0.39 , 8.6 )$
&
$( -4.0 , 11.2 )$
&
$( -6.2 , 12.4 )$
%&
%$( -10.3 , 14.6 )$
&
$( -14.1 , 16.1 )$
\\
\cline{2-5}
\multirow{2}{*}{$\langle{r}_{\nu_{\mu}}^2\rangle$}
&
$( -54.7 , -51.6 )$
&
$( -57.6 , -48.5 )$
&
$( -58.9 , -47.2 )$
%&
%$( -60.9 , -45.0 )$
&
$( -62.4 , -43.3 )$
\\
%$\langle{r}_{\nu_{\mu}}^2\rangle$
&
$( -5.9 , 0.74 )$
&
$( -8.1 , 2.9 )$
&
$( -9.4 , 4.3 )$
%&
%$( -11.4 , 6.2 )$
&
$( -13.1 , 7.8 )$
\\
\hline
\end{tabular}

}
\end{center}
\caption{ \label{tab:tab22-chr2-tot}
Bounds on the diagonal neutrino charge radii in units of $10^{-32}~\text{cm}^2$
obtained from the combined analysis the COHERENT CsI and Ar data
and the Dresden-II data
assuming the HMVE, HMK, or EFK reactor antineutrino flux and
the Fef or YBe quenching
in the absence of transition charge radii.
}
\end{table}

The contours of the 90\% C.L. allowed regions in the
$(\langle{r}_{\nu_{ee}}^2\rangle,\langle{r}_{\nu_{\mu\mu}}^2\rangle)$
plane obtained from the analysis of the COHERENT
CsI and Ar data, and from the combined analysis of the COHERENT data and
Dresden-II data assuming the HMVE, HMK, or EFK reactor antineutrino flux and
the Fef or YBe quenching are shown in Fig.~\ref{fig:chr5-fig22-contour-ee-mm} together with the SM values
in Eqs.~(\ref{reSM}) and (\ref{rmSM}) and the 90\% bounds on
$\langle{r}_{\nu_{ee}}^2\rangle$ and $\langle{r}_{\nu_{\mu\mu}}^2\rangle$ obtained, respectively, in the
TEXONO~\cite{Deniz:2009mu} and BNL-E734~\cite{Ahrens:1990fp} experiments. In Fig.~\ref{fig:chr5-fig22-chi2-ee} we show the marginal $\Delta\chi^2$'s for $\langle{r}_{\nu_{ee}}^2\rangle$ together with the SM value in Eq.~(\ref{reSM}) and the lower and upper 90\% bounds on $\langle{r}_{\nu_{ee}}^2\rangle$
obtained in the TEXONO~\cite{Deniz:2009mu}
experiment. As visible, the point corresponding to the SM
values of the diagonal CR lies at the edge of the 1$\sigma$ allowed region and very close to the best fit value for $\langle{r}_{\nu_{ee}}^2\rangle$ in the combined CsI+Ar+Dresden-II fit. 
For a better comparison, in Tab.~\ref{tab:limits} we report a summary of the most recent and precise bounds on $\langle{r}_{\nu_{ee}}^2\rangle$ and $\langle{r}_{\nu_{\mu\mu}}^2\rangle$. Please note that some of these limits have been corrected by a factor of two due to a different convention, see Ref.~\cite{Cadeddu:2018dux} for a detailed explanation. In Tab.~\ref{tab:limits} we also summarized the results found in this work from the combined Dresden-II + COHERENT analysis when considering non-null transition CR. Interestingly, we are able to improve the best upper bound limit for $\langle{r}_{\nu_{ee}}^2\rangle$ previously set by TEXONO.
%The latter, is also in good agreement with the TEXONO bounds. In this latter case, we can slightly restrict the TEXONO upper bound on the electron neutrino charge radius.
Finally, in Fig.~\ref{fig:chr5-fig22-chi2-em-et} we show the marginal $\Delta\chi^2$'s
for $|\langle{r}_{\nu_{e\mu}}^2\rangle|$
and
$|\langle{r}_{\nu_{e\tau}}^2\rangle|$, for which, especially in the latter case, the Fef QF permits to obtain significantly more stringent bounds.

We also assumed the absence of transition CR, fitting thus only for the diagonal charge radii
$\langle{r}_{\nu_{e}}^2\rangle \equiv \langle{r}_{\nu_{ee}}^2\rangle$
and
$\langle{r}_{\nu_{\mu}}^2\rangle \equiv \langle{r}_{\nu_{\mu\mu}}^2\rangle$.
In this way we
probe the values of the neutrino CR in the SM. However, since it is also possible that BSM physics generates off-diagonal neutrino CR that are much smaller than the diagonal ones and that can thus be neglected in a first approximation, also new physics models can be tested in this scenario. The bounds are shown in Tabs.~\ref{tab:tab22-chr2-CsI},~\ref{tab:tab22-chr2-DII} and~\ref{tab:tab22-chr2-tot}  from the analysis of COHERENT data only, Dresden-II data only for the different reactor antineutrino fluxes and
germanium QFs (only $\langle{r}_{\nu_{e}}^2\rangle$ can be tested in this case) and their combinations, respectively.
The corresponding contours of the 90\% C.L. allowed regions in the
$(\langle{r}_{\nu_{e}}^2\rangle,\langle{r}_{\nu_{\mu}}^2\rangle)$
plane  are shown in Fig.~\ref{fig:chr2-fig22-contour-ee-mm}.
One can see that the contribution of
the Dresden-II data leads to a considerable restriction of the allowed regions, especially when using the Fef QF. Here, we also show the SM values in Eqs.~(\ref{reSM}) and (\ref{rmSM}) and the 90\% bounds on
$\langle{r}_{\nu_{e}}^2\rangle$
and
$\langle{r}_{\nu_{\mu}}^2\rangle$
obtained, respectively, in the
TEXONO~\cite{Deniz:2009mu}
and
BNL-E734~\cite{Ahrens:1990fp}
experiments.
In Fig.~\ref{fig:chr2-fig22-chi2-ee} we also show the marginal $\Delta\chi^2$'s
for
$\langle{r}_{\nu_{e}}^2\rangle$. As summarized in Tab.~\ref{tab:limits}, assuming the absence of the transition CR we obtain a very competitive limit at 90\% C.L. with respect to that set by TEXONO when using the Fef QF, namely
\begin{equation}
    -7.1<\langle{r_{\nu_{e}}^{2}} \rangle<5,
    \label{eq:CR_limit}
\end{equation}
in units of $10^{-32} \, \text{cm}^2$.
In particular, we are able to restrict the upper bound limit from $6.6\times10^{-32} \, \text{cm}^2$ to $5\times10^{-32} \, \text{cm}^2$. When using the YBe QF, the limit becomes $-4.1<\langle{r_{\nu_{e}}^{2}} \rangle<10.8$ in units of $10^{-32} \, \text{cm}^2$, with a slightly better lower bound with respect to that set by TEXONO. In both cases, the limits obtained are practically independent of the particular reactor antineutrino flux used.

We repeated all of the above bound calculations including also the ES contribution for the CsI and Dresden-II data set. However, no effect is found due to ES on the neutrino CR, thus the results are independent of its inclusion.

%%%%%%%%%%%%%%%%%%%%%%%%%%%%%%%%%%%%%%%%%%%%%%%%%%%%%%%%%%%%%%%%%%%

\newpage
\subsection{Neutrino electric charge}

There are five electric charges that can be determined with the COHERENT
CE$\nu$NS data:
the two diagonal EC
$q_{\nu_{ee}}$
and
$q_{\nu_{\mu\mu}}$,
and the absolute values of the three transition EC
$q_{\nu_{e\mu}}=q_{\nu_{\mu e}}^{*}$,
$q_{\nu_{e\tau}}$, and
$q_{\nu_{\mu\tau}}$. Using the Dresden-II data instead, only $q_{\nu_{ee}}$, $|q_{\nu_{e\mu}}|$,
$|q_{\nu_{e\tau}}|$ can be tested.

\begin{table}
\begin{center}
%\resizebox{\textwidth}{!}
{
\begin{tabular}{ccccc}
&
$1\sigma$
&
$90\%$
&
$2\sigma$
%&
%$99\%$
&
$3\sigma$
\\
\hline
\multicolumn{5}{c}{\bf CsI (CEvNS)}
\\
%\hline
$q_{\nu_{ee}}$
&
$( -1.6 , 45.2 )\times10^{ -8 }$
&
$( -1.6 , 5.8 )\times10^{ -7 }$
&
$( -1.9 , 6.2 )\times10^{ -7 }$
%&
%$( -2.4 , 6.7 )\times10^{ -7 }$
&
$( -2.6 , 7.0 )\times10^{ -7 }$
\\
%\hline
$q_{\nu_{\mu\mu}}$
&
$( -8.0 , 136.0 )\times10^{ -9 }$
&
$( -3.2 , 25.2 )\times10^{ -8 }$
&
$( -4.4 , 30.8 )\times10^{ -8 }$
%&
%$( -6.8 , 38.0 )\times10^{ -8 }$
&
$( -8.4 , 43.2 )\times10^{ -8 }$
\\
%\hline
$|q_{\nu_{e\mu}}|$
&
$< 1.8 \times10^{ -7 }$
&
$< 2.3 \times10^{ -7 }$
&
$< 2.5 \times10^{ -7 }$
%&
%$< 2.7 \times10^{ -7 }$
&
$< 2.9 \times10^{ -7 }$
\\
%\hline
$|q_{\nu_{e\tau}}|$
&
$( 1.5 , 4.0 )\times10^{ -7 }$
&
$< 4.3 \times10^{ -7 }$
&
$< 4.6 \times10^{ -7 }$
%&
%$< 4.9 \times10^{ -7 }$
&
$< 5.2 \times10^{ -7 }$
\\
%\hline
$|q_{\nu_{\mu\tau}}|$
&
$< 1.8 \times10^{ -7 }$
&
$< 2.3 \times10^{ -7 }$
&
$< 2.5 \times10^{ -7 }$
%&
%$< 2.8 \times10^{ -7 }$
&
$< 3.0 \times10^{ -7 }$
\\
\hline
\multicolumn{5}{c}{\bf CsI (CEvNS+ES)}
\\
%\hline
$q_{\nu_{ee}}$
&
$( -3.6 , 3.6 )\times10^{ -10 }$
&
$( -5.0 , 5.0 )\times10^{ -10 }$
&
$( -5.6 , 5.6 )\times10^{ -10 }$
%&
%$( -6.7 , 6.7 )\times10^{ -10 }$
&
$( -7.5 , 7.5 )\times10^{ -10 }$
\\
%\hline
$q_{\nu_{\mu\mu}}$
&
$( -1.2 , 1.2 )\times10^{ -10 }$
&
$( -1.9 , 1.9 )\times10^{ -10 }$
&
$( -2.2 , 2.2 )\times10^{ -10 }$
%&
%$( -2.8 , 2.8 )\times10^{ -10 }$
&
$( -3.2 , 3.2 )\times10^{ -10 }$
\\
%\hline
$|q_{\nu_{e\mu}}|$
&
$< 1.2 \times10^{ -10 }$
&
$< 1.8 \times10^{ -10 }$
&
$< 2.2 \times10^{ -10 }$
%&
%$< 2.7 \times10^{ -10 }$
&
$< 3.1 \times10^{ -10 }$
\\
%\hline
$|q_{\nu_{e\tau}}|$
&
$< 3.5 \times10^{ -10 }$
&
$< 5.0 \times10^{ -10 }$
&
$< 5.6 \times10^{ -10 }$
%&
%$< 6.8 \times10^{ -10 }$
&
$< 7.5 \times10^{ -10 }$
\\
%\hline
$|q_{\nu_{\mu\tau}}|$
&
$< 1.2 \times10^{ -10 }$
&
$< 1.9 \times10^{ -10 }$
&
$< 2.2 \times10^{ -10 }$
%&
%$< 2.8 \times10^{ -10 }$
&
$< 3.2 \times10^{ -10 }$
\\
\hline
\multicolumn{5}{c}{\bf Ar (CEvNS)}
\\
%\hline
$q_{\nu_{ee}}$
&
$( -1.3 , 1.7 )\times10^{ -7 }$
&
$( -1.7 , 3.2 )\times10^{ -7 }$
&
$( -2.0 , 3.5 )\times10^{ -7 }$
%&
%$( -2.4 , 4.0 )\times10^{ -7 }$
&
$( -2.7 , 4.4 )\times10^{ -7 }$
\\
%\hline
$q_{\nu_{\mu\mu}}$
&
$( -4.4 , 10.0 )\times10^{ -8 }$
&
$( -6.8 , 21.6 )\times10^{ -8 }$
&
$( -8.0 , 24.4 )\times10^{ -8 }$
%&
%$( -1.0 , 2.8 )\times10^{ -7 }$
&
$( -1.2 , 3.0 )\times10^{ -7 }$
\\
%\hline
$|q_{\nu_{e\mu}}|$
&
$< 1.0 \times10^{ -7 }$
&
$< 1.4 \times10^{ -7 }$
&
$< 1.5 \times10^{ -7 }$
%&
%$< 1.7 \times10^{ -7 }$
&
$< 1.8 \times10^{ -7 }$
\\
%\hline
$|q_{\nu_{e\tau}}|$
&
$< 2.0 \times10^{ -7 }$
&
$< 2.5 \times10^{ -7 }$
&
$< 2.8 \times10^{ -7 }$
%&
%$< 3.2 \times10^{ -7 }$
&
$< 3.6 \times10^{ -7 }$
\\
%\hline
$|q_{\nu_{\mu\tau}}|$
&
$< 1.1 \times10^{ -7 }$
&
$< 1.5 \times10^{ -7 }$
&
$< 1.7 \times10^{ -7 }$
%&
%$< 1.9 \times10^{ -7 }$
&
$< 2.1 \times10^{ -7 }$
\\
\hline
\multicolumn{5}{c}{\bf CsI (CEvNS) + Ar (CEvNS)}
\\
%\hline
$q_{\nu_{ee}}$
&
$( -12.4 , 8.0 )\times10^{ -8 }$
&
$( -1.6 , 1.7 )\times10^{ -7 }$
&
$( -1.7 , 2.2 )\times10^{ -7 }$
%&
%$( -2.0 , 3.0 )\times10^{ -7 }$
&
$( -2.2 , 3.5 )\times10^{ -7 }$
\\
%\hline
$q_{\nu_{\mu\mu}}$
&
$( -1.2 , 7.6 )\times10^{ -8 }$
&
$( -3.2 , 11.2 )\times10^{ -8 }$
&
$( -4.0 , 12.8 )\times10^{ -8 }$
%&
%$( -5.6 , 16.0 )\times10^{ -8 }$
&
$( -6.8 , 18.4 )\times10^{ -8 }$
\\
%\hline
$|q_{\nu_{e\mu}}|$
&
$< 1.1 \times10^{ -7 }$
&
$< 1.4 \times10^{ -7 }$
&
$< 1.5 \times10^{ -7 }$
%&
%$< 1.7 \times10^{ -7 }$
&
$< 1.9 \times10^{ -7 }$
\\
%\hline
$|q_{\nu_{e\tau}}|$
&
$< 2.4 \times10^{ -7 }$
&
$< 2.9 \times10^{ -7 }$
&
$< 3.1 \times10^{ -7 }$
%&
%$< 3.4 \times10^{ -7 }$
&
$< 3.7 \times10^{ -7 }$
\\
%\hline
$|q_{\nu_{\mu\tau}}|$
&
$< 1.2 \times10^{ -7 }$
&
$< 1.5 \times10^{ -7 }$
&
$< 1.6 \times10^{ -7 }$
%&
%$< 1.8 \times10^{ -7 }$
&
$< 2.0 \times10^{ -7 }$
\\
\hline
\multicolumn{5}{c}{\bf CsI (CEvNS+ES) + Ar (CEvNS)}
\\
%\hline
$q_{\nu_{ee}}$
&
$( -3.5 , 3.5 )\times10^{ -10 }$
&
$( -5.0 , 5.0 )\times10^{ -10 }$
&
$( -5.6 , 5.6 )\times10^{ -10 }$
%&
%$( -6.7 , 6.7 )\times10^{ -10 }$
&
$( -7.5 , 7.5 )\times10^{ -10 }$
\\
%\hline
$q_{\nu_{\mu\mu}}$
&
$( -1.2 , 1.2 )\times10^{ -10 }$
&
$( -1.9 , 1.9 )\times10^{ -10 }$
&
$( -2.2 , 2.2 )\times10^{ -10 }$
%&
%$( -2.8 , 2.8 )\times10^{ -10 }$
&
$( -3.2 , 3.2 )\times10^{ -10 }$
\\
%\hline
$|q_{\nu_{e\mu}}|$
&
$< 1.2 \times10^{ -10 }$
&
$< 1.8 \times10^{ -10 }$
&
$< 2.2 \times10^{ -10 }$
%&
%$< 2.7 \times10^{ -10 }$
&
$< 3.1 \times10^{ -10 }$
\\
%\hline
$|q_{\nu_{e\tau}}|$
&
$< 3.6 \times10^{ -10 }$
&
$< 5.0 \times10^{ -10 }$
&
$< 5.6 \times10^{ -10 }$
%&
%$< 6.8 \times10^{ -10 }$
&
$< 7.5 \times10^{ -10 }$
\\
%\hline
$|q_{\nu_{\mu\tau}}|$
&
$< 1.2 \times10^{ -10 }$
&
$< 1.9 \times10^{ -10 }$
&
$< 2.2 \times10^{ -10 }$
%&
%$< 2.8 \times10^{ -10 }$
&
$< 3.2 \times10^{ -10 }$
\\
\hline
\end{tabular}

}
\end{center}
\caption{ \label{tab:tab22-ech5-CsI}
Bounds on the neutrino electric charges in units of the elementary charge $e$
obtained from the analysis of the COHERENT CsI and Ar data.
We show the results of the analyses of CsI data with CE$\nu$NS only interactions and with CE$\nu$NS+ES interactions. 
}
\end{table}

In this section, we present the constraints on the neutrino EC. The results of our analyses are shown in Tab.~\ref{tab:tab22-ech5-CsI} and Tab.~\ref{tab:tab22-ech5-DII} for the COHERENT CsI and Ar data set and for the Dresden-II data, respectively.
Focusing on the results shown in Tab.~\ref{tab:tab22-ech5-CsI}, differently from the analysis of the neutrino CR, the contribution of Ar data is dominant in the combined COHERENT analysis of the neutrino electric charges, although the CsI data set has more statistics. It follows from the enhancement of the neutrino electric charge effect
in CE$\nu$NS at low $q^2$,
because of the denominator in Eq.~(\ref{Qech}).
However, the expected enhancement due to the different CsI and Ar masses, is mitigated by the different
sizes of the energy bins:
in the Ar experiment the first bin includes energies from the threshold, of about 5 $\mathrm{keV}_{\mathrm{nr}}$,
to about 36 $\mathrm{keV}_{\mathrm{nr}}$,
whereas the first CsI energy bin have a much smaller size. 
Therefore, the enhancement of the EC effect
occurs only in the first energy bin of the Ar experiment.
Nevertheless, this enhancement is sufficient to achieve
a slightly better performance of the Ar data
in constraining the neutrino EC
in spite of the larger uncertainties. 
In Tab.~\ref{tab:tab22-ech5-CsI} we also explicitly show the impact of including the ES in the CsI analysis, also when combining it with  Ar. Thanks to the presence of the $q^2$ term in the denominator of Eq.~(\ref{Qech}), a large improvement of more than 2 orders of magnitude with respect to the limits derived ignoring the ES contribution is obtained.

\begin{table}
\begin{center}
%\resizebox{\textwidth}{!}
{
\begin{tabular}{ccccc}
&
$1\sigma$
&
$90\%$
&
$2\sigma$
%&
%$99\%$
&
$3\sigma$
\\
\hline
\multicolumn{5}{c}{\bf Dresden-II (HMVE-Fef CEvNS)}
\\
%\hline
$q_{\nu_{ee}}$
&
$( -1.5 , 10.1 )\times10^{ -10 }$
&
$( -3.4 , 12.5 )\times10^{ -10 }$
&
$( -4.3 , 13.6 )\times10^{ -10 }$
%&
%$( -5.6 , 15.0 )\times10^{ -10 }$
&
$( -6.5 , 16.0 )\times10^{ -10 }$
\\
%\hline
$|q_{\nu_{e\mu}}|$, $|q_{\nu_{e\tau}}|$
&
$< 6.0 \times10^{ -10 }$
&
$< 8.2 \times10^{ -10 }$
&
$< 9.1 \times10^{ -10 }$
%&
%$< 1.0 \times10^{ -9 }$
&
$< 1.1 \times10^{ -9 }$
\\
\hline
\multicolumn{5}{c}{\bf Dresden-II (HMVE-Fef CEvNS+ES)}
\\
%\hline
$q_{\nu_{ee}}$
&
$( -7.3 , 7.6 )\times10^{ -12 }$
&
$( -9.3 , 9.5 )\times10^{ -12 }$
&
$( -1.0 , 1.0 )\times10^{ -11 }$
%&
%$( -1.2 , 1.2 )\times10^{ -11 }$
&
$( -1.2 , 1.3 )\times10^{ -11 }$
\\
%\hline
$|q_{\nu_{e\mu}}|$, $|q_{\nu_{e\tau}}|$
&
$< 7.4 \times10^{ -12 }$
&
$< 9.4 \times10^{ -12 }$
&
$< 1.0 \times10^{ -11 }$
%&
%$< 1.2 \times10^{ -11 }$
&
$< 1.3 \times10^{ -11 }$
\\
\hline
\multicolumn{5}{c}{\bf Dresden-II (HMK-Fef CEvNS+ES)}
\\
%\hline
$q_{\nu_{ee}}$
&
$( -6.6 , 7.0 )\times10^{ -12 }$
&
$( -8.6 , 8.7 )\times10^{ -12 }$
&
$( -9.4 , 9.5 )\times10^{ -12 }$
%&
%$( -1.1 , 1.1 )\times10^{ -11 }$
&
$( -1.1 , 1.2 )\times10^{ -11 }$
\\
%\hline
$|q_{\nu_{e\mu}}|$, $|q_{\nu_{e\tau}}|$
&
$< 6.8 \times10^{ -12 }$
&
$< 8.6 \times10^{ -12 }$
&
$< 9.4 \times10^{ -12 }$
%&
%$< 1.1 \times10^{ -11 }$
&
$< 1.2 \times10^{ -11 }$
\\
\hline
\multicolumn{5}{c}{\bf Dresden-II (EFK-Fef CEvNS+ES)}
\\
%\hline
$q_{\nu_{ee}}$
&
$( -7.4 , 7.7 )\times10^{ -12 }$
&
$( -9.2 , 9.4 )\times10^{ -12 }$
&
$( -1.0 , 1.0 )\times10^{ -11 }$
%&
%$( -1.1 , 1.2 )\times10^{ -11 }$
&
$( -1.2 , 1.2 )\times10^{ -11 }$
\\
%\hline
$|q_{\nu_{e\mu}}|$, $|q_{\nu_{e\tau}}|$
&
$< 7.5 \times10^{ -12 }$
&
$< 9.4 \times10^{ -12 }$
&
$< 1.0 \times10^{ -11 }$
%&
%$< 1.1 \times10^{ -11 }$
&
$< 1.2 \times10^{ -11 }$
\\
\hline
\multicolumn{5}{c}{\bf Dresden-II (HMVE-YBe CEvNS)}
\\
%\hline
$q_{\nu_{ee}}$
&
$( -4.8 , 12.4 )\times10^{ -10 }$
&
$( -6.6 , 15.2 )\times10^{ -10 }$
&
$( -7.5 , 16.3 )\times10^{ -10 }$
%&
%$( -8.9 , 17.9 )\times10^{ -10 }$
&
$( -9.8 , 18.9 )\times10^{ -10 }$
\\
%\hline
$|q_{\nu_{e\mu}}|$, $|q_{\nu_{e\tau}}|$
&
$< 8.9 \times10^{ -10 }$
&
$< 1.1 \times10^{ -9 }$
&
$< 1.2 \times10^{ -9 }$
%&
%$< 1.3 \times10^{ -9 }$
&
$< 1.4 \times10^{ -9 }$
\\
\hline
\multicolumn{5}{c}{\bf Dresden-II (HMVE-YBe CEvNS+ES)}
\\
%\hline
$q_{\nu_{ee}}$
&
$( -1.1 , 1.1 )\times10^{ -11 }$
&
$( -1.2 , 1.3 )\times10^{ -11 }$
&
$( -1.3 , 1.3 )\times10^{ -11 }$
%&
%$( -1.4 , 1.4 )\times10^{ -11 }$
&
$( -1.5 , 1.5 )\times10^{ -11 }$
\\
%\hline
$|q_{\nu_{e\mu}}|$, $|q_{\nu_{e\tau}}|$
&
$< 1.1 \times10^{ -11 }$
&
$< 1.2 \times10^{ -11 }$
&
$< 1.3 \times10^{ -11 }$
%&
%$< 1.4 \times10^{ -11 }$
&
$< 1.5 \times10^{ -11 }$
\\
\hline
\multicolumn{5}{c}{\bf Dresden-II (HMK-YBe CEvNS+ES)}
\\
%\hline
$q_{\nu_{ee}}$
&
$( -9.9 , 10.2 )\times10^{ -12 }$
&
$( -1.1 , 1.2 )\times10^{ -11 }$
&
$( -1.2 , 1.2 )\times10^{ -11 }$
%&
%$( -1.3 , 1.3 )\times10^{ -11 }$
&
$( -1.4 , 1.4 )\times10^{ -11 }$
\\
%\hline
$|q_{\nu_{e\mu}}|$, $|q_{\nu_{e\tau}}|$
&
$< 1.0 \times10^{ -11 }$
&
$< 1.1 \times10^{ -11 }$
&
$< 1.2 \times10^{ -11 }$
%&
%$< 1.3 \times10^{ -11 }$
&
$< 1.4 \times10^{ -11 }$
\\
\hline
\multicolumn{5}{c}{\bf Dresden-II (EFK-YBe CEvNS+ES)}
\\
%\hline
$q_{\nu_{ee}}$
&
$( -1.0 , 1.1 )\times10^{ -11 }$
&
$( -1.2 , 1.2 )\times10^{ -11 }$
&
$( -1.3 , 1.3 )\times10^{ -11 }$
%&
%$( -1.4 , 1.4 )\times10^{ -11 }$
&
$( -1.4 , 1.4 )\times10^{ -11 }$
\\
%\hline
$|q_{\nu_{e\mu}}|$, $|q_{\nu_{e\tau}}|$
&
$< 1.1 \times10^{ -11 }$
&
$< 1.2 \times10^{ -11 }$
&
$< 1.3 \times10^{ -11 }$
%&
%$< 1.4 \times10^{ -11 }$
&
$< 1.4 \times10^{ -11 }$
\\
\hline
\end{tabular}

}
\end{center}
\caption{ \label{tab:tab22-ech5-DII}
Bounds on the neutrino electric charges in units of the elementary charge $e$
obtained from the analysis of the Dresden-II data
assuming the HMVE, HMK, or EFK reactor antineutrino flux and
the Fef or YBe quenching.
For the HMVE reactor antineutrino flux
we show the results obtained with CE$\nu$NS only interactions and with CE$\nu$NS+ES interactions.
}
\end{table}

In Tab.~\ref{tab:tab22-ech5-DII} we show the bounds on the EC found using the Dresden-II data. As for the neutrino CR limits discussed above, the different flux parameterizations cause only negligible differences in the obtained bounds. Thus, for the case in which we fit exclusively for the \cenns contribution, we show only the results obtained with the HMVE flux, while when we include the ES contribution we show all the three different parameterizations. 
As already stated in Sec.~\ref{sec:nuec}, the $|q^2|$ corresponding to ES is much smaller than the \cenns $|q^2|$, resulting in improved sensitivity when the ES contribution is included with respect to \cenns only. 
Namely, with \cenns only there is an improvement with respect to COHERENT \cenns only of about 2 orders of magnitude, while with \cenns+ ES the improvement is of about 4 orders of magnitude.

\begin{figure}
\begin{center}
\includegraphics*[height=0.48\textwidth]{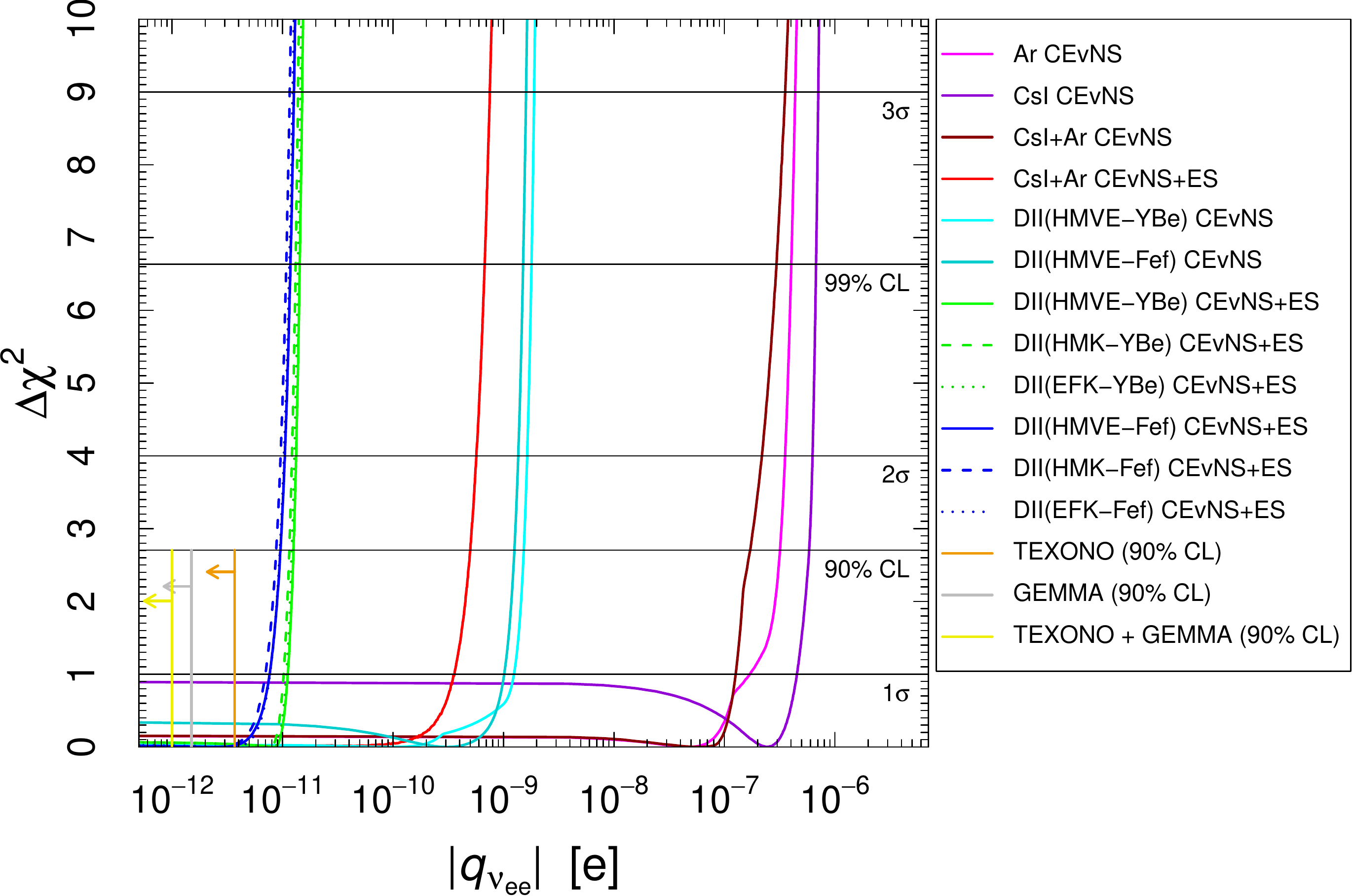}
\caption{ \label{fig:ech5-fig22-chi2-ee}
Marginal $\Delta\chi^2$'s
for
$|q_{\nu_{ee}}|$
obtained from:
the separate analyses of the COHERENT
Ar (magenta) and CsI (darkviolet)
data with \cenns interactions;
the combined analyses of the COHERENT
Ar and CsI data
with \cenns interactions only (dark red) and with \cenns+ES interactions (red);
the \cenns-only analyses of
Dresden-II data
assuming the HMVE reactor antineutrino flux and
the YBe (cyan) or Fef (dark cyan) quenching;
the \cenns+ES analyses of
Dresden-II data
assuming the HMVE, HMK, or EFK reactor antineutrino flux and
the YBe (green) or Fef (blue) quenching.
The short vertical orange, gray, and yellow lines show
the 90\% C.L. upper bounds on
$|q_{\nu_{ee}}|$
obtained,
respectively,
in Ref.~\cite{Gninenko:2006fi} from TEXONO data~\cite{TEXONO:2002pra},
in Ref.~\cite{Studenikin:2013my} from the GEMMA~\cite{Beda:2012zz}
bound on $|\mu_{\nu_{e}}|$,
and
in Ref.~\cite{Chen:2014dsa}
from TEXONO data~\cite{TEXONO:2006xds} and GEMMA data ~\cite{Beda:2012zz}.
}
\end{center}
\end{figure}

In Fig.~\ref{fig:ech5-fig22-chi2-ee} we show the marginal $\Delta\chi^2$'s
for $|q_{\nu_{ee}}|$
obtained from the separate analyses of the COHERENT
Ar and CsI data and their combinations, with \cenns interactions only and with the ES contribution, as well as the \cenns-only analyses of
Dresden-II data assuming the HMVE reactor antineutrino flux and the YBe or Fef QF. Moreover, also the \cenns+ ES analysis of
Dresden-II data assuming the HMVE, HMK, or EFK reactor antineutrino flux and the YBe or Fef QF is drawn. We also show the 90\% C.L. upper bounds on
$|q_{\nu_{ee}}|$
obtained, respectively, in Ref.~\cite{Gninenko:2006fi} from TEXONO data~\cite{TEXONO:2002pra},
in Ref.~\cite{Studenikin:2013my} from the GEMMA~\cite{Beda:2012zz} bound on $|\mu_{\nu_{e}}|$,
and in Ref.~\cite{Chen:2014dsa} from TEXONO data~\cite{TEXONO:2006xds} and GEMMA data~\cite{Beda:2012zz}. Intriguingly, the bounds on $|q_{\nu_{ee}}|$ obtained from the combination of COHERENT with the Dresden-II \cenns+ ES data set are much more stringent than the COHERENT ones and the \cenns only fit, namely at 90\% C.L. and using the Fef quenching factor
\begin{equation}
    -9.3<q_{\nu_{ee}}<9.5,
\end{equation}
in units of $10^{-12}\, e$.
This limit is competitive with respect to the other aforementioned bounds,
that are at the level of $10^{-12} \, e$, the best limit being $|q_{\nu_{ee}}|< 1.0\times10^{-12} \, e$~\cite{Chen:2014dsa}. However, when comparing these limits one has to keep in mind that, differently from this work, the limits in Ref.~\cite{Chen:2014dsa} have been derived using for the neutrino-electron cross section the MCRRPA theory~\cite{PhysRevA.25.634,PhysRevA.26.734,Chen:2013lba}. As discussed in Sec.~\ref{sec:essec}, this becomes relevant for data from Ge detectors at sub-keV sensitivities and allows them to achieve more stringent limits with respect to FEA in particular for the neutrino EC. Thus, the limits obtained in this work can be considered as very conservative and we will investigate the impact of using a random-phase approximation theory in a future work. \\

\begin{figure}
\begin{center}
\includegraphics*[width=0.9\textwidth, viewport=40 539 856 598]{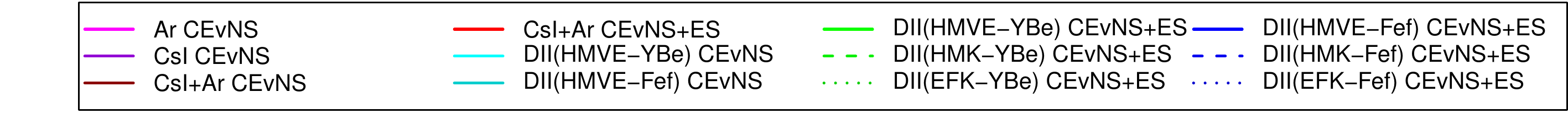}
\\
\includegraphics*[height=0.48\textwidth]{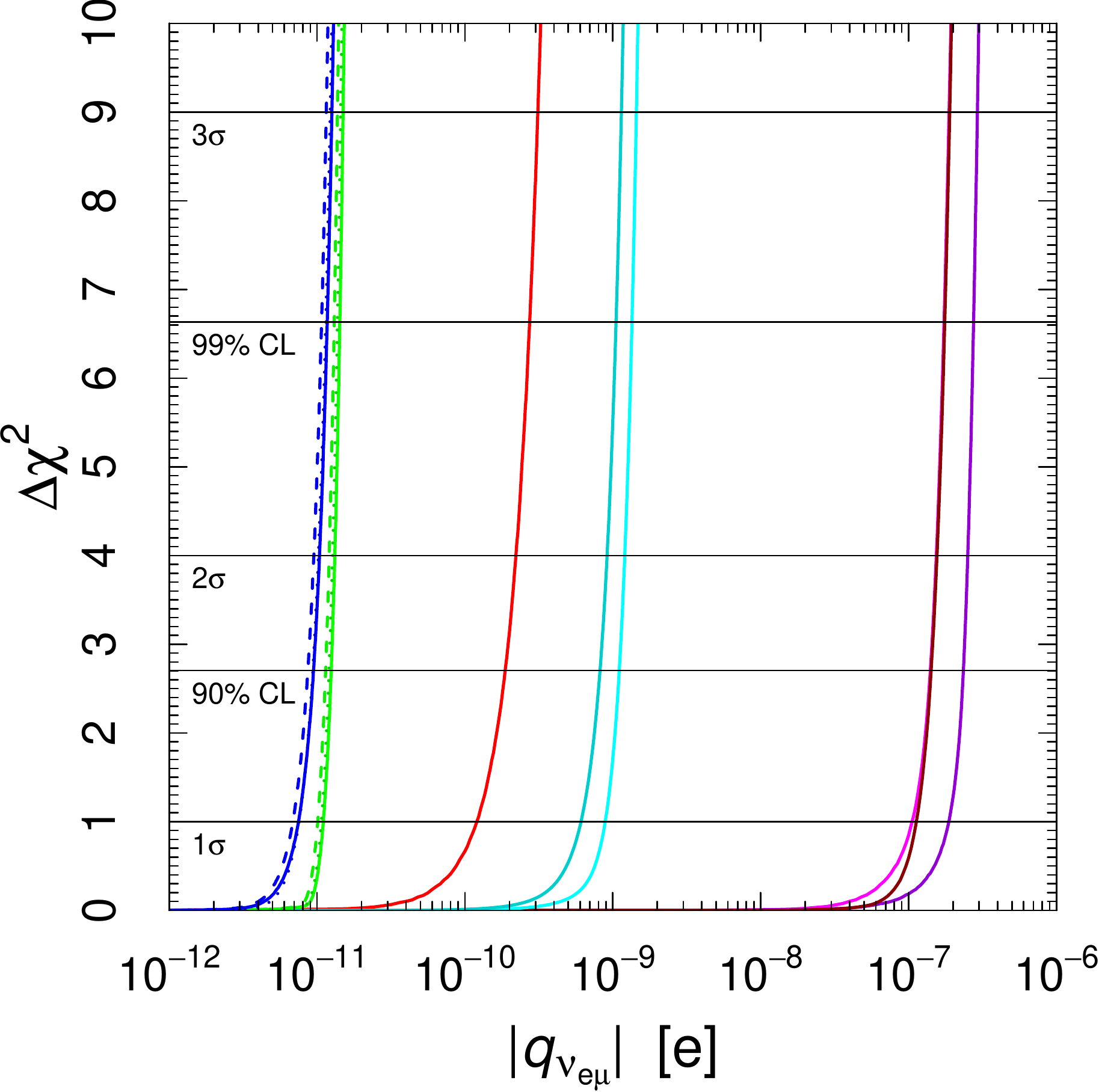}
\hfill
\includegraphics*[height=0.48\textwidth]{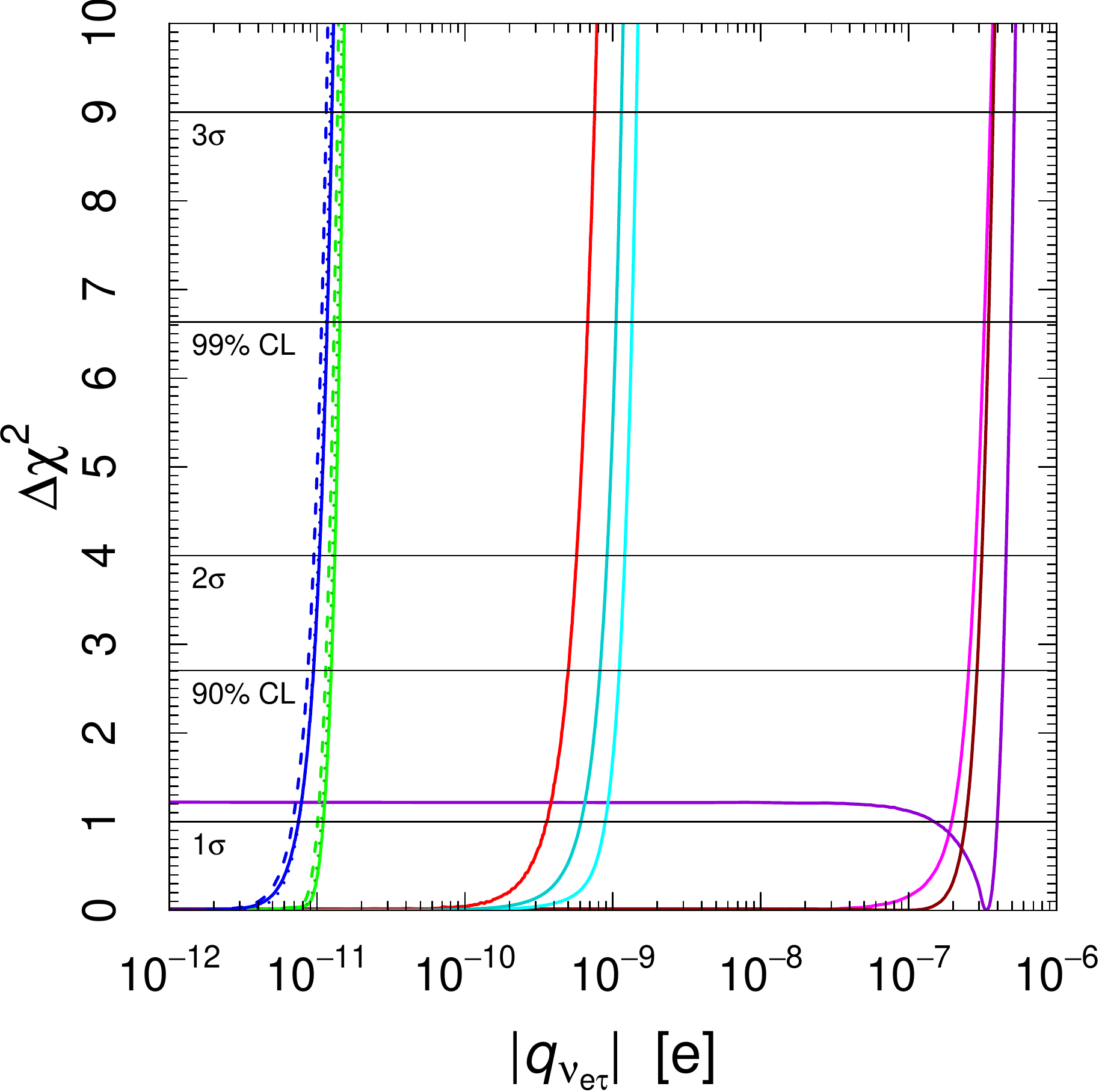}
\caption{ \label{fig:ech5-fig22-chi2-em-et}
Marginal $\Delta\chi^2$'s
for
$|q_{\nu_{e\mu}}|$
and
$|q_{\nu_{e\tau}}|$
obtained from:
the separate analyses of the COHERENT
Ar (magenta) and CsI (darkviolet)
data with \cenns interactions;
the combined analyses of the COHERENT
Ar and CsI data
with \cenns interactions only (dark red) and with \cenns+ ES interactions (red);
the \cenns-only analyses of
Dresden-II data
assuming the HMVE reactor antineutrino flux and
the YBe (cyan) or Fef (dark cyan) quenching;
the \cenns+ ES analyses of
Dresden-II data
assuming the HMVE, HMK, or EFK reactor antineutrino flux and
the YBe (green) or Fef (blue) quenching.
}
\end{center}
\end{figure}

In Fig.~\ref{fig:ech5-fig22-chi2-em-et}(a) and (b) we show the marginal $\Delta\chi^2$'s
for $|q_{\nu_{e\mu}}|$
and $|q_{\nu_{e\tau}}|$, respectively, 
obtained from the separate analyses of the COHERENT
Ar and CsI data with \cenns interactions and the combined analyses of the COHERENT
Ar and CsI data with \cenns interactions only and with the ES contribution, as well as
the \cenns-only analyses of the Dresden-II data
assuming the HMVE reactor antineutrino flux and the two QFs, and 
the \cenns+ ES analyses of
Dresden-II data assuming the HMVE, HMK, or EFK reactor antineutrino flux and
the two QFs. Also in this case it is possible to see that the different fluxes result in negligible differences, while the impact of the QF is visible. Again, the inclusion of the ES contribution significantly improves the bounds obtained for both Dresden-II and COHERENT.

\begin{figure}
\begin{center}
\includegraphics*[height=0.48\textwidth]{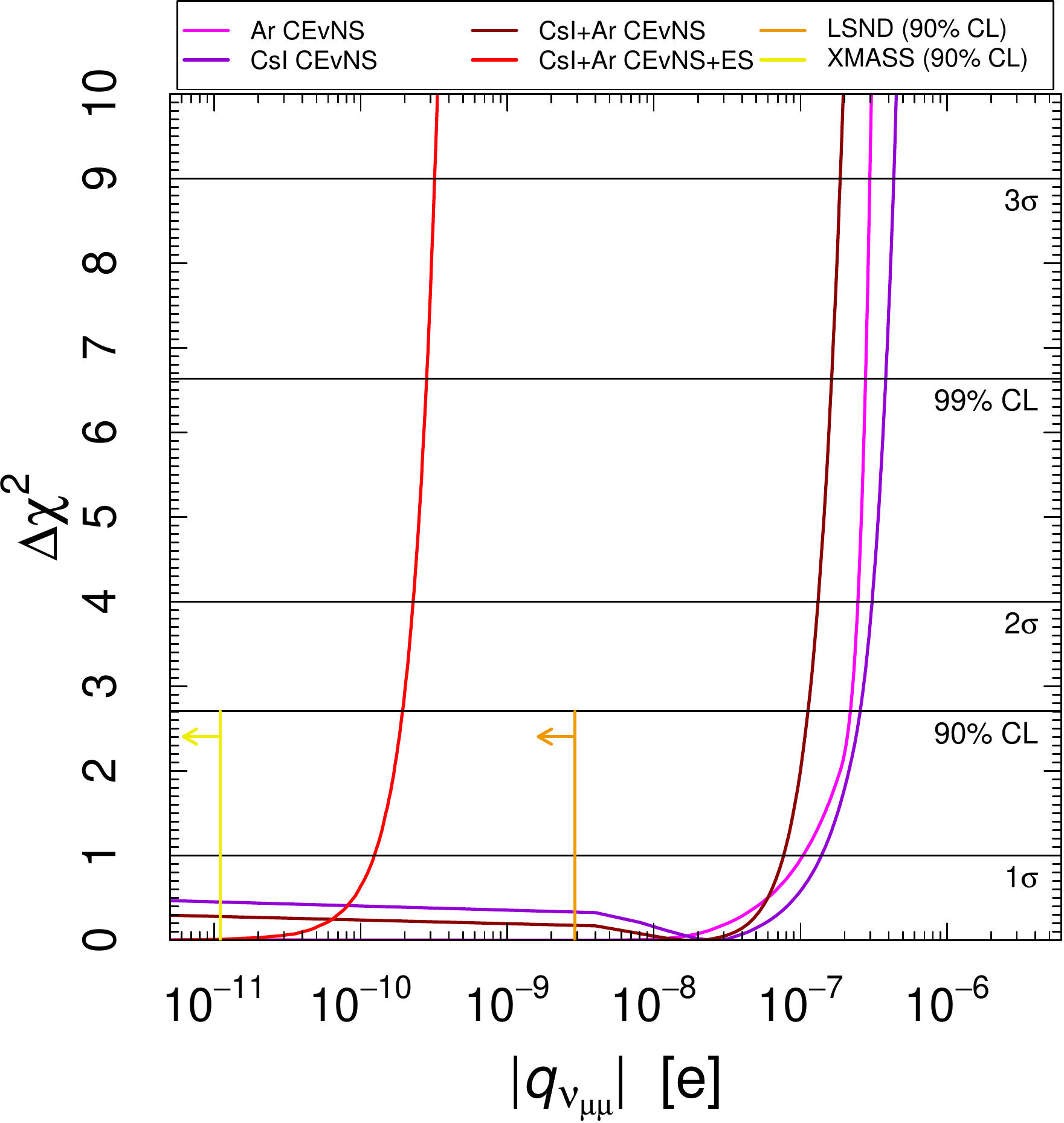}
\hfill
\includegraphics*[height=0.48\textwidth]{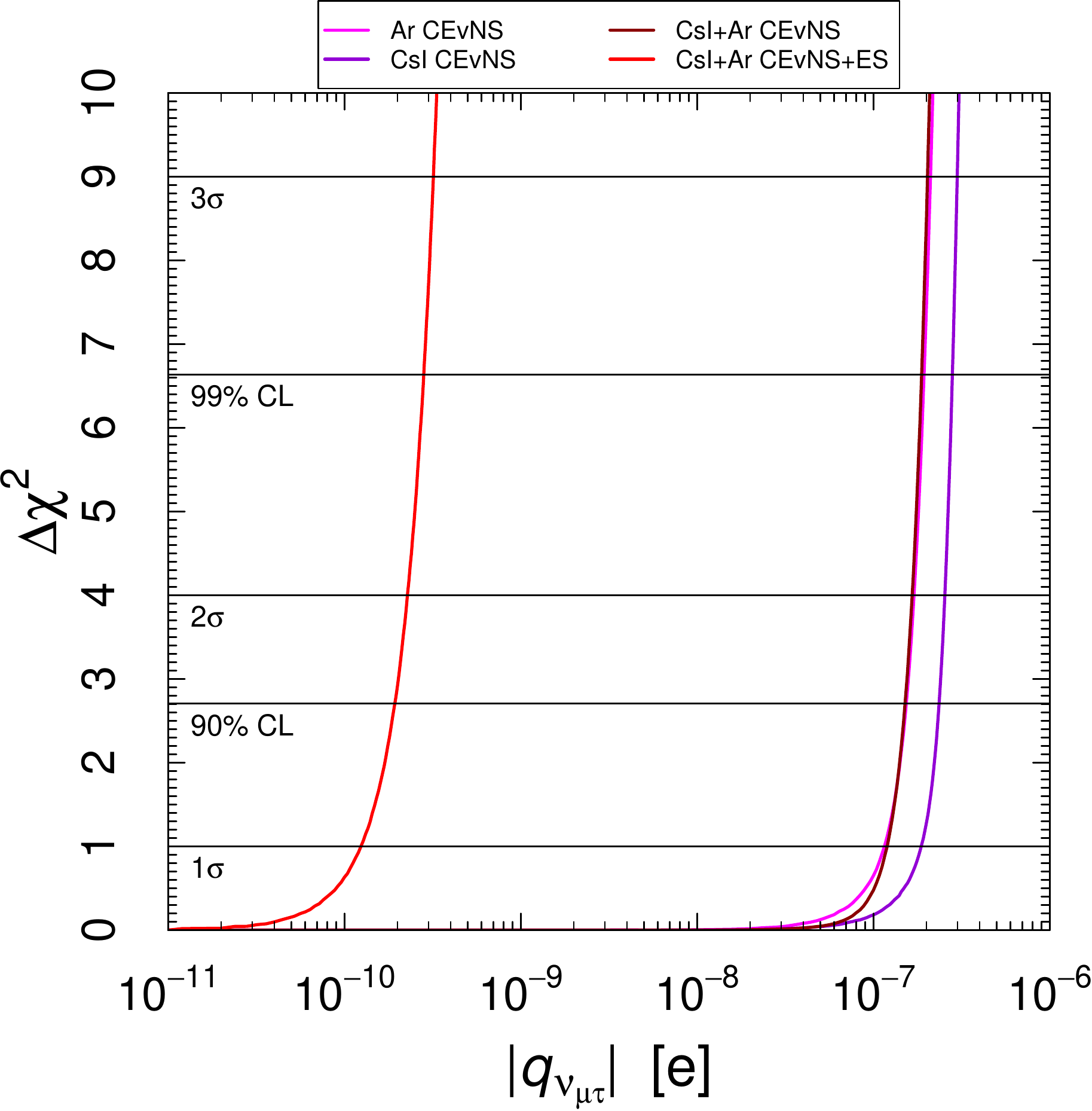}
\caption{ \label{fig:ech5-fig22-chi2-mm-mt}
Marginal $\Delta\chi^2$'s
for
$|q_{\nu_{\mu\mu}}|$
and
$|q_{\nu_{\mu\tau}}|$
obtained from:
the separate analyses of the COHERENT
Ar (magenta) and CsI (darkviolet)
data with CE$\nu$NS interactions;
the combined analyses of the COHERENT
Ar and CsI data
with CE$\nu$NS interactions only (dark red) and with CE$\nu$NS+ES interactions (red).
The short vertical orange and yellow lines show
the 90\% C.L. upper bounds on
$|q_{\nu_{\mu\mu}}|$
obtained,
respectively,
in Ref.~\cite{Das:2020egb}
from the LSND~\cite{Auerbach:2001wg} bound on $|\mu_{\nu_{\mu}}|$
and
in the XMASS-I experiment~\cite{XMASS:2020zke} from solar neutrino ES.
}
\end{center}
\end{figure}

Finally, in Fig.~\ref{fig:ech5-fig22-chi2-mm-mt}(a) and (b) we show similar marginal $\Delta\chi^2$'s
$|q_{\nu_{\mu\mu}}|$
and
$|q_{\nu_{\mu\tau}}|$, respectively, using COHERENT data only. Here, together with the various bounds obtained in this work we also show the 90\% C.L. upper bounds on
$|q_{\nu_{\mu\mu}}|$
obtained,
respectively,
in Ref.~\cite{Das:2020egb}
from the LSND~\cite{Auerbach:2001wg} bound on $|\mu_{\nu_{\mu}}|$
and
in the XMASS-I experiment~\cite{XMASS:2020zke} from solar neutrino ES.\footnote{Also in the case of the XMASS-I limit, that is the most stringent one for $|q_{\nu_{\mu\mu}}|$, the electron-neutrino cross section is derived using an ab-initio multi-configuration relativistic random phase approximation~\cite{XMASS:2020zke} that allows them to set more stringent limits.} Also in this case, the inclusion of the ES contribution significantly improves the bounds obtained for COHERENT, superseding the existing bounds from LSND concerning $|q_{\nu_{e\mu}}|$, while our bounds represent the only existing laboratory bounds for $|q_{\nu_{e\tau}}|$.

%%%%%%%%%%%%%%%%%%%%%%%%%%%%%%%%%%%%%%%%%%%%%%%%%%%%%%%%%%%%%%%%%%%

\subsection{Neutrino magnetic moment}

Finally, we study the bounds on the neutrino MM, namely on $|\mu_{\nu_{e}}|$ and $|\mu_{\nu_{\mu}}|$ using the COHERENT data and $|\mu_{\nu_{e}}|$ only using the Dresden-II data.
The results of our analysis for the neutrino MM are shown in Tab.~\ref{tab:tab22-mag2-CsI} and Tab.~\ref{tab:tab22-mag2-DII} for COHERENT CsI and Ar data set and for the Dresden-II data, respectively.
In both cases, we separate the scenarios in which ES is not considered, from those in which the ES contribution is added in the COHERENT CsI and the Dresden-II data set analyses. In the latter case, the different antineutrino fluxes and QFs are also considered.

By comparing Tab.~\ref{tab:tab22-mag2-CsI} and Tab.~\ref{tab:tab22-mag2-DII}, it is clear that the Dresden-II data allow us to significantly reduce the bound on $|\mu_{\nu_{e}}|$ with respect to COHERENT by more than one order of magnitude. Also in this case, the different antineutrino fluxes result in a negligible difference, while the two QFs produce a much more noticeable effect, with the Fef QF limits being almost a factor of two more precise. Finally, the inclusion of ES results in a marginal improvement of the Dresden-II limits of about $10\%$. At $90\%$ C.L., the bounds on the neutrino MM obtained in this work are
\begin{eqnarray}
    |\mu_{\nu_{e}}|& < & 2.13 \times 10^{-10} \, \mu_{\text{B}} \quad \mathrm{Dresden-II\, (CE\nu NS + ES)},\\
    |\mu_{\nu_{\mu}}| & < & 18 \times 10^{-10} \, \mu_{\text{B}} \quad \mathrm{CsI\, (CE\nu NS + ES) + Ar\, (CE\nu NS)},
\end{eqnarray}
where for the Dresden-II data the Fef QF has been considered. These limits can be compared with the bounds obtained in accelerator experiments with
$\nu_{\mu}-e$ scattering (see Table~IV of Ref.~\cite{Giunti:2014ixa}).
The most stringent is the LSND bound
$ |\mu_{\nu_{\mu}}| < 6.8 \times 10^{-10} \, \mu_{\text{B}}$
at 90\% CL~\cite{Auerbach:2001wg}, and that on $|\mu_{\nu_{e}}|$
established in reactor neutrino experiments, namely $
|\mu_{\nu_{e}}| < 2.9 \times 10^{-11} \, \mu_{\text{B}}
$~\cite{,Giunti:2014ixa,Tanabashi:2018oca}.

\begin{figure}
\centering
\subfigure[]{\label{fig:mag-fig22-chi2-ee}
\includegraphics[width=0.48\textwidth]{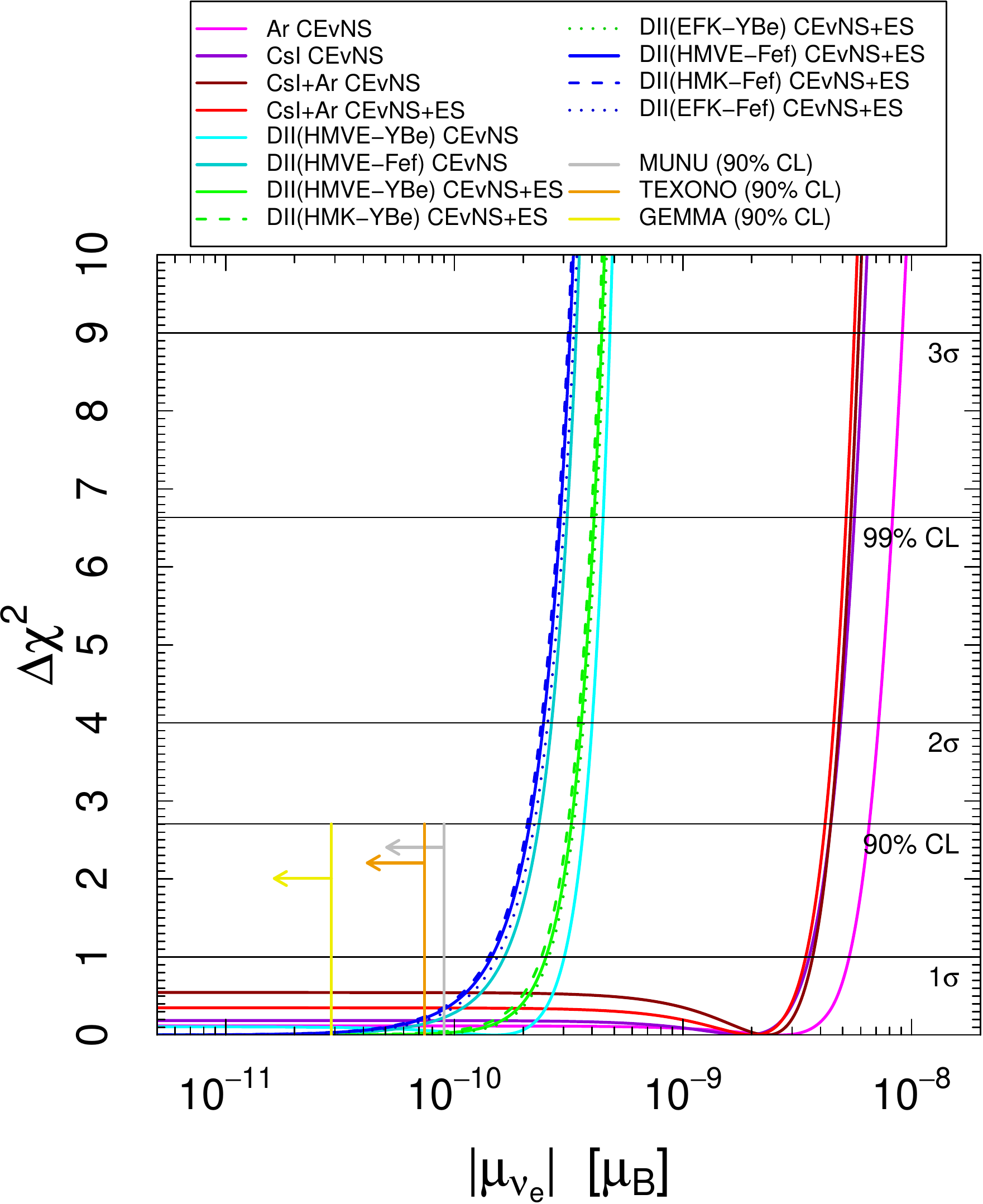}}
\subfigure[]{\label{fig:mag-fig22-chi2-mm}
\includegraphics[width=0.48\textwidth]{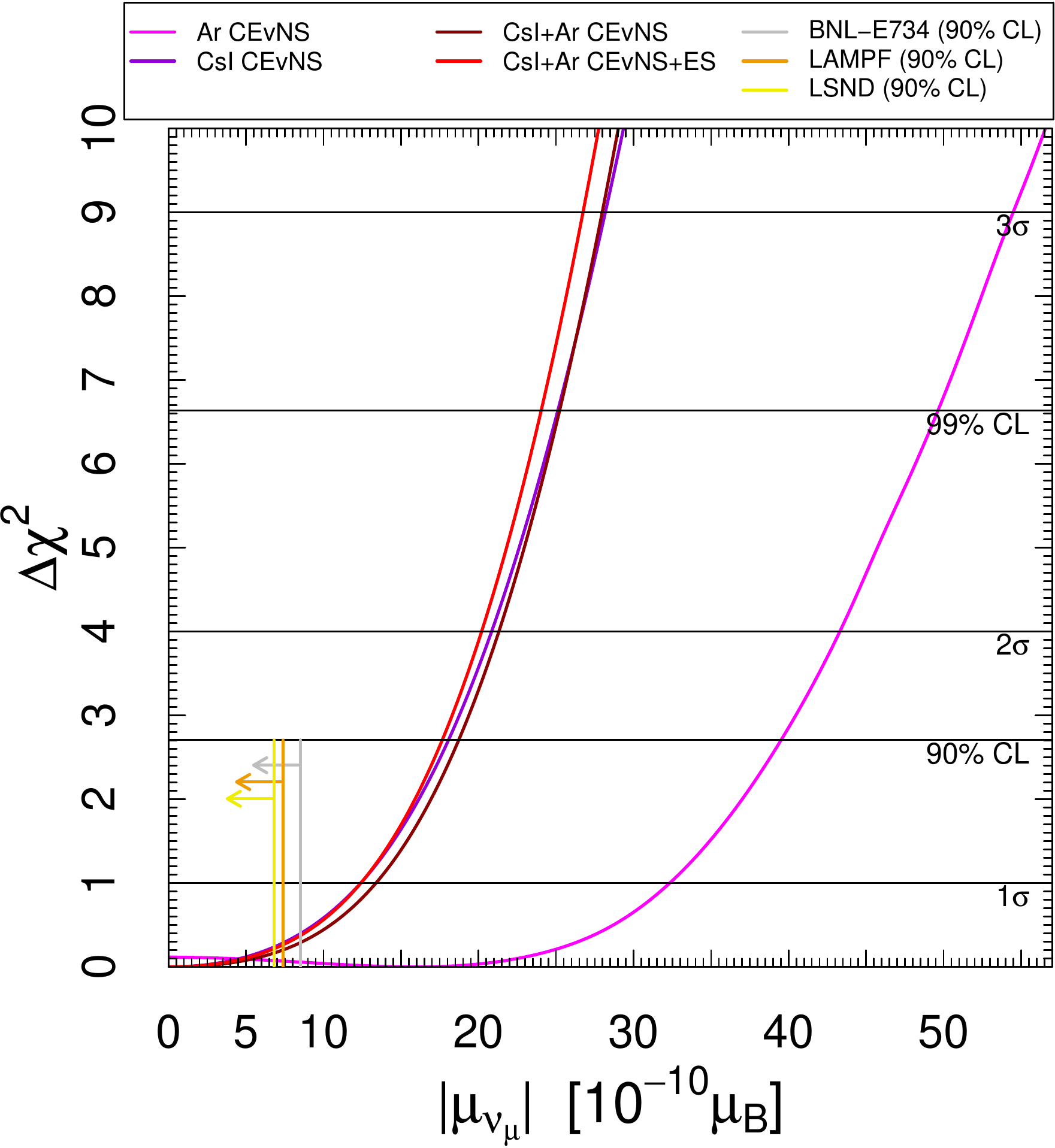}}
\caption{ \label{fig:mag-fig22-chi2}
Marginal $\Delta\chi^2$'s
for
\subref{fig:mag-fig22-chi2-ee}
$|\mu_{\nu_{e}}|$
and
\subref{fig:mag-fig22-chi2-mm}
$|\mu_{\nu_{\mu}}|$
obtained from:
the separate analyses of the COHERENT
Ar (magenta) and CsI (darkviolet)
data with CE$\nu$NS interactions;
the combined analyses of the COHERENT
Ar and CsI data
with CE$\nu$NS interactions only (dark red) and with CE$\nu$NS+ES interactions (red);
the CE$\nu$NS-only analyses of
Dresden-II data
assuming the HMVE reactor antineutrino flux and
the YBe (cyan) or Fef (dark cyan) quenching;
the CE$\nu$NS+ES analyses of
Dresden-II data
assuming the HMVE, HMK, or EFK reactor antineutrino flux and
the YBe (green) or Fef (blue) quenching.
The short vertical gray, orange, and yellow lines show, respectively,
the 90\% C.L. upper bounds on:
\subref{fig:mag-fig22-chi2-ee}
$|\mu_{\nu_{e}}|$
obtained
in the
MUNU~\cite{MUNU:2005xnz},
TEXONO~\cite{TEXONO:2006xds}, and
GEMMA~\cite{Beda:2012zz}
experiments;
\subref{fig:mag-fig22-chi2-mm}
$|\mu_{\nu_{\mu}}|$
obtained
in the
BNL-E734~\cite{Ahrens:1990fp},
LAMPF~\cite{Allen:1992qe}, and
LSND~\cite{Auerbach:2001wg}
experiments.
}
\end{figure}

In Fig.~\ref{fig:mag-fig22-chi2} we show the marginal $\Delta\chi^2$'s
for $|\mu_{\nu_{e}}|$
and $|\mu_{\nu_{\mu}}|$
obtained from the COHERENT
Ar and CsI data as well as their combination with the CE$\nu$NS-only analyses of
Dresden-II data
assuming the HMVE reactor antineutrino flux and
the YBe or Fef QF. We also show the impact of the ES contribution assuming the HMVE, HMK, or EFK reactor antineutrino flux and
the YBe or Fef QF.
For comparison, we also show
the 90\% C.L. upper bounds on
$|\mu_{\nu_{e}}|$
obtained
in the
MUNU~\cite{MUNU:2005xnz},
TEXONO~\cite{TEXONO:2006xds}, and
GEMMA~\cite{Beda:2012zz}
experiments; and 
$|\mu_{\nu_{\mu}}|$
obtained
in the
BNL-E734~\cite{Ahrens:1990fp},
LAMPF~\cite{Allen:1992qe}, and
LSND~\cite{Auerbach:2001wg}
experiments.

Before the completion of this work, other analyses also studying the \cenns impact on the neutrino MM appeared on the arXiv~\cite{Coloma:2022avw,Liao:2022hno}. Similar bounds to those found in this work for $|\mu_{\nu_{e}}|$ have been obtained, although with some differences among the various data analyses. Namely, in Ref.~\cite{Liao:2022hno} a bound at 90\% C.L. of $|\mu_{\nu_{e}}| < 2.7 \times 10^{-10} \, \mu_{\text{B}}$ is found when using a modified Lindahrd model for the QF with $k=0.157$ and ignoring the ES contribution. Similarly to Ref.~\cite{AristizabalSierra:2022axl} only the \cenns Dresden-II residuals after the subtraction of the background are fitted, with no background uncertainty propagated in the analysis. In Ref.~\cite{Coloma:2022avw}, a bound at 90\% C.L. of $|\mu_{\nu_{e}}| < 2.2 \times 10^{-10} \, \mu_{\text{B}}$ is found when using the Dresden-II data in combination with ES as in this work, also using the Fef QF. In this latter case, a very similar treatment of the Dresden-II data with respect to this work has been followed by the authors, with only minimal differences in the antineutrino flux treatment and least-squares function definition. 

\begin{table}
\begin{center}
%\resizebox{\textwidth}{!}
{
\setlength{\tabcolsep}{10pt}
\begin{tabular}{ccccc}
&
$1\sigma$
&
$90\%$
&
$2\sigma$
%&
%$99\%$
&
$3\sigma$
\\
\hline
\multicolumn{5}{c}{\bf CsI (CEvNS)}
\\
%\hline
$|\mu_{\nu_{e}}|$
&
$ < 36 $
&
$ < 44 $
&
$ < 49 $
%&
%$ < 56 $
&
$ < 62 $
\\
%\hline
$|\mu_{\nu_{\mu}}|$
&
$ < 12 $
&
$ < 18 $
&
$ < 21 $
%&
%$ < 25 $
&
$ < 28 $
\\
\hline
\multicolumn{5}{c}{\bf CsI (CEvNS+ES)}
\\
%\hline
$|\mu_{\nu_{e}}|$
&
$ < 32 $
&
$ < 41 $
&
$ < 46 $
%&
%$ < 53 $
&
$ < 58 $
\\
%\hline
$|\mu_{\nu_{\mu}}|$
&
$ < 11 $
&
$ < 17 $
&
$ < 19 $
%&
%$ < 24 $
&
$ < 27 $
\\
\hline
\multicolumn{5}{c}{\bf Ar (CEvNS)}
\\
%\hline
$|\mu_{\nu_{e}}|$
&
$ < 53 $
&
$ < 65 $
&
$ < 72 $
%&
%$ < 82 $
&
$ < 91 $
\\
%\hline
$|\mu_{\nu_{\mu}}|$
&
$ < 32 $
&
$ < 39 $
&
$ < 43 $
%&
%$ < 50 $
&
$ < 54 $
\\
\hline
\multicolumn{5}{c}{\bf CsI (CEvNS) + Ar (CEvNS)}
\\
%\hline
$|\mu_{\nu_{e}}|$
&
$< 37 $
&
$< 44 $
&
$< 48 $
%&
%$< 54 $
&
$< 59 $
\\
%\hline
$|\mu_{\nu_{\mu}}|$
&
$< 13 $
&
$< 19 $
&
$< 21 $
%&
%$< 25 $
&
$< 28 $
\\
\hline
\multicolumn{5}{c}{\bf CsI (CEvNS+ES) + Ar (CEvNS)}
\\
%\hline
$|\mu_{\nu_{e}}|$
&
$< 34 $
&
$< 42 $
&
$< 46 $
%&
%$< 52 $
&
$< 56 $
\\
%\hline
$|\mu_{\nu_{\mu}}|$
&
$< 12 $
&
$< 18 $
&
$< 20 $
%&
%$< 24 $
&
$< 27 $
\\
\hline
\end{tabular}

}
\end{center}
\caption{ \label{tab:tab22-mag2-CsI}
Bounds on the neutrino magnetic moments in units of $10^{-10}~\mu_{\text{B}}$
obtained from the analysis of the COHERENT CsI and Ar data.
We show the results of the analyses of CsI data with CE$\nu$NS only interactions and with CE$\nu$NS+ES interactions. 
}
\end{table}

\begin{table}
\begin{center}
%\resizebox{\textwidth}{!}
{
\begin{tabular}{cccccc}
&
$1\sigma$
&
$90\%$
&
$2\sigma$
%&
%$99\%$
&
$3\sigma$
&
Interaction
\\
\hline
\multicolumn{6}{c}{\bf Dresden-II (HMVE-Fef)}
\\
%\hline
\multirow{2}{*}{$|\mu_{\nu_{e}}|$}
&
$< 1.65$
&
$< 2.34$
&
$< 2.66$
%&
%$< 3.11$
&
$< 3.41$
&
CEvNS
\\
%\hline
%$|\mu_{\nu_{e}}|$
&
$< 1.45$
&
$< 2.13$
&
$< 2.45$
%&
%$< 2.90$
&
$< 3.20$
&
CEvNS+ES
\\
\hline
\multicolumn{6}{c}{\bf Dresden-II (HMK-Fef)}
\\
%\hline
\multirow{2}{*}{$|\mu_{\nu_{e}}|$}
&
$< 1.64$
&
$< 2.32$
&
$< 2.64$
%&
%$< 3.09$
&
$< 3.38$
&
CEvNS
\\
%\hline
%$|\mu_{\nu_{e}}|$
&
$< 1.41$
&
$< 2.08$
&
$< 2.40$
%&
%$< 2.85$
&
$< 3.15$
&
CEvNS+ES
\\
\hline
\multicolumn{6}{c}{\bf Dresden-II (EFK-Fef)}
\\
%\hline
\multirow{2}{*}{$|\mu_{\nu_{e}}|$}
&
$< 1.79$
&
$< 2.49$
&
$< 2.81$
%&
%$< 3.27$
&
$< 3.57$
&
CEvNS
\\
%\hline
%$|\mu_{\nu_{e}}|$
&
$< 1.54$
&
$< 2.23$
&
$< 2.56$
%&
%$< 3.02$
&
$< 3.32$
&
CEvNS+ES
\\
\hline
\multicolumn{6}{c}{\bf Dresden-II (HMVE-YBe)}
\\
%\hline
\multirow{2}{*}{$|\mu_{\nu_{e}}|$}
&
$< 3.02$
&
$< 3.68$
&
$< 4.00$
%&
%$< 4.47$
&
$< 4.79$
&
CEvNS
\\
%\hline
%$|\mu_{\nu_{e}}|$
&
$< 2.51$
&
$< 3.25$
&
$< 3.58$
%&
%$< 4.07$
&
$< 4.41$
&
CEvNS+ES
\\
\hline
\multicolumn{6}{c}{\bf Dresden-II (HMK-YBe)}
\\
%\hline
\multirow{2}{*}{$|\mu_{\nu_{e}}|$}
&
$< 2.98$
&
$< 3.64$
&
$< 3.96$
%&
%$< 4.42$
&
$< 4.75$
&
CEvNS
\\
%\hline
%$|\mu_{\nu_{e}}|$
&
$< 2.39$
&
$< 3.14$
&
$< 3.49$
%&
%$< 3.98$
&
$< 4.30$
&
CEvNS+ES
\\
\hline
\multicolumn{6}{c}{\bf Dresden-II (EFK-YBe)}
\\
%\hline
\multirow{2}{*}{$|\mu_{\nu_{e}}|$}
&
$< 3.16$
&
$< 3.84$
&
$< 4.16$
%&
%$< 4.63$
&
$< 4.94$
&
CEvNS
\\
%\hline
%$|\mu_{\nu_{e}}|$
&
$< 2.59$
&
$< 3.33$
&
$< 3.67$
%&
%$< 4.17$
&
$< 4.51$
&
CEvNS+ES
\\
\hline
\end{tabular}

}
\end{center}
\caption{ \label{tab:tab22-mag2-DII}
Bounds on the electron neutrino magnetic moment $|\mu_{\nu_{e}}|$
in units of $10^{-10}~\mu_{\text{B}}$
obtained from the analysis of the Dresden-II data
assuming the HMVE, HMK, or EFK reactor antineutrino flux and
the Fef or YBe quenching.
We show the results obtained with CE$\nu$NS only interactions and with CE$\nu$NS+ES interactions.
}
\end{table}

\section{Conclusions}\label{sec:conclusions}

In this paper, we describe the results of a combined analysis of all the \cenns data set so far available, profiting from the first observation of \cenns recently obtained with electron antineutrinos from the Dresden-II reactor site, using the NCC-1701 germanium detector.
Thanks to the much lower energy of reactor antineutrinos and the
low energy threshold of semiconductor detectors, these data provide complementary information
with respect to \cenns processes observed with neutrinos produced at spallation neutron sources, with a negligible dependence on the neutron distribution inside the target nuclei.

Following closely the instructions provided in the various data releases, we analysed the data collected with the CsI and Ar detectors by the COHERENT Collaboration and the recent data set provided by the Dresden-II reactor \cenns measurement. We focused in particular on the constraints on electroweak and neutrino electromagnetic properties, namely on the determination of the weak mixing angle and the neutrino magnetic moments, charge radii and millicharges. In the analysis of the Dresden-II reactor data we employ three different antineutrino fluxes, denoted as HMVE, HMK and EFK. We have also studied the dependence
of the results on the germanium quenching factor by considering two models: one based on the use of
iron-filtered monochromatic neutrons, indicated as Fef, and another one based on photo-neutron source measurements, indicated as YBe.
The impact of the various antineutrino fluxes on the results obtained is negligible, while the two quenching factors always result in visible differences in the obtained measurements and limits. This observation clearly underline the necessity of accurate measurements of the germanium quenching factor at low energies. Related to this, during the completion of this work an interesting study appeared~\cite{nuGeN:2022bmg} in which \cenns processes are searched for by the $\nu$GEN Collaboration using antineutrinos from the Kalinin Nuclear Power Plant and a germanium detector. While no \cenns excess is observed, assuming the SM the authors set an upper limit on the quenching parameter $k$ of the standard Lindhard model to be less than 0.177 at 90\% confidence level. 

Finally, in the analysis of both COHERENT and Dresden-II data, we evaluate the impact of the inclusion of the elastic neutrino-electron scattering contribution. Although in the SM this process contributes in a negligible way to the total event rate at low recoil energies, in certain scenarios beyond the SM the electron scattering contribution could increase significantly, making it important to consider. In particular, given that no electron-recoil discrimination is possible in the CsI and Dresden-II data set, we include the electron scattering contribution in both of them.

From an analysis of the Dresden-II data set alone, we are able to derive a new measurement of the weak mixing angle at low energies. The different antineutrino fluxes have a negligible impact, while the Fef and YBe quenching factors produce different results, namely
\begin{equation}\nonumber
  \sin^2{\vartheta_{\text{W}}}(\mathrm{Dresden-II\,Fef}) = 0.219^{+0.06}_{-0.05}\,(1\sigma),
  ^{+0.11}_{-0.08}\,(90\%),
  ^{+0.14}_{-0.09}\,(2\sigma),
\end{equation}
\begin{equation}\nonumber
  \sin^2{\vartheta_{\text{W}}}(\mathrm{Dresden-II\,YBe}) = 0.286^{+0.08}_{-0.07}\,(1\sigma),
  ^{+0.16}_{-0.11}\,(90\%),
  ^{+0.22}_{-0.13}\,(2\sigma),
\end{equation}
focusing thus only on the HMVE flux.

Thanks to a combined Dresden-II and COHERENT analysis, we are able to constrain different neutrino charge radii, namely $\langle{r_{\nu_{ee}}^{2}} \rangle$, $\langle{r_{\nu_{\mu\mu}}^{2}} \rangle$, $|\langle{r_{\nu_{e\mu}}^{2}} \rangle|$, $|\langle{r_{\nu_{e\tau}}^{2}} \rangle|$, and $|\langle{r_{\nu_{\mu\tau}}^{2}} \rangle|$.
Assuming the absence of the transition charge radii, we obtain a very competitive limit at 90\% C.L. with respect to that set by TEXONO when using the Fef quenching factor, namely
\begin{equation}\nonumber
    -7.1\times 10^{-32} \, \text{cm}^2<\langle{r_{\nu_{ee}}^{2}} \rangle<5\times 10^{-32} \, \text{cm}^2.
\end{equation}
In particular, we are able to restrict the upper bound limit from $6.6\times10^{-32} \, \text{cm}^2$ to $5\times10^{-32} \, \text{cm}^2$. No effect due to the inclusion of the electron scattering contribution is observed when fitting for the neutrino charge radii.

Furthermore, we set limits on five neutrino electric charges, namely $q_{\nu_{ee}}$, $q_{\nu_{\mu\mu}}$, $|q_{\nu_{e\mu}}|$, $|q_{\nu_{e\tau}}|$, and $|q_{\nu_{\mu \tau}}|$. In this case, the inclusion of the neutrino-electron scattering allows us to significantly improve the bounds obtained with both COHERENT and Dresden-II data. 
Thanks to the fact that the $|q|^2$ corresponding to neutrino-electron elastic scattering is much smaller than the \cenns $|q|^2$, the inclusion of the ES contribution allows us to achieve more stringent constraints with respect to \cenns only. 
Namely, with \cenns only there is an improvement with respect to COHERENT \cenns only of about 2 orders of magnitude, while with CE$\nu$NS+ES the improvement is of about 4 orders of magnitude. Intriguingly, the bounds on $q_{\nu_{ee}}$ obtained from the combination of COHERENT with the Dresden-II CE$\nu$NS+ES data set are much more stringent than the COHERENT ones and the \cenns only fit, namely at 90\% C.L. and using the Fef quenching factor
\begin{equation}\nonumber
    -9.3\times10^{-12}\, e<q_{\nu_{ee}}<9.5\times10^{-12}\, e.
\end{equation}
This limit is competitive with respect to the other existing bounds,
that are also at the level of $10^{-12} \, e$.

Finally, we presented the bounds on the neutrino magnetic moments, namely on $|\mu_{\nu_{e}}|$ and $|\mu_{\nu_{\mu}}|$ using the COHERENT data and $|\mu_{\nu_{e}}|$ only using the Dresden-II data.
At $90\%$ C.L., the bounds on the neutrino magnetic moments obtained in this work are
\begin{eqnarray}\nonumber
    |\mu_{\nu_{e}}|& < & 2.13 \times 10^{-10} \, \mu_{\text{B}} \quad \mathrm{Dresden-II\, (CE\nu NS + ES)},\\\nonumber
    |\mu_{\nu_{\mu}}| & < & 18 \times 10^{-10} \, \mu_{\text{B}} \quad \mathrm{CsI\, (CE\nu NS + ES) + Ar\, (CE\nu NS)},
\end{eqnarray}
where for the Dresden-II data the Fef QF has been considered. These limits are still less stringent than the bounds obtained in reactor and accelerator neutrino experiments.
\\

As evident from the results described in this work, the \cenns process proved to be once again a spectacular window to test many and diverse sectors, with precision that are competitive to, if not better than, the existing ones. Thus, we strongly encourage all existing and foreseen experimental efforts in this sector, using neutrinos and antineutrinos both from spallation neutron sources and reactor sites.

\begin{acknowledgements}
The work of C. Giunti  and C.A. Ternes is supported by the research grant "The Dark Universe: A Synergic Multimessenger Approach" number 2017X7X85K under the program PRIN 2017 funded by the Ministero dell'Istruzione, Universit\`a e della Ricerca (MIUR).
The work of Y.F. Li and Y.Y. Zhang is supported in part by the National Natural Science Foundation of China under Grant Nos.~12075255, 12075254 and 11835013, and by the Key Research Program of the Chinese Academy of Sciences under Grant No.~XDPB15.
The work of Y.Y. Zhang is also supported by China Postdoctoral Science Foundation under Grant No.~2021T140669.
%Y.F. Li is also grateful for the support by the CAS Center for Excellence in Particle Physics (CCEPP).
%The work of Y.Y. Zhang is supported by the National Natural Science Foundation of China under Grant No.~12075254 and also No.~11835013.
\end{acknowledgements}

%\newpage
%\bibliographystyle{apsrev4-1}
\bibliography{ref}
%BIB-2

\end{document}